C. J. PAPACHRISTOU

# ASPECTS OF RELATIVITY
# IN FLAT SPACETIME

# Aspects of Relativity in Flat Spacetime

**Costas J. Papachristou**

*Department of Physical Sciences*
*Hellenic Naval Academy*

*To the memory of*

*B. Kent Harrison,*

*relativist and inspiring teacher*

# PREFACE

In the world around us, as well as in the arts, *symmetry* is a property that enhances beauty. In Physics, however, symmetry is not just a matter of aesthetics! Indeed, it is a dynamical aspect of most (if not all) physical theories, where it is realized as *invariance* under certain sets of transformations. In some cases the requirement of symmetry leads to physical principles such as conservation laws, as Emmy Noether's beautiful theorem has shown. Also, the *covariance* (form-invariance) of physical laws upon passing from one inertial frame of reference to another opens the gate to high-energy physics. The main aspects of this latter kind of symmetry, which is at the heart of *Special Relativity* (SR), is the subject of this short book.

Symmetry transformations constitute the most important topic in group theory. So, a proper study of SR should include at least an elementary study of the *Lorentz group*. After the basic "philosophical" ideas regarding SR have been briefly discussed in Chapter 1, the Lorentz group is presented in Chap. 2 in its own right, as a mathematical entity not yet directly associated with SR. The connection of this group with relativistic mechanics is the subject of Chap. 3, while Chap. 4 deals with Lorentz covariance in Maxwell's theory of electrodynamics.

Several special topics are discussed in Chap. 5. Section 5.1 serves as a brief introduction to Lie groups and Lie algebras, some knowledge of which is a prerequisite for Chap. 2. Section 5.2 focuses on the Lorentz group and its homomorphism with the group $SL(2,C)$, while Sec. 5.3 discusses the concept of the metric in flat and curved (Riemannian) spaces. Section 5.4 presents some relatively recently published ideas supporting the view that Maxwell's equations, seen as a Bäcklund transformation, form a system of independent equations. This independence is particularly significant with regard to the coherence of the covariant formulation of the Maxwell system.

To make the book suitable for self-study, all problems at the end of Chaps. 3 and 4 are accompanied by detailed solutions.

The reader is assumed to have some acquaintance with SR at the basic (undergraduate) level. Knowledge of basic electrodynamics is also necessary at the level, e.g., of this author's textbook *"Introduction to Electromagnetic Theory and the Physics of Conducting Solids"* (Springer, 2020). In fact, the present book may be considered as a supplement to the aforementioned one, albeit at a somewhat more advanced level. Finally, some previous knowledge of group theory may be helpful but is not required for reading this book, given that, as mentioned above, the necessary ideas regarding groups and algebras are presented in Sec. 5.1.

<div align="right">Costas J. Papachristou<br>July 2025</div>

# CHAPTER 1

# OVERVIEW OF THE BASIC IDEAS

## 1.1 Why Relativity?

Einstein's *Special Theory of Relativity* (SR) assumes the existence of an underlying 4-dimensional *flat* spacetime – i.e., a spacetime admitting a *globally* constant (albeit non-Euclidean) metric – which has a preferred class of frames of reference, namely, *inertial frames*. (This is analogous to the existence of global Cartesian systems of coordinates in a standard Euclidean space.) According to the *special* (or *restricted*) *principle of relativity*, all inertial frames are equivalent to each other with regard to describing physical phenomena. This suggests that all physical laws must be expressed in *covariant forms*, i.e., in forms that are invariant upon passing from one inertial frame to another. We are thus in search of two things:

(*a*) the proper coordinate transformations that relate inertial frames to one another;

(*b*) the proper mathematical statements of physical laws so that these laws be form-invariant under the aforementioned transformations.

In classical mechanics, invariance of mechanical laws is established by means of the *Galilean transformation* (GT). The GT, however, fails to satisfy the invariance requirement for Maxwell's equations of electromagnetism. In particular, the value of the speed of light, which is a direct consequence of these equations, is not an invariant quantity under the GT. It is found experimentally, however, that the speed of light *is* an invariant, the same for all inertial observers. This suggests that either Maxwell's equations are not correct – thus need to be corrected in order to comply with the GT – or the GT itself is not correct and a new transformation is needed that makes the Maxwell equations frame-invariant and, in particular, treats the speed of light as a constant of the theory, independent of any particular frame of reference.

Einstein chose to accept the second possibility, which eventually leads to the replacement of the GT with the *Lorentz transformation* (LT) and to the requirement that all physical laws be expressible in Lorentz-invariant (or covariant) forms. The Maxwell equations are already consistent with this requirement, by construction of the LT. The laws of mechanics, however, which are Galilean-invariant, need to be reformulated in order to comply with the LT. For example, a redefinition of momentum in relativistic form is required in order for the law of conservation of momentum to be Lorentz-invariant.

As for the LT itself, it is defined as a *linear* transformation that preserves the value of the elementary spacetime interval

$$ds^2 = c^2 dt^2 - dx^2 - dy^2 - dz^2$$

where $(x, y, z, t)$ are the spacetime coordinates and where $c$ is the (Lorentz-invariant) speed of light in empty space. In particular, for $ds^2 = 0$, the invariance of $ds^2$ under a LT is equivalent to the invariance of the speed of light. Moreover, the invariant spacetime interval $ds^2$ endows the spacetime of SR with a metric, represented by the diagonal (4×4) matrix $g = [g_{\mu\nu}] = diag(1, -1, -1, -1)$, where $\mu, \nu = 0,1,2,3$.

Since the metric elements $g_{\mu\nu}$ are constant quantities having the same values at all spacetime points, the underlying spacetime is flat (see Sec. 5.3). This is true as long as gravity may be ignored. In the presence of gravity the flat spacetime of SR must be replaced by the *curved* spacetime of *General Relativity* (GR). In GR the notion of a global inertial frame of reference has no meaning and the *general principle of relativity* requires invariance of physical laws under all spacetime transformations (not just a restricted class such as the LTs, which are specifically associated with the inertial frames of SR).

## 1.2 Inertial Frames of Reference

In SR, as in classical mechanics, an *inertial frame of reference* is any system of coordinates or axes – say, $(x, y, z)$ – relative to which a *free* particle (i.e., a particle subject to no interactions) remains at rest or moves uniformly (that is, with constant velocity and hence with no acceleration). An observer using an inertial frame (relative to which she is at rest) is called an *inertial observer* and is herself subject to no net external interaction.

Let us look at this last statement in more detail: Consider two observers $O_1$ and $O_2$ located at the corresponding origins of two inertial frames. Consider also a free particle $P$. Then $P$ will move with constant velocity relative to both $O_1$ and $O_2$ . It follows that $O_1$ and $O_2$ will move uniformly relative to each other. In particular, since the observer $O_2$ is moving with constant velocity relative to an inertial frame, she must be a free "particle". By the same token, observer $O_1$ will also be a free "particle" given that he moves uniformly relative to the frame of $O_2$ . As a corollary, two free particles move with constant velocities (are not accelerating) relative to each other. Classically, this is one way to express Newton's first law of mechanics [1].

Time in Galilean relativity has a universal meaning, independent of any particular observer. On the contrary, in SR time is relative and depends on the motion of one observer relative to another. Thus the space $(x, y, z)$ of classical mechanics is enhanced to a 4-dimensional spacetime with coordinates $(x, y, z, t)$ or, for the purpose of dimensional homogeneity, $(x, y, z, ct)$. The coordinates $(x, y, z)$ correspond to the spatial axes of the frame of reference used by an inertial observer, while $t$ is the time of occurrence of events as determined by that observer.

The trajectory of a particle in spacetime is called the *worldline* of the particle. This line describes the position of the particle as a function of time and is mathematically expressed by the functions $x(t)$, $y(t)$, $z(t)$. Geometrically, a worldline is the plot of position versus time in a coordinate system $(x, y, z, ct)$. The graph of the system of equations $\{x=x(t), y=y(t), z=z(t)\}$ in a system of axes $(x, y, z, ct)$ is a curve in 4-dimensional spacetime.

A free particle moves with constant velocity in any inertial frame. The coordinates $(x, y, z)$ of the particle are therefore linear functions of $t$ and the particle's worldline is a *straight* line. An inertial frame may thus be defined as a system of spacetime coordinates (or axes) in which the worldlines of free particles are straight lines.

The Lorentz transformation (LT), which transforms *both* space and time coordinates of a particle, ensures that uniform motion in one inertial frame transforms into uniform motion in any other inertial frame. Since uniform motions are described geometrically by straight worldlines, it follows that a LT transforms straight worldli-nes in a spacetime coordinate system $(x, y, z, ct)$ into straight lines in some other coordinate system $(x', y', z', ct')$. This requires the LT connecting the two systems of coordinates to be a *linear* transformation [2].

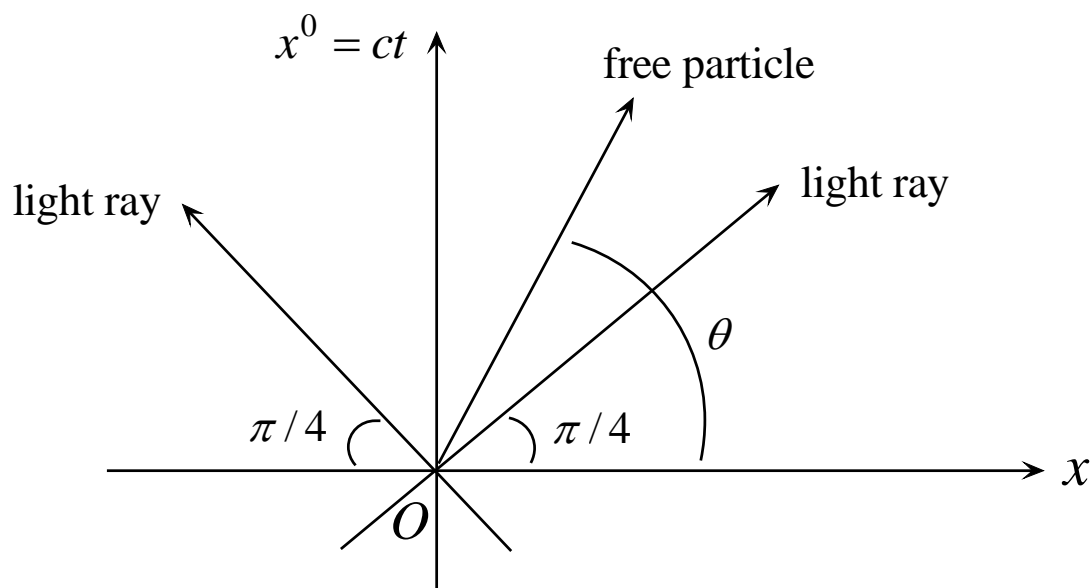

Fig. 1.1. Worldlines of two light rays and a free particle; the former define a 2-dimensional light cone.

The spacetime diagram in Fig. 1.1 shows the worldlines of two light rays and a free particle. We set $x^0 = ct$ and we assume that both the particle and the light rays are traveling along the $x$-axis (the two rays in opposite directions), so that $y$ and $z$ are constant. By making this choice we effectively reduce the spacetime dimensions from 4 to 2.

The velocity $v = dx/dt$ of a free particle is constant, where $v$ may be positive or negative in accordance with the direction of motion. For a light ray, $v = dx/dt = \pm c$. In either case,

$$\cot\theta = \frac{dx}{dx^0} = \frac{dx}{d(ct)} = \frac{v}{c}$$

where $\theta$ is the angle formed by the worldline and the $x$-axis. For a light ray, $\cot\theta = \pm 1$ $\Rightarrow$ $\theta = \pi/4$ or $\theta = 3\pi/4$. For a free particle, $|v| < c \Rightarrow |\cot\theta| < 1 \Rightarrow \pi/4 < \theta < 3\pi/4$.

The latter result, which follows from the well-known fact that no speed in Nature can exceed the speed of light, has the following geometrical interpretation: The worldline of a free particle – and, indeed, the worldline of *any* massive particle – must lie in the interior of the *light cone* formed by the possible worldlines of a light ray (see Fig. 1.1). Of course, for an *accelerating* (thus non-free) particle the angle $\theta$ is not constant (since $v$ is not constant) and the associated worldline cannot be a straight line (it is generally curved). At all points of the worldline, however, the tangent line must be such that its angle $\theta$ with the $x$-axis conforms to the condition $\pi/4 < \theta < 3\pi/4$. This result can be generalized for light cones in higher spacetime dimensions.

## CHAPTER 2

# THE LORENTZ GROUP

### 2.1 The Group $SO(3,1)\uparrow$

In this chapter we discuss the Lorentz group in its own right, independently of its role in Relativity. Before we proceed, let us note that the reader who is not familiar with group theory – and, in particular, with Lie groups and Lie algebras – may find it useful to consult Sec. 5.1. Additional material on the Lorentz group can be found in Sec. 5.2.

Let $a$ be a vector in $R^4$, with components $a^\mu$ ($\mu$=1,2,3,4). This vector may be represented by a (4×1) matrix (column vector)

$$a \equiv [a^\mu] = \begin{bmatrix} a^1 \\ a^2 \\ a^3 \\ a^4 \end{bmatrix}.$$

Then $a^t$ (transpose of $a$) is the row vector

$$a^t = \begin{bmatrix} a^1 & a^2 & a^3 & a^4 \end{bmatrix}.$$

We introduce the symmetric (4×4) matrix

$$g \equiv [g_{\mu\nu}] = \begin{bmatrix} 1 & 0 & 0 & 0 \\ 0 & 1 & 0 & 0 \\ 0 & 0 & 1 & 0 \\ 0 & 0 & 0 & -1 \end{bmatrix} = diag\,(1,1,1,-1) \qquad (2.1)$$

where $\mu,\nu$=1,2,3,4 and where we have used a standard notation for diagonal matrices. Given two vectors $a$ and $b$ in $R^4$, we define the *scalar product* $(a,b)$ of $a$ and $b$ by the relation

$$(a,b) = a^t g b \qquad (2.2)$$

In terms of components, and by using the familiar summation convention of summing from 1 to 4 over repeated up and down indices, we rewrite (2.2) as

$$(a,b) = a^\mu g_{\mu\nu} b^\nu = g_{\mu\nu} a^\mu b^\nu = a^1 b^1 + a^2 b^2 + a^3 b^3 - a^4 b^4 \qquad (2.3)$$

As a consequence of the symmetry of $g$, i.e., since $g^t$=$g \Leftrightarrow g_{\mu\nu}$= $g_{\nu\mu}$, the scalar product is symmetric: $(a,b) = (b,a)$, as is obvious from the right-hand-side of (2.3).

Now, let $\Lambda \equiv [\Lambda^{\mu}{}_{\nu}]$ be a constant (4×4) matrix with real elements $\Lambda^{\mu}{}_{\nu}$ ($\mu, \nu$=1,2,3,4). For any vector $a \equiv [a^{\mu}] \in R^4$ we consider the homogeneous linear transformation

$$a \to a' = \Lambda a \quad \Leftrightarrow \quad a^{\mu\prime} = \Lambda^{\mu}{}_{\nu}\, a^{\nu} \tag{2.4}$$

We assume that $\Lambda$ is such that, for any vectors $a, b \in R^4$, the transformation (2.4) leaves the scalar product $(a,b)$ invariant. That is, by (2.2) and (2.4),

$$(a', b') = (\Lambda a, \Lambda b) = (a,b) \;\Rightarrow\; a^t (\Lambda^t g \Lambda) b = a^t g b \,.$$

For this to be true for all $a, b$ we must have

$$\Lambda^t g \Lambda = g \quad \Leftrightarrow \quad \Lambda^{\mu}{}_{\lambda}\, g_{\mu\nu}\, \Lambda^{\nu}{}_{\rho} = g_{\lambda\rho} \tag{2.5}$$

with the understanding that $\Lambda^{\mu}{}_{\lambda} = (\Lambda^t)_{\lambda}{}^{\mu}$.

We notice that $\det(\Lambda^t g \Lambda) = \det g \;\Rightarrow\; (\det\Lambda)^2 = 1$. This can be satisfied in two ways:

$\det\Lambda = +1 \;\Rightarrow$ the transformation (2.4) is a *proper* transformation; or

$\det\Lambda = -1 \;\Rightarrow$ the transformation (2.4) is an *improper* transformation.

Now, assume that all transformation matrices $\Lambda$ can be obtained from the identity transformation $\Lambda=1 \Leftrightarrow \Lambda^{\mu}{}_{\nu} = \delta_{\mu\nu}$ by continuously varying a certain set of parameters on which these matrices depend. Clearly, only *proper* transformations can be connected to the identity in this fashion. These transformations form a *group* [1,2] named $SO(3,1)$, in accordance with the number of plus and minus signs in the diagonal elements of the matrix $g$ in (2.1).

To verify the group property of $SO(3,1)$, let us note the following:

1. The set of $SO(3,1)$ matrices is closed under the operation of matrix multiplication. Indeed, let $\Lambda_1, \Lambda_2 \in SO(3,1)$ and call $\Lambda = \Lambda_1\Lambda_2$. Both $\Lambda_1$ and $\Lambda_2$ satisfy the condition (2.5) and we must show that the same is true for $\Lambda$. We have:

$$\Lambda^t g \Lambda = (\Lambda_1\Lambda_2)^t g\, (\Lambda_1\Lambda_2) = (\Lambda_2)^t\, [(\Lambda_1)^t g\, \Lambda_1]\, \Lambda_2 = (\Lambda_2)^t g\, \Lambda_2 = g \,.$$

Moreover, $\det\Lambda = \det(\Lambda_1\Lambda_2) = (\det\Lambda_1)(\det\Lambda_2) = (+1)(+1) = +1$.

2. If $\Lambda \in SO(3,1)$, then $\Lambda^{-1} \in SO(3,1)$ also. Indeed, by using (2.5) and the fact that $g^{-1} = g$, we have:

$$(\Lambda^t g \Lambda)^{-1} = g \;\Rightarrow\; \Lambda^{-1} g\, (\Lambda^{-1})^t = g \;\Rightarrow\; g\, (\Lambda^{-1})^t = \Lambda g \;\Rightarrow$$
$$(\Lambda^{-1})^t = g \Lambda g \;\Rightarrow\; (\Lambda^{-1})^t g = g\Lambda \;\Rightarrow\; (\Lambda^{-1})^t g\, \Lambda^{-1} = g \,.$$

Moreover, $\det(\Lambda^{-1}) = (\det\Lambda)^{-1} = 1$.

However, not all elements of the group $SO(3,1)$ can be connected to the identity $\Lambda=1$ in a continuous way. Indeed, an additional condition must be satisfied. Setting $\lambda=\rho=4$ in (2.5), we have:

$$\Lambda^{\mu}{}_{4}\, g_{\mu\nu}\, \Lambda^{\nu}{}_{4} = g_{44} = -1$$

from which we get

$$\left(\Lambda^4{}_4\right)^2 = 1+\sum_{i=1}^{3}\left(\Lambda^i{}_4\right)^2$$

(show this). This suggests that

$$|\Lambda^4{}_4| \geq 1 \;\;\Rightarrow\;\; \Lambda^4{}_4 \geq 1 \;\text{ or }\; \Lambda^4{}_4 \leq -1 \;.$$

Since $\Lambda^4{}_4 = 1$ for the identity transformation, connectivity with the identity requires that all *proper* transformations have $\Lambda^4{}_4 \geq 1$.

It can be shown [2] that the set of proper transformations $\Lambda$ having the additional property $\Lambda^4{}_4 \geq 1$ is a *subgroup* of $SO(3,1)$, called the *restricted Lorentz group* and denoted $SO(3,1)\uparrow$. For the purpose of notational simplicity, in what follows we will denote this group by $L$.

## 2.2 The Lie Algebra of the Lorentz Group

The restricted Lorentz group $L$ is a *Lie group*, the elements of which depend on 6 real parameters. The associated *Lie algebra*, named $so(3,1)$, is thus 6-dimensional. An element $\Lambda\in L$ can be written as

$$\Lambda = e^{\omega} \equiv \exp\omega \tag{2.6}$$

for some (4×4) matrix $\omega\in so(3,1)$, where

$$\exp\omega \equiv \sum_{n=0}^{\infty}\frac{\omega^n}{n!} = 1+\omega+\frac{\omega^2}{2}+\cdots \;.$$

By (2.6) the matrix $\omega$ inherits certain properties from $\Lambda$:

1. From the fact that $\det\Lambda=1$, and by using the matrix property $\det(e^{\omega}) = e^{tr\omega}$, we have that $\det\Lambda= \det(e^{\omega}) = e^{tr\omega} = 1 \;\Rightarrow$

$$tr\,\omega = 0 \tag{2.7}$$

That is, the matrix $\omega\in so(3,1)$ is *traceless*.

2. From (2.5) and (2.6) we have:

$$(e^{\omega})^t g\, e^{\omega} \equiv e^{\omega^t} g\, e^{\omega}= g \Rightarrow e^{\omega^t} = g\, e^{-\omega} g^{-1} \equiv e^{-g\omega g^{-1}} \Rightarrow \omega^t = -\,g\omega g^{-1} \Rightarrow$$

$$\omega^t g + g\,\omega = 0 \tag{2.8}$$

Note that, since $g^t = g$, the above relation is written

$$(g\,\omega)^t + g\,\omega = 0 \;\;\Leftrightarrow\;\; (g\,\omega)^t = -\,g\,\omega \tag{2.9}$$

That is, the matrix $g\,\omega$ is *antisymmetric*.

Matrices $\omega$ satisfying the properties (2.7) - (2.9) can be expressed in the following parametric form:

$$\omega = \begin{bmatrix} 0 & -\alpha_3 & \alpha_2 & \beta_1 \\ \alpha_3 & 0 & -\alpha_1 & \beta_2 \\ -\alpha_2 & \alpha_1 & 0 & \beta_3 \\ \beta_1 & \beta_2 & \beta_3 & 0 \end{bmatrix} \tag{2.10}$$

where $\alpha_i$ , $\beta_i$ ($i$=1,2,3) are 6 real parameters. The matrix $\omega$ can be written as a linear combination

$$\omega = \sum_{i=1}^{3} (\alpha_i A_i + \beta_i B_i) \tag{2.11}$$

where the 6 matrices $A_i$ , $B_i$ ($i$=1,2,3) form the basis of the Lie algebra *so*(3,1) of the group *L*. These matrices can be read-off from (2.10), by using (2.11). For example,

$$A_1 = \begin{bmatrix} 0 & 0 & 0 & 0 \\ 0 & 0 & -1 & 0 \\ 0 & 1 & 0 & 0 \\ 0 & 0 & 0 & 0 \end{bmatrix}, \text{ etc.}; \quad B_1 = \begin{bmatrix} 0 & 0 & 0 & 1 \\ 0 & 0 & 0 & 0 \\ 0 & 0 & 0 & 0 \\ 1 & 0 & 0 & 0 \end{bmatrix}, \text{ etc.}$$

*Exercise:* Write the remaining 4 matrices $A_2$ , $A_3$ , $B_2$ , $B_3$..

The *commutation relations* of the algebra *so*(3,1) are

$$[A_i , A_j] = \sum_{k=1}^{3} \varepsilon_{ijk} A_k \tag{2.12$a$}$$

$$[B_i , B_j] = -\sum_{k=1}^{3} \varepsilon_{ijk} A_k \tag{2.12$b$}$$

$$[A_i , B_j] = -\sum_{k=1}^{3} \varepsilon_{ijk} B_k \tag{2.12$c$}$$

where $\varepsilon_{123} = \varepsilon_{231} = \varepsilon_{312} = 1$ , $\varepsilon_{213} = \varepsilon_{132} = \varepsilon_{321} = -1$ , and $\varepsilon_{ilk} = 0$ in all other cases. By $[M, N]=MN-NM$ we denote the *commutator* of two matrices $M, N$.

In the context of Relativity (to be studied in the next chapter) the *generators* $A_i$ , $B_i$ of transformations $\Lambda\in L$ admit a certain geometrical interpretation. The $\{A_i\}$ generate *rotations* of the system of spatial axes. (Careful: by "rotation" we mean redefinition of orientation of the system in space, *not* any kind of rotational motion!) The $\{B_i\}$ generate *boosts*, which physically represent uniform motions (without change of orientation) of the system of spatial axes, along the corresponding three axes. We note the following:

1. According to (2.12$a$), rotations are closed and form a subgroup [namely, *SO*(3)] of the Lorentz group $L=SO(3,1)\uparrow$.

2. According to (2.12*b*), boosts are not closed and do *not* form a subgroup of $L$ (except for boosts along the same axis, which *do* form a 1-parameter subgroup of $L$).

3. According to (2.12*c*), the rotation group $SO(3)$ is not an *invariant* subgroup of $L$ since the Lie algebra $so(3)$, with basis $\{A_i\}$, is not an invariant subalgebra (or *ideal*; cf. Sec. 5.1) of $so(3,1)$. This is related to the fact that the commutators $[A_i, B_j]$ are not linear combinations of the basis vectors $\{A_k\}$. Technically speaking, the absence of an ideal suggests that the Lie algebra $so(3,1)$ is *simple* [1].

Finally, we note that, for infinitesimal values of the parameters $\alpha_i, \beta_i$ appearing in (2.10), the matrix $\omega$ is infinitesimal and $e^{\omega} \simeq 1+\omega$. Then, by using (2.11), relation (2.6) reduces to the *infinitesimal Lorentz transformation*

$$\Lambda \simeq 1+\omega = 1+\sum_{i=1}^{3}(\alpha_i A_i + \beta_i B_i) \ .$$

## CHAPTER 3

# RELATIVISTIC TRANSFORMATIONS

### 3.1 Lorentz Transformations in Relativistic Spacetime

The group-theoretical ideas presented in the previous chapter will now be applied to Special Relativity (SR), albeit with some notational revision regarding the numbering of vector components.

In Chapter 1 we defined an *inertial frame of reference* as a system of coordinates (or axes) relative to which a free particle moves with constant velocity (thus no acceleration). This reference frame is chosen to be a 3-dimensional Cartesian system of axes $(x, y, z)$. An observer using this frame (relative to which she/he is at rest) is called an *inertial observer*.

An *event* is something that occurs at a certain point $(x, y, z)$ at a certain time $t$, as measured by an inertial observer in her/his own frame. The set of all possible events constitutes 4-dimensional *spacetime* with coordinates $(x, y, z, t)$ or $(x, y, z, ct)$, the latter choice being made for the purpose of dimensional homogeneity (the constant $c$ is, of course, the speed of light in empty space, the value of which speed is independent of any particular frame of reference).

Vectors in spacetime are 4-component objects. In SR it is customary to rename the fourth component of a vector $a$ as "zero" component and write

$$a^{\mu} \equiv (a^0, a^1, a^2, a^3) \equiv (a^0, \vec{a}) \tag{3.1}$$

where $\mu$=0,1,2,3 and where $\vec{a} \equiv (a_x, a_y, a_z)$ is a vector in $R^3$. As for the spacetime coordinates, we write

$$x^{\mu} \equiv (x^0, x^1, x^2, x^3) \equiv (ct, x, y, z) \tag{3.2}$$

The $a^{\mu}$ and, likewise, the $x^{\mu}$ may be regarded as elements of column vectors $a \equiv [a^{\mu}]$ and $X \equiv [x^{\mu}]$, respectively. Then $a^t$ and $X^t$ are the corresponding row vectors.

The matrix $g$ introduced in Chapter 2, which in SR plays the role of a *metric tensor* (cf. Sec. 5.3), will be rewritten here as

$$g \equiv [g_{\mu\nu}] = \begin{bmatrix} 1 & 0 & 0 & 0 \\ 0 & -1 & 0 & 0 \\ 0 & 0 & -1 & 0 \\ 0 & 0 & 0 & -1 \end{bmatrix} = diag\,(1,-1,-1,-1) \tag{3.3}$$

$(\mu, \nu$=0,1,2,3$)$. For any two vectors $a \equiv [a^{\mu}]$ and $b \equiv [b^{\mu}]$ we define the *scalar product*

$$(a,b) = a^t g\, b = g_{\mu\nu}\, a^{\mu} b^{\nu} = a^0 b^0 - a^1 b^1 - a^2 b^2 - a^3 b^3 \tag{3.4}$$

(note the use of the summation convention in $\mu$ and $\nu$). In particular, for the infinitesimal vector $dX \equiv [dx^{\mu}]$ with $dx^{\mu} \equiv (cdt, dx, dy, dz)$ we define the *spacetime interval* $ds^2 = (dX, dX)$:

$$ds^2 = g_{\mu\nu}\, dx^\mu dx^\nu = (dx^0)^2 - (dx^1)^2 - (dx^2)^2 - (dx^3)^2$$
$$= c^2 dt^2 - dx^2 - dy^2 - dz^2 \tag{3.5}$$

Note that $ds^2$ may be positive, negative or zero. The sign of the spacetime interval has profound significance for *causality* (see Problem 3).

As in Chapter 2, given two vectors $a$ and $b$ we consider the transformation

$$a \to a' = \Lambda a \quad \Leftrightarrow \quad a^{\mu\,\prime} = \Lambda^\mu{}_\nu\, a^\nu \tag{3.6}$$

and similarly for $b$, where $\Lambda \equiv [\Lambda^\mu{}_\nu]$ is a real (4×4) matrix such that

$$(a', b') = (\Lambda a, \Lambda b) = (a, b) \;\Rightarrow\; a^t (\Lambda^t g \Lambda) b = a^t g b\,.$$

This will be true for all $a$ and $b$ if

$$\Lambda^t g \Lambda = g \quad \Leftrightarrow \quad \Lambda^\mu{}_\lambda\, g_{\mu\nu}\, \Lambda^\nu{}_\rho = g_{\lambda\rho} \tag{3.7}$$

where $\Lambda^\mu{}_\lambda = (\Lambda^t)_\lambda{}^\mu$. It follows from (3.7) that $(\det\Lambda)^2 = 1 \Rightarrow \det\Lambda = \pm 1$. Moreover, by setting $\lambda = \rho = 0$ we have that

$$\left(\Lambda^0{}_0\right)^2 = 1 + \sum_{i=1}^{3} \left(\Lambda^i{}_0\right)^2$$

(explain this) so that $|\Lambda^0{}_0| \geq 1 \Rightarrow \Lambda^0{}_0 \geq 1$ or $\Lambda^0{}_0 \leq -1$. As argued in Chap. 2, connectivity with the identity transformation ($\Lambda$=1) dictates that we choose

$$\det\Lambda = +1\,, \quad \Lambda^0{}_0 \geq 1 \tag{3.8}$$

Notice that, by (3.7), the transformation matrix $\Lambda$ is unaffected if we choose our metric to be $-g$ instead of $g$.

The transformation (3.6) that leaves the scalar product $(a, b)$ invariant for all vectors $a$, $b$, and which satisfies the matrix relation (3.7) with the additional constraints (3.8), is called a *proper orthochronous Lorentz transformation* (LT) and, as we saw in the previous chapter, is represented by the matrix group $SO(3,1)\uparrow$. Four-component objects $a^\mu \equiv (a^0, a^1, a^2, a^3)$ or, equivalently, column vectors $a \equiv [a^\mu]$, transforming according to (3.6), are called *4-vectors*. The *"magnitude"* of a 4-vector:

$$(a, a) = a^t g a = g_{\mu\nu}\, a^\mu a^\nu = (a^0)^2 - (a^1)^2 - (a^2)^2 - (a^3)^2 \tag{3.9}$$

is invariant under a LT: $(a', a') = (\Lambda a, \Lambda a) = (a, a)$. Thus, $g_{\mu\nu}\, a^{\mu\,\prime} a^{\nu\,\prime} = g_{\mu\nu}\, a^\mu a^\nu \Rightarrow$

$$(a^{0\,\prime})^2 - (a^{1\,\prime})^2 - (a^{2\,\prime})^2 - (a^{3\,\prime})^2 = (a^0)^2 - (a^1)^2 - (a^2)^2 - (a^3)^2 \tag{3.10}$$

In particular, for the vector $dX \equiv [dx^\mu]$ the invariant magnitude is the spacetime interval $ds^2 = (dX, dX)$, given in explicit form by Eq. (3.5). We thus have that, under a LT, $g_{\mu\nu}\, dx^{\mu\,\prime} dx^{\nu\,\prime} = g_{\mu\nu}\, dx^\mu dx^\nu \Rightarrow$

$$c^2 (dt')^2 - (dx')^2 - (dy')^2 - (dz')^2 = c^2 dt^2 - dx^2 - dy^2 - dz^2 \tag{3.11}$$

## 3.2 Rotations and Boosts

Special types of LTs are rotations and boosts. *Spatial rotations* redefine the orientation of the 3-dimensional Cartesian system of axes of an inertial frame (no physical motion of the axes is implied!) and, in group-theoretical terms, are represented by (3×3) orthogonal matrices $[R^i{}_j]$ with unit determinant, which belong to the 3-parameter subgroup $SO(3)$ of the Lorentz group (see, e.g., [1]). The LT matrix for rotations is of the form

$$\Lambda = [\Lambda^\mu{}_\nu] = \begin{bmatrix} 1 & 0 & 0 & 0 \\ 0 & & & \\ 0 & & [R^i{}_j] & \\ 0 & & & \end{bmatrix} \quad \text{where } [R^i{}_j] \in SO(3) \tag{3.12}$$

Given a 4-vector $a^\mu \equiv (a^0, a^1, a^2, a^3) \equiv (a^0, a^k)$ ($k$=1,2,3) its transformation under $\Lambda$ is, according to (3.6), $a^{\mu\,\prime} = \Lambda^\mu{}_\nu\, a^\nu$. For $\mu$=0 we have:

$$a^{0\,\prime} = \Lambda^0{}_\nu\, a^\nu = \delta^0{}_\nu\, a^\nu = a^0 \,.$$

Thus a rotation of spatial axes does not affect the zero-component of a 4-vector. For $\mu$=$k$ ($k$=1,2,3) we have:

$$a^{k\,\prime} = \Lambda^k{}_\nu\, a^\nu = \Lambda^k{}_0\, a^0 + \Lambda^k{}_l\, a^l = 0 + \Lambda^k{}_l\, a^l = R^k{}_l\, a^l \,.$$

Thus the $k$-components ($k$=1,2,3) of a 4-vector transform according to the $SO(3)$ matrix $[R^i{}_j]$ . Finally, the LT (3.12) leaves the magnitude (3.9) of a 4-vector invariant, i.e., guarantees that (3.10) is satisfied. This follows from the invariance of $a^0$, as well as the invariance of $(a^1)^2+(a^2)^2+(a^3)^2$ under $SO(3)$ transformations $a^{k\,\prime} = R^k{}_l\, a^l$.

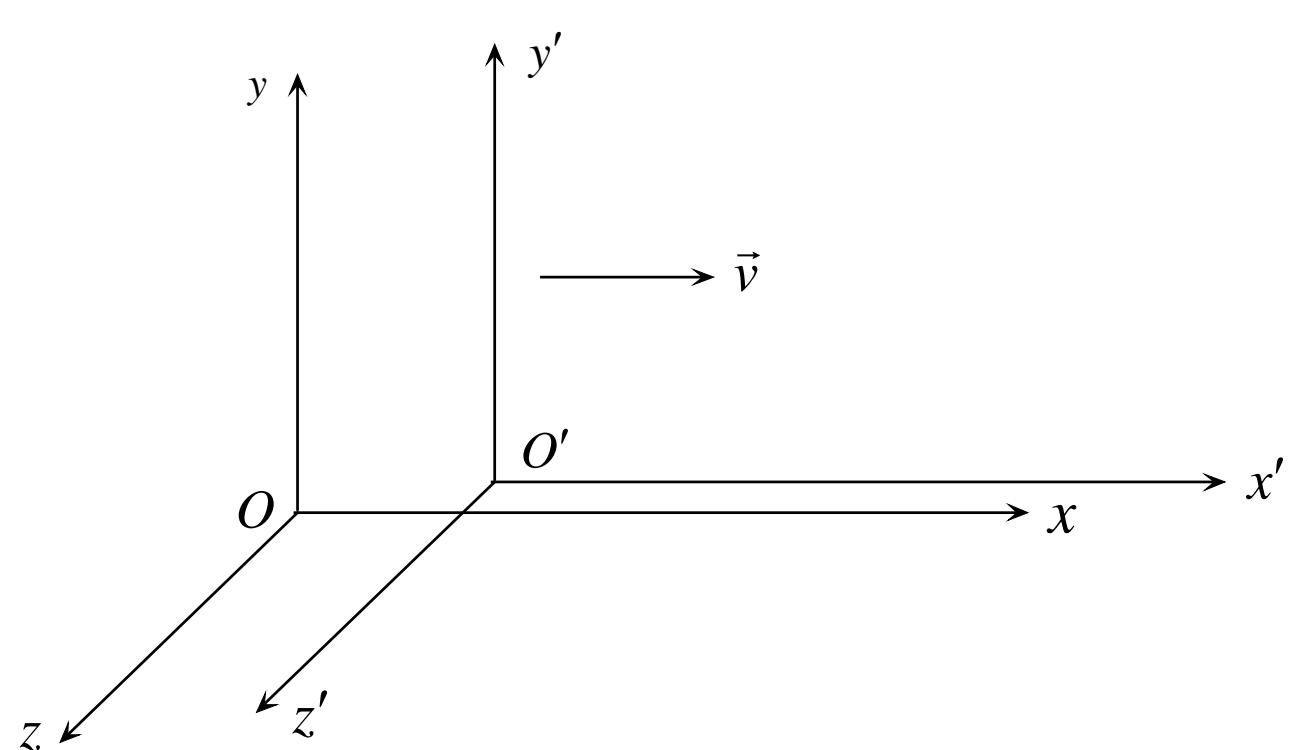

Fig. 3.1. An $x$-boost. The origins $O$ and $O'$ coincide for $t$=$t'$=0.

The $SO(3)$ matrix $[R^i{}_j]$, which constitutes the essential part of the LT matrix $\Lambda$ in (3.12), contains 3 of the total 6 parameters that parametrize a general LT. Three more parameters come from the 3 *boosts* along the 3 spatial axes. Consider two inertial observers using inertial frames $S$ and $S'$ with Cartesian systems of axes $(x, y, z)$ and $(x', y', z')$, respectively, as shown in Fig. 3.1. The frame $S'$ is moving with velocity $v$ relative to $S$, along the common $x$-axis of the two frames. (Note that $v$ is an *algebraic* *value* that may be positive or negative, depending on the direction of motion of $S'$.) At the moment when the origins $O$ and $O'$ of the frames coincide, the two observers arrange their clocks so that they show the same time, namely, $t=t'=0$. The LT (3.6) from the system $(x, y, z)$ to the system $(x', y', z')$ is called an *x-boost*. It is a one-parameter transformation parametrized by the velocity $v$ of $S'$ in a direction parallel to the $x$-axis. Similar boosts along the $y$- and $z$-directions will introduce two more velocities, hence two more parameters.

We introduce the constants

$$\beta = v/c \ , \quad \gamma = (1-\beta^2)^{-1/2} = (1-v^2/c^2)^{-1/2} \tag{3.13}$$

(note that $|\beta| \leq 1$). Consider a 4-vector with components $a^\mu \equiv (a^0, a^1, a^2, a^3)$ in the inertial frame $S$. In the case of an $x$-boost, the matrix of the LT $a^{\mu\prime} = \Lambda^\mu{}_\nu \, a^\nu$ of this vector from the frame $S$ to the frame $S'$ is [2]

$$\Lambda = [\Lambda^\mu{}_\nu] = \begin{bmatrix} \gamma & -\gamma\beta & 0 & 0 \\ -\gamma\beta & \gamma & 0 & 0 \\ 0 & 0 & 1 & 0 \\ 0 & 0 & 0 & 1 \end{bmatrix} \tag{3.14}$$

and the transformation equations are

$$\begin{aligned} a^{0\prime} &= \gamma(a^0 - \beta a^1) = \frac{a^0 - (v/c)a^1}{(1-v^2/c^2)^{1/2}} \\ a^{1\prime} &= \gamma(a^1 - \beta a^0) = \frac{a^1 - (v/c)a^0}{(1-v^2/c^2)^{1/2}} \\ a^{2\prime} &= a^2 \, , \quad a^{3\prime} = a^3 \end{aligned} \tag{3.15}$$

*Exercise:* Check the invariance of the magnitude $(a, a) = g_{\mu\nu} a^\mu a^\nu$ of a 4-vector, under the transformation (3.15); i.e., show that (3.10) is satisfied. [*Hint:* Take into account (3.13).]

In particular, the 4-vector $x^\mu \equiv (x^0, x^1, x^2, x^3) \equiv (ct, x, y, z)$ transforms under an $x$-boost as follows:

$$t' = \frac{t - (v/c^2)x}{(1-v^2/c^2)^{1/2}} \ , \quad x' = \frac{x - vt}{(1-v^2/c^2)^{1/2}} \ , \quad y' = y \, , \quad z' = z \tag{3.16}$$

Now, since the matrix $\Lambda$ in (3.14), representing a certain type of LT, is an element of a group (namely, the Lorentz group), its inverse $\Lambda^{-1}$ must also be in the group, thus it must also be a LT. Obviously, $\Lambda^{-1}$ takes us from the frame $S'$ back to the frame $S$. So, let $a^{\mu\prime} = \Lambda^\mu{}_\nu \, a^\nu$ be the LT of a 4-vector, produced by the matrix $\Lambda$ in (3.14). In matrix form, $a' = \Lambda a$, where $a \equiv [a^\mu]$ and $a' \equiv [a^{\mu\prime}]$ are column vectors. Then,

$$a = \Lambda^{-1} a' \iff a^\mu = (\Lambda^{-1})^\mu{}_\nu \, a^{\nu\prime}.$$

An easy way to construct the matrix $\Lambda^{-1}$ is to think as follows: If the frame $S'$ moves with velocity $v$ relative to the frame $S$, the latter frame moves with velocity $-v$ relative to the former. Thus all we have to do is replace $v$ and $\beta$ by $-v$ and $-\beta$, respectively, in Eqs. (3-14) - (3.16). We then have:

$$\Lambda^{-1} = [(\Lambda^{-1})^{\mu}{}_{\nu}] = \begin{bmatrix} \gamma & \gamma\beta & 0 & 0 \\ \gamma\beta & \gamma & 0 & 0 \\ 0 & 0 & 1 & 0 \\ 0 & 0 & 0 & 1 \end{bmatrix} \tag{3.17}$$

$$\begin{aligned} a^0 &= \gamma(a^{0\prime} + \beta a^{1\prime}) = \frac{a^{0\prime} + (v/c)a^{1\prime}}{(1-v^2/c^2)^{1/2}} \\ a^1 &= \gamma(a^{1\prime} + \beta a^{0\prime}) = \frac{a^{1\prime} + (v/c)a^{0\prime}}{(1-v^2/c^2)^{1/2}} \\ a^2 &= a^{2\prime}, \quad a^3 = a^{3\prime} \end{aligned} \tag{3.18}$$

$$t = \frac{t' + (v/c^2)x'}{(1-v^2/c^2)^{1/2}}, \quad x = \frac{x' + vt'}{(1-v^2/c^2)^{1/2}}, \quad y = y', \quad z = z' \tag{3.19}$$

*Exercise:* Verify that the matrix $\Lambda^{-1}$ in (3.17) is indeed the inverse of the matrix $\Lambda$ in (3.14); i.e., that $\Lambda^{-1}\Lambda=1$. [*Hint:* Take into account (3.13).]

*Exercise:* Show that, in the limit of very small velocities ($v \ll c \Rightarrow v/c \to 0$) the LT (3.16) reduces to the *Galilean transformation* of classical mechanics:

$$x' = x - vt, \quad y' = y, \quad z' = z, \quad t' = t.$$

## 3.3 Physical 4-Vectors

Let us now see some of the most common 4-vectors appearing in relativistic mechanics.

*1. The spacetime coordinate 4-vector*

In the previous section we treated the spacetime coordinate vector $x^{\mu} \equiv (ct, x, y, z)$ as a 4-vector, i.e., an object transforming according to the LT (3.6): $x^{\mu\prime} = \Lambda^{\mu}{}_{\nu}\, x^{\nu}$, with matrix $\Lambda$ belonging to the Lorentz group. But, how is this assumed property of the $x^{\mu}$ justified?

Consider two infinitesimally separated points $P$ and $Q$ in 3-space. Assume that a light ray proceeds from $P$ to $Q$. From the point of view of an observer using an inertial frame $S$, the passage of the ray from $P$ and $Q$ at definite times constitutes two separate spacetime *events* ($x^{\mu}$) and ($x^{\mu}+dx^{\mu}$). If the ray traveled a distance

$$dl = (dx^2 + dy^2 + dz^2)^{1/2}$$

within a time interval $dt$, then, according to this observer, the speed of light is

$$c = dl/dt = (dx^2 + dy^2 + dz^2)^{1/2}/dt \,.$$

For a different inertial observer, using a frame $S'$, the corresponding events associated with the passage of the light ray from $P$ and $Q$ are $(x^{\mu}{}')$ and $(x^{\mu}{}' + dx^{\mu}{}')$, respectively, while the speed of light is

$$c' = [(dx')^2 + (dy')^2 + (dz')^2]^{1/2}/dt' \,.$$

It is an experimental fact, however, that the speed of light in empty space is the same in all inertial frames. Thus $c = c'$, and hence

$$c^2 dt^2 - dx^2 - dy^2 - dz^2 = c^2 (dt')^2 - (dx')^2 - (dy')^2 - (dz')^2 = 0 \,.$$

That is, *for the propagation of a light ray* we must have

$$g_{\mu\nu}\, dx^{\mu} dx^{\nu} = g_{\mu\nu}\, dx^{\mu}{}' dx^{\nu}{}' = 0 \qquad (3.20)$$

where $[g_{\mu\nu}]$ is the matrix $g = diag(1, -1, -1, -1)$. Four-dimensional spacetime endowed with a metric equal to $g \equiv [g_{\mu\nu}]$ is known as *Minkowski space*.

We now *demand* that, more generally, the spacetime interval between *any* two infinitesimally separated events $(x^{\mu})$ and $(x^{\mu} + dx^{\mu})$, given by (3.5):

$$ds^2 \equiv g_{\mu\nu}\, dx^{\mu} dx^{\nu} = c^2 dt^2 - dx^2 - dy^2 - dz^2 \,,$$

is the same for all inertial observers. Consider two such observers that are using inertial frames $S$ and $S'$ with spacetime coordinates $(x^{\mu})$ and $(x^{\mu}{}')$, respectively. Assume that the two coordinate systems are related by $x^{\mu}{}' = \Lambda^{\mu}{}_{\nu}\, x^{\nu}$, for some (yet unspecified) constant matrix $\Lambda \equiv [\Lambda^{\mu}{}_{\nu}]$. Then, $dx^{\mu}{}' = \Lambda^{\mu}{}_{\nu}\, dx^{\nu}$. If $dX \equiv [dx^{\mu}]$ (column vector) and $(dX)^t \equiv [dx^{\mu}]^t$ (row vector), we can write: $ds^2 = (dX)^t g\, dX$ and express the invariance of $ds^2$ under the coordinate change $(x^{\mu}) \to (x^{\mu}{}')$ as follows:

$$(dX')^t g\, dX' = (dX)^t g\, dX \;\Rightarrow\; (\Lambda dX)^t g\, \Lambda dX = (dX)^t g\, dX \;\Rightarrow$$
$$(dX)^t (\Lambda^t g \Lambda)\, dX = (dX)^t g\, dX \,.$$

For this to be satisfied for any $dX$, we must have: $\Lambda^t g \Lambda = g$, which suggests that $\Lambda$ is in fact a matrix belonging to the Lorentz group. Then, from

$$x^{\mu}{}' = \Lambda^{\mu}{}_{\nu}\, x^{\nu} \;\Leftrightarrow\; [x^{\mu}{}'] = \Lambda\, [x^{\mu}]$$

we infer that the 4-component object $[x^{\mu}]$ is indeed a 4-vector.

Note that spacetime coordinate transformations $(x^{\mu}) \to (x^{\mu}{}')$ are *linear* transformations of the form $x^{\mu}{}' = \Lambda^{\mu}{}_{\nu}\, x^{\nu}$. This must be assumed *a priori* [2] in order that inertial (thus uniform) motions transform into inertial motions under a LT. In other words, the LT must guarantee that a free particle, subject to no net interaction, moves with constant velocity relative to *all* inertial observers.

As we have said, the spacetime interval $ds^2$ is invariant under LT, i.e., is a *Lorentz scalar*:

$$ds^2 = g_{\mu\nu}\, dx^\mu dx^\nu = g_{\mu\nu}\, dx^{\mu\,\prime} dx^{\nu\,\prime} \tag{3.21}$$

In particular, by (3.20) the invariance of $ds^2 = 0$ is equivalent to the invariance of the speed $c$ of light upon passing from one inertial reference frame to another. From the Lorentz scalars $c$ and $ds^2$ we can now construct a new scalar:

$$d\tau^2 \equiv ds^2 / c^2 = (1/c^2)\, g_{\mu\nu}\, dx^\mu dx^\nu \tag{3.22}$$

If $ds^2$ is a *timelike* interval (see Problem 3), i.e., if $ds^2 > 0$, then we may write

$$ds = +(ds^2)^{1/2} \quad , \quad d\tau = +(d\tau^2)^{1/2}$$

and define *proper time* $d\tau$ by

$$d\tau = ds / c \quad \Leftrightarrow \quad ds = c\, d\tau \tag{3.23}$$

Clearly, $d\tau$ is a Lorentz-invariant quantity, its value being the same for all inertial observers.

Assume now that $ds^2 > 0$ is an element of spacetime distance along the *worldline* (spacetime trajectory; see Sec. 1.2) of a particle, as viewed by an inertial observer using a frame $S$. The speed of the particle is, according to $S$,

$$u = dl / dt = (dx^2 + dy^2 + dz^2)^{1/2} / dt \,.$$

Given that $dl = u\, dt$ and that

$$ds^2 = c^2 dt^2 - dx^2 - dy^2 - dz^2 = c^2 dt^2 - dl^2 = (c^2 - u^2)\, dt^2 \,,$$

by (3.23) we have

$$d\tau = \left(1 - \frac{u^2}{c^2}\right)^{1/2} dt \quad \Leftrightarrow \quad dt = \left(1 - \frac{u^2}{c^2}\right)^{-1/2} d\tau \equiv \gamma(u)\, d\tau \tag{3.24}$$

where $\gamma(u) = (1 - u^2/c^2)^{-1/2}$. In particular, for a frame $S'$ momentarily at rest relative to the particle, $u' = 0$, $\gamma(u') = 1$ and, by (3.24), $dt' = \gamma(u')\, d\tau = d\tau$. So, according to $S'$, $d\tau$ is a purely time interval.[1] For any other frame $S$, relative to which $u \neq 0$, we have that $\gamma(u) > 1$ and so $dt > d\tau$, hence $dt > dt'$. This result expresses the familiar relativistic effect of *time dilation*. (The analogous effect of *length contraction* will be studied in Problem 2.)

*2. Four-velocity*

Consider the spacetime coordinate 4-vector $x^\mu \equiv (x^0, x^1, x^2, x^3) \equiv (ct, x, y, z)$. Under a general LT with (constant) matrix $\Lambda$, the differential $dx^\mu$ transforms as a 4-vector ($dx^{\mu\,\prime} = \Lambda^\mu{}_\nu dx^\nu$) while $d\tau$ is invariant. Hence, $dx^\mu / d\tau$ must transform as a 4-vector. This suggests defining *4-velocity* $U^\mu$ by

[1] In general, a *proper time interval* is the time interval between two events occurring at a point of space that is *at rest* relative to an observer.

$$U^{\mu} = \frac{dx^{\mu}}{d\tau} \equiv \left( \frac{dx^0}{d\tau}\,,\, \frac{dx^k}{d\tau} \right) \quad (k=1,2,3) \tag{3.25}$$

Let $x^k \equiv (x^1, x^2, x^3) = (x, y, z)$ be the spatial coordinates of a particle at time $t$, relative to some frame $S$. Then, $x^k = x^k(t)$. The 3-velocity of the particle in $S$ is

$$\vec{u} = \left( \frac{dx^k}{dt} \right) = \left( \frac{dx}{dt}\,,\, \frac{dy}{dt}\,,\, \frac{dz}{dt} \right) = \left( u_x\,, u_y\,, u_z \right).$$

According to (3.24), $dt = \gamma(u)\, d\tau$, where $\gamma(u) = (1 - u^2/c^2)^{-1/2}$ and where $u$ is the speed of the particle: $u = (u_x^{\,2} + u_y^{\,2} + u_z^{\,2})^{1/2}$. In terms of derivatives,

$$\frac{d}{d\tau} = \frac{dt}{d\tau}\frac{d}{dt} = \gamma(u)\frac{d}{dt} \ .$$

We then have:

$$U^0 = dx^0/d\tau = \gamma(u)\, c\ , \quad U^1 = dx^1/d\tau = \gamma(u)\, u_x\ ,$$

$$U^2 = dx^2/d\tau = \gamma(u)\, u_y\ , \quad U^3 = dx^3/d\tau = \gamma(u)\, u_z\ .$$

Therefore,

$$U^{\mu} \equiv \left( \gamma(u)c,\, \gamma(u)u_x\,,\, \gamma(u)u_y\,,\, \gamma(u)u_z \right) = \gamma(u)\left(c, \vec{u}\right) \tag{3.26}$$

The Lorentz-invariant magnitude of the 4-velocity is

$$g_{\mu\nu}\, U^{\mu} U^{\nu} = (U^0)^2 - (U^1)^2 - (U^2)^2 - (U^3)^2 = c^2 \tag{3.27}$$

*Exercise:* Prove Eq. (3.27).

Under an $x$-boost the 4-vector $U^{\mu}$ transforms according to Eqs. (3.15). We thus have:

$$U^{0\,\prime} = \gamma(v)\,[U^0 - (v/c)\, U^1] \quad \text{where} \quad \gamma(v) = (1 - v^2/c^2)^{-1/2}\ .$$

By setting $U^0 = \gamma(u)c$ , $U^{0\,\prime} = \gamma(u')c$ , $U^1 = \gamma(u)u_x$ , we get:

$$\gamma(u') = \gamma(v)\gamma(u)\,[1 - (v/c^2)\, u_x] \tag{3.28}$$

We also have:

$$U^{1\,\prime} = \gamma(v)\,[U^1 - (v/c)\, U^0]\ , \quad U^{2\,\prime} = U^2\ , \quad U^{3\,\prime} = U^3\ .$$

By using (3.26) in primed and unprimed forms [$U^1 = \gamma(u)\, u_x$, $U^{1\,\prime} = \gamma(u')\, u_x{}'$, etc.] and by taking into account (3.28), we find:

$$u_x{}' = \frac{u_x - v}{1 - v u_x / c^2}\ , \quad u_y{}' = \frac{u_y}{\gamma(v)(1 - v u_x / c^2)}\ , \quad u_z{}' = \frac{u_z}{\gamma(v)(1 - v u_x / c^2)} \tag{3.29}$$

*Exercise:* Prove Eqs. (3.29) and show that they are consistent with the Lorentz invariance of the speed of light. (Consider a light ray propagating in the $x$-direction.)

Equations (3.29) express the LT for velocity. An alternative way to derive these relations is to use the LT of $dx^\mu$ directly. We have that

$$dx' = \gamma(v)(dx - vdt) = \gamma(v)(u_x - v)\,dt\,, \quad dy' = dy\,, \quad dz' = dz\,,$$
$$dt' = \gamma(v)[dt - (v/c^2)\,dx] = \gamma(v)[1 - (v/c^2)\,u_x]\,dt\,.$$

By using the fact that $u_x = dx/dt$, $u_x' = dx'/dt'$, etc., relations (3.29) follow.

*3. Four-acceleration*

We define *4-acceleration* by

$$A^\mu = dU^\mu / d\tau \tag{3.30}$$

where $U^\mu$ is the 4-velocity. Clearly, $A^\mu$ is a 4-vector. Then, by (3.27),

$$g_{\mu\nu}\,U^\mu U^\nu = c^2 \;\Rightarrow\; d\,(g_{\mu\nu}\,U^\mu U^\nu)/d\tau = 0 \;\Rightarrow\; g_{\mu\nu}\,U^\mu A^\nu + g_{\mu\nu}\,U^\nu A^\mu = 0\,.$$

But,

$$g_{\mu\nu}\,U^\nu A^\mu = g_{\nu\mu}\,U^\mu A^\nu = g_{\mu\nu}\,U^\mu A^\nu \quad \text{(explain!)}\,.$$

Therefore, $2\,g_{\mu\nu}\,U^\mu A^\nu = 0 \;\Rightarrow$

$$(U, A) \equiv g_{\mu\nu}\,U^\mu A^\nu = 0 \tag{3.31}$$

This result generalizes the familiar 3-dimensional principle of mechanics, according to which, if the magnitude of the velocity is constant then the acceleration is normal to the velocity (see, e.g., [3]).

More on the issue of acceleration in SR will be said in Problem 5.

*4. Energy-momentum 4-vector*

Consider a particle of *rest mass m*. By definition, $m$ is the mass measured by an inertial observer relative to whom the particle is (momentarily) at rest. Like the speed $c$ of light and like proper time $\tau$, the rest mass $m$ is a Lorentz scalar.

We define the *4-momentum* of the particle $m$ by

$$P^\mu = mU^\mu \equiv \left(\gamma(u)mc\,,\; \gamma(u)m\vec{u}\right) \tag{3.32}$$

where $\gamma(u) = (1 - u^2/c^2)^{-1/2}$ and where $\vec{u} \equiv (u_x, u_y, u_z)$ is the 3-velocity of $m$. It can be shown (see Problem 8) that the *total relativistic energy* of a particle (*excluding* external potential energy) is given by

$$E = \gamma(u)\,mc^2 \tag{3.33}$$

(the most famous equation in Relativity and, probably, in Physics!). Moreover, the proper relativistic expression for 3-momentum, required by the Lorentz-invariance of the principle of conservation of momentum, is [2,4]

$$\vec{p} = \gamma(u)\,m\vec{u} \tag{3.34}$$

In view of (3.33) and (3.34), Eq. (3.32) is written:

$$P^{\mu} \equiv \left( \frac{E}{c} \, , \, \vec{p} \right) \tag{3.35}$$

The invariant magnitude of $P^{\mu}$ is

$$g_{\mu\nu} P^{\mu} P^{\nu} = (E/c)^2 - p^2 = m^2 c^2 \tag{3.36}$$

[*Exercise:* Prove (3.36).] From (3.36) we get the familiar energy-momentum relation

$$E^2 = m^2 c^4 + c^2 p^2 \tag{3.37}$$

Consider now a *free* particle, subject to no net interaction. Its energy and momentum must be constant in *any* inertial frame, which suggests that the principle of conservation of energy *and* momentum (see Problem 7) must be a Lorentz-invariant concept. That is,

$$\text{if} \quad dE/dt = 0 \quad \text{and} \quad d\vec{p}/dt = 0 \quad (\text{inertial frame } S),$$

$$\text{then} \quad dE'/dt' = 0 \quad \text{and} \quad d\vec{p}'/dt' = 0 \quad (\text{inertial frame } S').$$

Now, $E$ and $\vec{p}$ form the energy-momentum 4-vector $P^{\mu}$, defined in (3.35). We can thus write:

$$dP^{\mu}/dt = 0 \;\;\Leftrightarrow\;\; dP^{\mu\prime}/dt' = 0 \quad (\mu=0,1,2,3)\,.$$

In other words, the relation $dP^{\mu}/dt = 0$ is Lorentz-invariant. The truth of this statement can be proven as follows:

Let $\tau$ be proper time, i.e., time as measured by a clock in an inertial frame attached to the free particle. According to (3.24), time intervals in the frames $S$ and $S'$ are related to $d\tau$ by $dt = \gamma(u)\, d\tau$ and $dt' = \gamma(u')\, d\tau$ (remember that $d\tau$ is Lorentz-invariant). We have:

$$dP^{\mu}/dt = \gamma(u)^{-1}\, dP^{\mu}/d\tau\,, \quad dP^{\mu\prime}/dt' = \gamma(u')^{-1}\, dP^{\mu\prime}/d\tau\,.$$

Under a LT, $dP^{\mu\prime} = \Lambda^{\mu}{}_{\nu}\, dP^{\nu}$. Therefore,

$$dP^{\mu\prime}/dt' = \gamma(u')^{-1}\, \Lambda^{\mu}{}_{\nu}\, dP^{\nu}/d\tau = \gamma(u)\,\gamma(u')^{-1}\, \Lambda^{\mu}{}_{\nu}\, dP^{\nu}/dt\,.$$

So, if $dP^{\mu}/dt = 0$ in the $S$-frame, then $dP^{\mu\prime}/dt' = 0$ in the $S'$-frame, and vice versa. This result can be generalized [2,4] for any *isolated* system of particles, subject to no net *external* interaction.

Note that, in contrast to the situation in classical mechanics, conservation of energy and conservation of momentum are not separate matters in Relativity! The reason for this will become clear in Problem 7.

## 3.4 Transformation of Derivatives

As we know, the spacetime coordinates $x^\mu \equiv (x^0, x^1, x^2, x^3)$ transform like components of a 4-vector under a LT. We now want to see the manner in which partial derivatives with respect to the $x^\mu$, denoted $\partial_\mu \equiv \partial / \partial x^\mu$, transform.

As we did before, we consider the column vector $X \equiv [x^\mu]$. Under a LT represented by a matrix $\Lambda$,

$$X' = \Lambda X \quad \Leftrightarrow \quad x^{\mu\,\prime} = \Lambda^\mu{}_\nu \, x^\nu \, .$$

The inverse transformation is

$$X = \Lambda^{-1} X' \quad \Leftrightarrow \quad x^\mu = (\Lambda^{-1})^\mu{}_\nu \, x^{\nu\,\prime} \, .$$

We notice that

$$\Lambda^\mu{}_\nu = \partial x^{\mu\,\prime} / \partial x^\nu \, , \quad (\Lambda^{-1})^\mu{}_\nu = \partial x^\mu / \partial x^{\nu\,\prime} \tag{3.38}$$

Then, by the chain rule of differentiation,

$$\partial_\mu{}' = \partial / \partial x^{\mu\,\prime} = (\partial x^\lambda / \partial x^{\mu\,\prime}) \, \partial / \partial x^\lambda \;\Rightarrow$$

$$\partial_\mu{}' = (\Lambda^{-1})^\lambda{}_\mu \, \partial_\lambda \tag{3.39}$$

Therefore, partial derivatives with respect to the spacetime coordinates transform according to the *inverse* LT.

## 3.5 Transformation of Covariant Vectors

Let $a^\mu$ and $b^\mu$ be 4-vectors, represented by the column vectors $a \equiv [a^\mu]$, $b \equiv [b^\mu]$. Their scalar product is

$$(a, b) = a^t g \, b = g_{\mu\nu} \, a^\mu b^\nu$$

where $a^t \equiv [a^\mu]^t$ is a row vector. We define the quantities $a_\mu$ and $b_\mu$ by

$$a_\mu = g_{\mu\nu} \, a^\nu \, , \quad b_\mu = g_{\mu\nu} \, b^\nu \tag{3.40}$$

Then, by taking into account that $g_{\mu\nu} = g_{\nu\mu}$, we have:

$$(a, b) = (b, a) = a_\mu \, b^\mu = a^\mu b_\mu \tag{3.41}$$

From (3.40) we have that

$$a_0 = a^0 \, , \quad a_k = -\, a^k \;\; (k{=}1,2,3)$$

and similarly for $b_\mu$.

The quantities $a^\mu$ ($\mu$=0,1,2,3) are called the *contravariant components* of the 4-vector $a$, while the $a_\mu$ are called the *covariant components* of $a$. The symmetric scalar product (3.41) is, of course, invariant under a LT, i.e., is a Lorentz scalar.

Under a LT with matrix $\Lambda \equiv [\Lambda^{\mu}{}_{\nu}]$ the contravariant components $a^{\mu}$ transform as

$$a^{\mu\prime} = \Lambda^{\mu}{}_{\nu}\, a^{\nu} = (\partial x^{\mu\prime} / \partial x^{\nu})\, a^{\nu} \tag{3.42}$$

where we have used (3.38). We now want to see how the covariant components $a_{\mu}$ transform under this LT.

Consider the column vectors $a \equiv [a^{\mu}]$, $a^{\prime} \equiv [a^{\mu\prime}]$. In matrix form the LT is written as $a^{\prime} = \Lambda a$. The Lorentz-invariant magnitude of $a$ is

$$(a, a) = a^{t} g\, a = a_{\mu}\, a^{\mu} \tag{3.43}$$

Since $a$ is a column vector with components $a^{\mu}$, the product $a^{t}g$ must be a row vector with components $a_{\mu}$. We write: $a^{t}g \equiv [a_{\mu}]_{R}$, where $R$ stands for "row". Under a LT,

$$a^{t}g \equiv [a_{\mu}]_{R} \;\rightarrow\; (a^{\prime})^{t}g \equiv [a_{\mu}{}^{\prime}]_{R}\,.$$

We have:

$$(a^{\prime})^{t}g = (\Lambda a)^{t}g = a^{t}\Lambda^{t}\, g\,.$$

But, since $\Lambda$ belongs to the Lorentz group,

$$\Lambda^{t} g\, \Lambda = g \;\Rightarrow\; \Lambda^{t} g = g\Lambda^{-1}$$

and so

$$(a^{\prime})^{t}g = a^{t}g\Lambda^{-1} \;\;\Leftrightarrow\;\; [a_{\mu}{}^{\prime}]_{R} = [a_{\mu}]_{R}\, \Lambda^{-1}\,.$$

In terms of components,

$$a_{\mu}{}^{\prime} = a_{\lambda}\, (\Lambda^{-1})^{\lambda}{}_{\mu} = (\partial x^{\lambda} / \partial x^{\mu\prime})\, a_{\lambda} \tag{3.44}$$

where again we have used (3.38). By comparing (3.39) and (3.44) we notice that the derivatives $\partial_{\mu} \equiv \partial / \partial x^{\mu}$ with respect to the spacetime coordinates transform like covariant components of a 4-vector.

*Exercise:* By using Eqs. (3.42) and (3.44) demonstrate the Lorentz invariance of the magnitude $(a, a)$ of a 4-vector $a$, defined in (3.43).

*Exercise:* By using Eqs. (3.42) and (3.39) show the Lorentz invariance of the *4-divergence* $\partial_{\mu} a^{\mu}$ of a 4-vector $a$. Note that this property rests critically on the constancy (coordinate-independence) of the $\Lambda^{\mu}{}_{\nu}$, which reflects the *flatness* of spacetime in SR. It is no longer true that $\partial_{\mu} a^{\mu}$ is an invariant quantity under the more general coordinate transformations in the *curved* spacetime of General Relativity. In this latter case, ordinary derivatives must be replaced by *covariant* ones [2,5,6].

*Exercise:* By using (3.17) for the matrix inverse $\Lambda^{-1}$ show that, in an $x$-boost, the transformation equations for the covariant components $a_{\mu}$ of a 4-vector are given by

$$a_{0}{}^{\prime} = \gamma\, (a_{0} + \beta a_{1})\,, \quad a_{1}{}^{\prime} = \gamma\, (a_{1} + \beta a_{0})\,, \quad a_{2}{}^{\prime} = a_{2}\,, \quad a_{3}{}^{\prime} = a_{3} \tag{3.45}$$

where $\beta = v/c$, $\gamma = (1 - \beta^{2})^{-1/2} = (1 - v^{2}/c^{2})^{-1/2}$.

## 3.6 Transformation of Antisymmetric Tensors

An *antisymmetric tensor* $T^{\mu\nu} = -T^{\nu\mu}$ has only 6 independent components. In matrix form,

$$[T^{\mu\nu}] = \begin{bmatrix} 0 & T^{01} & T^{02} & T^{03} \\ -T^{01} & 0 & T^{12} & T^{13} \\ -T^{02} & -T^{12} & 0 & T^{23} \\ -T^{03} & -T^{13} & -T^{23} & 0 \end{bmatrix} \tag{3.46}$$

Under a LT, $T^{\mu\nu}$ follows the general tensor transformation law [5,6]

$$T^{\mu\nu\,\prime} = \Lambda^{\mu}{}_{\lambda}\,\Lambda^{\nu}{}_{\rho}\,T^{\lambda\rho} \tag{3.47}$$

The transformed tensor $T^{\mu\nu\,\prime}$ also is an antisymmetric tensor. Indeed,

$$T^{\nu\mu\,\prime} = \Lambda^{\nu}{}_{\lambda}\,\Lambda^{\mu}{}_{\rho}\,T^{\lambda\rho} = -\,\Lambda^{\mu}{}_{\rho}\,\Lambda^{\nu}{}_{\lambda}\,T^{\rho\lambda} = -\,T^{\mu\nu\,\prime}\ .$$

In particular, if the LT is an $x$-boost with matrix $\Lambda$ given by (3.14), then

$$\begin{aligned} &T^{01\,\prime} = T^{01}\ , \qquad T^{02\,\prime} = \gamma\,(T^{02} - \beta T^{12})\ , \qquad T^{03\,\prime} = \gamma\,(T^{03} - \beta T^{13})\ , \\ &T^{23\,\prime} = T^{23}\ , \qquad T^{13\,\prime} = \gamma\,(T^{13} - \beta T^{03})\ , \qquad T^{12\,\prime} = \gamma\,(T^{12} - \beta T^{02}) \end{aligned} \tag{3.48}$$

where, as always, $\beta = v/c$ , $\gamma = (1-\beta^2)^{-1/2} = (1-v^2/c^2)^{-1/2}$ .

Given two 4-vectors $a^{\mu}$, $b^{\mu}$ we may construct an antisymmetric tensor by the expression

$$T^{\mu\nu} = a^{\mu}b^{\nu} - a^{\nu}b^{\mu} = -\,T^{\nu\mu} \tag{3.49}$$

Then,

$$\begin{aligned} T^{\mu\nu\,\prime} &= a^{\mu\,\prime}b^{\nu\,\prime} - a^{\nu\,\prime}b^{\mu\,\prime} = (\Lambda^{\mu}{}_{\lambda}\,a^{\lambda})(\Lambda^{\nu}{}_{\rho}\,b^{\rho}) - (\Lambda^{\nu}{}_{\rho}\,a^{\rho})(\Lambda^{\mu}{}_{\lambda}\,b^{\lambda}) \\ &= \Lambda^{\mu}{}_{\lambda}\,\Lambda^{\nu}{}_{\rho}\,(a^{\lambda}b^{\rho} - a^{\rho}b^{\lambda}) = \Lambda^{\mu}{}_{\lambda}\,\Lambda^{\nu}{}_{\rho}\,T^{\lambda\rho}\ . \end{aligned}$$

## 3.7 The d'Alembert Operator

Given the spacetime coordinates $x^{\mu} \equiv (x^0, x^1, x^2, x^3) \equiv (ct, x, y, z)$, the partial derivatives with respect to the $x^{\mu}$ are given by

$$\partial_{\mu} \equiv \partial/\partial x^{\mu} \equiv \left( \frac{1}{c}\frac{\partial}{\partial t}\,,\ \frac{\partial}{\partial x}\,,\ \frac{\partial}{\partial y}\,,\ \frac{\partial}{\partial z} \right)\ .$$

As shown in Sec. 3.4, these derivatives transform like the covariant components of a 4-vector under a LT $\Lambda = [\Lambda^{\mu}{}_{\nu}]$:

$$\partial_{\mu}{}' = (\Lambda^{-1})^{\lambda}{}_{\mu}\,\partial_{\lambda} \tag{3.50}$$

We now define the quantities

$$\partial^\mu = g^{\mu\nu}\,\partial_\nu \tag{3.51}$$

where

$$[g^{\mu\nu}] = [\,g_{\mu\nu}] = diag\,(1, -1, -1, -1) \tag{3.52}$$

(Clearly, $g^{\mu\nu} = g^{\nu\mu}$.) We thus have: $\partial^0 = \partial_0$ , $\partial^k = -\,\partial_k$ $(k = 1,2,3)$ $\Rightarrow$

$$\partial^\mu \equiv \left( \frac{1}{c}\frac{\partial}{\partial t}\,,\; -\frac{\partial}{\partial x}\,,\; -\frac{\partial}{\partial y}\,,\; -\frac{\partial}{\partial z} \right) \tag{3.53}$$

Let us call $g^\mu{}_\nu \equiv g^{\mu\lambda}\,g_{\lambda\nu}$. Then, $g^\mu{}_\nu = \delta^\mu{}_\nu$ (Kronecker delta). (*Exercise:* Prove this.) From (3.51) we have:

$$g_{\lambda\mu}\,\partial^\mu = g_{\lambda\mu}\,g^{\mu\nu}\,\partial_\nu = g^{\nu\mu}\,g_{\mu\lambda}\,\partial_\nu = g^\nu{}_\lambda\,\partial_\nu = \delta^\nu{}_\lambda\,\partial_\nu \;\Rightarrow$$

$$g_{\lambda\mu}\,\partial^\mu = \partial_\lambda \tag{3.54}$$

By comparing (3.54) with (3.40) and by taking into account that the $\partial_\lambda$ transform like covariant components, we come to the conclusion that the operators $\partial^\mu$ transform like the *contravariant* components of a 4-vector:

$$\partial^{\mu\,\prime} = \Lambda^\mu{}_\nu\,\partial^\nu \tag{3.55}$$

It follows that the second-order differential operator $\partial_\mu\,\partial^\mu$ is a *scalar operator*. To show this we consider a scalar function $\Phi(x^\mu)$, the value (but not necessarily the functional form) of which is invariant under a LT: $\Phi'(x^{\mu\,\prime}) = \Phi(x^\mu)$. Then, by using (3.50) and (3.55) we have:

$$\partial_\mu{}'\,\partial^{\mu\,\prime}\,\Phi'(x^{\rho\,\prime}) = (\Lambda^{-1})^\lambda{}_\mu\,\partial_\lambda\,[\Lambda^\mu{}_\nu\,\partial^\nu\,\Phi(x^\rho)]$$

$$= \delta^\lambda{}_\nu\,\partial_\lambda\,\partial^\nu\,\Phi(x^\rho) = \partial_\lambda\,\partial^\lambda\,\Phi(x^\rho)\;.$$

The operator $\partial_\mu\partial^\mu$ is called the *d'Alembert operator*:

$$\Box^2 \equiv \partial_\mu\,\partial^\mu = \frac{1}{c^2}\frac{\partial^2}{\partial t^2} - \nabla^2 \tag{3.56}$$

The *wave equation* for a scalar function $\Phi(x^\mu)$ is written as

$$\Box^2\,\Phi(x^\mu) \equiv \frac{1}{c^2}\frac{\partial^2\Phi}{\partial t^2} - \nabla^2\Phi = 0 \tag{3.57}$$

Since both the d'Alembert operator and the function $\Phi$ are Lorentz scalars, Eq. (3.57) is *covariant* under LTs. That is, if $\Phi(x^\mu)$ satisfies the wave equation in the "unprimed" frame, then $\Phi'(x^{\mu\,\prime})$ satisfies the wave equation in the "primed" frame.

Let now $A^{\mu}(x^{\lambda})$ be a *4-vector field*, which transforms as

$$A^{\mu\,\prime}(x^{\lambda\,\prime}) = \Lambda^{\mu}{}_{\nu}\, A^{\nu}(x^{\lambda}) \,.$$

Assume that each component of this field satisfies the wave equation:

$$\Box^2 A^{\mu}(x^{\lambda}) = 0 \quad (\mu = 0,1,2,3)\,.$$

Then the above equation is covariant under LTs. Indeed, notice that

$$\Box^2 A^{\mu\prime}(x^{\lambda\prime}) = \Box^2 \left[ \Lambda^{\mu}{}_{\nu}\, A^{\nu}(x^{\lambda}) \right] = \Lambda^{\mu}{}_{\nu}\, \Box^2 A^{\nu}(x^{\lambda}) = 0\,.$$

We emphasize once more that this property rests critically on the constancy (coordinate-independence) of the LT matrix elements $\Lambda^{\mu}{}_{\nu}$, i.e., on the *flatness* of Minkowski spacetime.

Another example of Lorentz covariance is the following. Consider a *tensor field* $T^{\mu\nu}(x^{\lambda})$. Under a LT this field transforms as $T^{\mu\nu}(x^{\lambda}) \rightarrow T^{\mu\nu\,\prime}(x^{\lambda\,\prime})$, with

$$T^{\mu\nu\,\prime} = \Lambda^{\mu}{}_{\lambda}\, \Lambda^{\nu}{}_{\rho}\, T^{\lambda\rho}.$$

*Proposition:* The expression $\partial_{\mu} T^{\mu\nu}$ (sum on $\mu$ !) transforms like a (contravariant) component of a 4-vector.

*Proof:* Set $\partial_{\mu} T^{\mu\nu} \equiv A^{\nu}$. We must show that $A^{\nu}$ transforms like a 4-vector component under a LT. We have:

$$\begin{aligned} A^{\nu\,\prime} = \partial_{\mu}{}^{\prime}\, T^{\mu\nu\,\prime} &= (\Lambda^{-1})^{\sigma}{}_{\mu}\, \partial_{\sigma} [\Lambda^{\mu}{}_{\lambda}\, \Lambda^{\nu}{}_{\rho}\, T^{\lambda\rho}] = [(\Lambda^{-1})^{\sigma}{}_{\mu}\, \Lambda^{\mu}{}_{\lambda}]\, \Lambda^{\nu}{}_{\rho}\, \partial_{\sigma} T^{\lambda\rho} \\ &= \delta^{\sigma}{}_{\lambda}\, \Lambda^{\nu}{}_{\rho}\, \partial_{\sigma} T^{\lambda\rho} = \Lambda^{\nu}{}_{\rho}\, \partial_{\lambda} T^{\lambda\rho} \equiv \Lambda^{\nu}{}_{\rho} A^{\rho} \end{aligned} \tag{3.58}$$

which is what we needed to prove.

Now, let us assume that, in the "unprimed" frame of reference, the following set of differential equations is valid:

$$\partial_{\mu} T^{\mu\nu}(x^{\lambda}) = 0 \quad (\nu{=}0,1,2,3) \tag{3.59}$$

It then follows from (3.58) that, in the "primed" frame,

$$\partial_{\mu}{}^{\prime}\, T^{\mu\nu\,\prime}(x^{\lambda\,\prime}) = 0 \,,$$

which is of the same form as (3.59). Therefore, the differential system (3.59) is covariant (invariant *in form*) under LTs.

Generally speaking, with regard to their mathematical structure, all properly formulated physical laws must exhibit covariance under LTs. This will guarantee that the validity of these laws is independent of the inertial frame of reference in which they are being tested.

## Problems

**1.** Show that the necessary and sufficient condition in order for two events to be simultaneous in *any* inertial frame is that these events occur at the same point in space, relative to *any* inertial frame.

***Solution:*** Consider two infinitesimally separated spacetime events $(x, y, z, t)$ and $(x+dx,\ y+dy,\ z+dz,\ t+dt)$. Consider also two inertial frames $S$ and $S'$, assuming that the LT from $S$ to $S'$ is an $x$-boost:

$$dx' = \gamma\,(dx - v dt)\ ,\ \ dy' = dy\ ,\ \ dz' = dz\ ,\ \ dt' = \gamma\,[dt - (v/c^2)\,dx]$$

where $\gamma(v) = (1 - v^2/c^2)^{-1/2}$ .

($a$) Assume that the two events are simultaneous relative to *both* frames $S$ and $S'$. Then, $dt = dt' = 0$ and, by the LT, $dx = dx' = 0$. By considering boosts in the $y$- and $z$-directions we find, by similar reasoning, that $dy = dy' = 0$ and $dz = dz' = 0$. Thus the events must occur at the same spatial point relative to both $S$ and $S'$.

($b$) Conversely, assume that the two events occur at the same point relative to both $S$ and $S'$. Then, $dx = dx' = 0$, $dy = dy' = 0$, $dz = dz' = 0$ and, by the LT, $dt = dt' = 0$. That is, the events are simultaneous relative to both $S$ and $S'$.

**2.** Show that a linear object appears shortened to an observer moving parallel to the object. This is the familiar *length contraction* effect of SR.

***Solution:*** Assume that the object is at rest on the $x'$-axis of an inertial frame $S'$ used by an observer $O'$. The ends of the object are at points $x'$ and $x'+dx'$. Hence the length of the object is $dl' = dx'$ ; it is called the *proper length* of the stationary object.

Let $O$ be another inertial observer whose frame $S$ has axes $(x, y, z)$ parallel to the corresponding axes of $S'$, and assume that $O'$ is moving in the $x$ direction with velocity $v$ relative to $O$. To measure the length of the object in her own frame $S$, observer $O$ must record the two ends of the object *simultaneously*. Thus, if $(x, t)$ and $(x+dx,\ t+dt)$ represent the events of assigning positions to the ends of the object at times $t$ and $t+dt$, respectively, we must demand that $dt=0$. (The observer $O'$, however, may record $x'$ and $x'+dx'$ at *any* times, given that the object is stationary in the $S'$ frame!)

Now, the LT from $S$ to $S'$ is an $x$-boost, so that

$$dx' = \gamma\,(dx - v dt) = \gamma\, dx \quad \text{where} \quad \gamma(v) = (1 - v^2/c^2)^{-1/2} .$$

Putting $dx = dl$ and $dx' = dl'$, we have that $dl' = \gamma\, dl \ \Rightarrow$

$$dl = \gamma^{-1} dl' = (1 - v^2/c^2)^{1/2}\, dl'.$$

Clearly, $dl < dl'$, which means that the moving object will appear *shorter* to $O$. Equivalently, since $O$ is moving with respect to the object with velocity $-v$, this observer will again measure a shorter length. In conclusion: Relative motion "shortens" the length of objects in the direction *parallel* to the motion, by a factor equal to $\gamma(v)^{-1}$ .

**3.** Show that, along the worldline (spacetime trajectory) of a massive particle the spacetime interval must be *timelike*, while along the worldline of a light signal the interval is *lightlike*. How are these observations related to the principle of causality?

***Solution:*** Let $(x, y, z, t)$ and $(x+dx,\ y+dy,\ z+dz,\ t+dt)$ be two infinitesimally separated points on a worldline, representing two infinitesimally separated events in spacetime. The spacetime interval between these events is

$$ds^2 = g_{\mu\nu}\, dx^\mu dx^\nu = c^2 dt^2 - dx^2 - dy^2 - dz^2 ,$$

which is a Lorentz invariant quantity. In particular, $ds^2$ retains its sign under a LT. For example, if $ds^2 > 0$ in some inertial frame, then $ds^2 > 0$ in *all* frames. There are three classes of spacetime intervals:

- *timelike* intervals where $ds^2 > 0$ ;
- *spacelike* intervals where $ds^2 < 0$ ; and
- *lightlike* intervals where $ds^2 = 0$ .

Suppose now a *massive* particle is at point $(x, y, z)$ at time $t$, and at point $(x+dx,\ y+dy,\ z+dz)$ at time $t+dt$. The particle's speed at time $t$ is

$$u = dl/dt = (dx^2 + dy^2 + dz^2)^{1/2} / dt .$$

Given that $u < c$, we have:

$$(dx^2 + dy^2 + dz^2) / dt^2 < c^2 \;\Rightarrow\; ds^2 = c^2 dt^2 - dx^2 - dy^2 - dz^2 > 0 .$$

That is, the spacetime interval between two infinitesimally separated points on the particle's worldline is *timelike*.

Next, suppose a light signal has reached points $(x, y, z)$ and $(x+dx,\ y+dy,\ z+dz)$ at respective times $t$ and $t+dt$. Here $u=c$, so that

$$(dx^2 + dy^2 + dz^2) / dt^2 = c^2 \;\Rightarrow\; ds^2 = 0 .$$

That is, the spacetime interval along the signal's worldline is *lightlike*.

In geometrical terms, the worldline of a massive particle must lie in the interior of the *light cone* (see Sec. 1.2 and Fig. 1.1), while the worldline of a light ray lies *on* the cone. No particle or photon worldline may lie in the exterior of the light cone, since along such a line the spacetime interval would be *spacelike*. In such a case we would have $ds^2 < 0$ and, therefore, $u > c$, which is impossible given that no matter or energy can travel faster than light!

The foregoing discussion can be related to the *principle of causality*. Consider two spacetime events $(x^\mu)$ and $(x^\mu + dx^\mu)$ and let $ds^2 = g_{\mu\nu}\, dx^\mu dx^\nu$ be the spacetime interval separating them. The fact that no speed in Nature can exceed the speed $c$ of light suggests that, if $ds^2 \geq 0$, it is possible to connect these events by a material signal such as a massive particle or a massless photon, while such a connection is impossible if $ds^2 < 0$ (the speed of the signal would have to be greater than $c$). This means that two spacelike-related events may not have influenced each other, i.e., may not be *causally related*.

The fact that causality cannot be violated for timelike-separated events also follows from the observation that, as can be proven [4], the time ordering ("before" and "after") for such events is absolute, i.e., Lorentz-invariant. On the contrary, the time ordering of spacelike-separated events is frame-dependent and can be reversed by a LT. Thus such events may not be causally related.

**4.** A moving clock describes a (generally curved) worldline in an inertial frame $S$ with coordinates $x^\mu \equiv (ct, x, y, z)$. Show that the time interval measured by this clock is given by

$$\tau = \frac{1}{c} \int \left( dx^\mu dx_\mu \right)^{1/2} = \frac{1}{c} \int \left( c^2 dt^2 - dx^2 - dy^2 - dz^2 \right)^{1/2} .$$

***Solution:*** If the clock describes a curved worldline in *flat* spacetime, it executes *non-inertial* (accelerated) motion. We may assume, however, that the worldline can be divided into an infinite number of infinitesimal *linear* segments, along each of which the motion of the clock may be considered inertial. Moreover, being a part of the worldline of a massive particle, each segment must be *timelike* (cf. Prob. 3). Hence, the spacetime interval along a segment is

$$ds^2 = c^2 dt^2 - dx^2 - dy^2 - dz^2 > 0 \quad \Rightarrow \quad ds = +(ds^2)^{1/2} \in R .$$

Now, $ds = c\, d\tau \Rightarrow d\tau = ds/c$, where $d\tau$ is the *proper time* of the segment, equal to the time measured by a *local inertial frame* relative to which the clock is *momentarily* at rest (obviously, an infinite number of such frames are needed, one for each momentary position of the clock). So, the time interval $d\tau$ in the instantaneous inertial frame moving with the clock is equal to the spacetime interval $ds/c$ measured in the frame $S$ with coordinates $x^\mu$ or $(x, y, z, t)$, relative to which frame the clock is moving. The total time measured by the clock is, therefore,

$$\tau = \int d\tau = (1/c) \int ds = (1/c) \int (c^2 dt^2 - dx^2 - dy^2 - dz^2)^{1/2}$$

or

$$\tau = (1/c) \int (g_{\mu\nu} dx^\mu dx^\nu)^{1/2} = (1/c) \int (dx^\mu dx_\mu)^{1/2}$$

where $dx_\mu = g_{\mu\nu}\, dx^\nu$. Notice that, if the clock traveled at the speed of light, then we would have $ds^2 = 0$ (Prob. 3) and $d\tau = 0$: the clock would measure no time at all!

**5.** As is well known, acceleration is Galilean-invariant in Newtonian mechanics. Is it Lorentz-invariant in SR? For simplicity, consider one-dimensional motion of a particle in the $x$ direction, and a LT in the form of an $x$-boost.

***Solution:*** Consider a particle viewed by two inertial observers $O$ and $O'$ using spacetime coordinates $(x, y, z, t)$ and $(x', y', z', t')$, respectively. As usual, $O'$ moves along the common $x$-axis of the two frames of reference, with velocity $v$. We call

$$(u_x, u_y, u_z) = (dx/dt, dy/dt, dz/dt) \quad \text{and} \quad (u_x', u_y', u_z') = (dx'/dt', dy'/dt', dz'/dt') .$$

As we have shown (Sec. 3.3), the LT for velocities is

$$u_x{}' = \frac{u_x - v}{1 - v u_x / c^2} \quad , \quad u_y{}' = \frac{u_y}{\gamma(v)(1 - v u_x / c^2)} \quad , \quad u_z{}' = \frac{u_z}{\gamma(v)(1 - v u_x / c^2)}$$

where $\gamma(v) = (1 - v^2/c^2)^{-1/2}$. For motion of the particle along the $x$-axis, we set $u_x = u$, $u_y = u_z = 0$, so that

$$u_x{}' = u' = \frac{u - v}{1 - v u / c^2} \quad , \quad u_y{}' = u_z{}' = 0 \tag{1}$$

We want to find the LT for acceleration, for motion along the $x$-axis. We have that $a = du/dt$, $a' = du'/dt'$. Now, from (1) $\Rightarrow$

$$du' = \frac{d}{du}\left(\frac{u - v}{1 - v u / c^2}\right) du = \frac{1 - v^2/c^2}{(1 - v u / c^2)^2}\, du \, .$$

Moreover,

$$dt' = \frac{dt - (v / c^2)\, dx}{(1 - v^2/c^2)^{1/2}} = \frac{1 - v u / c^2}{(1 - v^2/c^2)^{1/2}}\, dt \quad (\text{since } dx = u dt) \, .$$

Therefore,

$$a' = \frac{a\,(1 - v^2/c^2)^{3/2}}{(1 - v u / c^2)^3} \tag{2}$$

To find the inverse transformation we simply exchange $a$ with $a'$ and put $u'$ in place of $u$ and $-v$ in place of $v$.

We note the following:

1. Equation (1) reduces to the Galilean transformation for velocities, $u' = u - v$, if $v << c$ (small relative velocity of the two inertial frames) or/and $u << c$ (small particle velocity).

2. We recall that the relativistic momentum of a particle of mass $m$ moving with velocity $\vec{u}$ is

$$\vec{p} = \gamma(u)\, m \vec{u} = \frac{m \vec{u}}{(1 - u^2/c^2)^{1/2}} \, .$$

In the case of a *massless* particle ($m$=0), such as a photon, a non-vanishing momentum requires that $u$=$c$, which expresses the well-known fact that a massless particle travels at the speed of light in *any* inertial frame. Let us check this last assertion: According to an observer $O$, a photon's velocity is $u$=$c$. Then, by Eq. (1), the velocity of the photon according to another observer $O'$ is, again, $u' = c$, in accordance with the relativistic principle of frame-independence of the speed of light.

3. For $v << c \Leftrightarrow v/c \to 0$, Eq. (2) reduces to the Newtonian result $a' = a$. That is, according to Galilean relativity all inertial observers must measure the same acceleration of a moving particle. According to (2) this is *not* the case with SR. *It is still true,* however, that $a' = 0$ *if* $a$=0. This reflects the fact that inertial motions transform into inertial motions under LTs.

**6.** Show that, according to SR, the force on a particle is *not necessarily* proportional to the acceleration of the particle, although it is still true that the vanishing of the acceleration implies the vanishing of the force. Find the relationship between force and acceleration in the cases of rectilinear motion and general uniform motion. What do you observe?

***Solution:*** Consider a particle of mass *m*, moving with velocity $\vec{u}$ and acceleration $\vec{a} = d\vec{u}/dt$ relative to some inertial observer. The relativistic momentum of the particle is $\vec{p} = \gamma(u) m \vec{u}$, where $u = |\vec{u}|$ and $\gamma(u) = (1-u^2/c^2)^{-1/2}$. We define the total force on the particle at time *t* by $\vec{F} = d\vec{p}/dt$. We then have:

$$\vec{F} = \frac{d}{dt}(\gamma m \vec{u}) = m(d\gamma/dt)\vec{u} + \gamma m \vec{a}.$$

As we can show,

$$\frac{d\gamma}{dt} = \frac{u}{c^2}\gamma(u)^3 \frac{du}{dt}.$$

Therefore,

$$\vec{F} = \frac{mu}{c^2}\gamma(u)^3 \frac{du}{dt}\vec{u} + \gamma(u) m \vec{a} \qquad (1)$$

Clearly, in the general case, force is not proportional to acceleration.

As is well known, the velocity is a vector tangent to the trajectory of the particle and can be expressed as $\vec{u} = u\hat{\tau}$, where $\hat{\tau}$ is the unit tangent vector in the direction of motion and where, as defined previously, $u > 0$ is the speed of the particle. Let us now concentrate on two special cases of motion:

(*a*) In the case of *rectilinear* motion the direction of the unit vector $\hat{\tau}$ is constant and so $\vec{a} = d\vec{u}/dt = a\hat{\tau}$, where $a = du/dt = \pm|\vec{a}|$ (algebraic value!). Moreover, the total force on the particle cannot have a centripetal component since a normal component would produce curvilinear motion. Thus $\vec{F}$ is directed along the line of motion and can be expressed as $\vec{F} = F\hat{\tau}$, where $F = \pm|\vec{F}|$. From (1) we have, after eliminating the common factor $\hat{\tau}$ and by putting $du/dt = a$:

$$F = \gamma(u) m a\, [1 + (u^2/c^2)\gamma(u)^2].$$

But, $u^2/c^2 = 1 - \gamma(u)^{-2}$, so that, finally, $F = \gamma(u)^3 m a$. In rectilinear motion, total force *is* proportional to acceleration.

(*b*) In *uniform* (generally curvilinear) motion the speed *u* is constant and hence $du/dt = 0$. Equation (1) then reduces to $\vec{F} = \gamma(u) m \vec{a}$. As in the previous example, the total force is proportional to the acceleration.

Back to the general case (1) we note that $F \to \infty$ as $u \to c$ (explain). This means that an *infinite* force would be needed to accelerate a *massive* particle up to the speed of light (this is not a problem, however, for a *massless* particle such as a photon, which, as seen in Prob. 5, always travels at speed *c*). We also notice that, if $\vec{a} = 0$ (thus $\vec{u} = const.$ and so $u = |\vec{u}| = const.$) then $\vec{F} = 0$, in accordance with the requirement that a free particle should move with constant velocity.

**7.** Derive the LT ($x$-boost) for energy and momentum. Are conservation of energy and conservation of momentum independent issues in SR, like in classical mechanics?

***Solution:*** The energy-momentum 4-vector for a particle of mass $m$ and velocity $\vec{u}$ is

$$P^\mu = (E/c\,,\, p_x\,,\, p_y\,,\, p_z) \equiv (P^0, P^1, P^2, P^3)$$

where $E=\gamma(u)mc^2$ and $\vec{p}=\gamma(u)m\vec{u}$, with $\gamma(u)=(1-u^2/c^2)^{-1/2}$. The LT of $P^\mu$ is

$$P^{0\prime}=\frac{P^0-(v/c)P^1}{(1-v^2/c^2)^{1/2}}\,,\quad P^{1\prime}=\frac{P^1-(v/c)P^0}{(1-v^2/c^2)^{1/2}}\,,\quad P^{2\prime}=P^2,\quad P^{3\prime}=P^3.$$

Substituting for the $P^\mu$ and $P^{\mu\,\prime}$ we have:

$$E'=\frac{E-vp_x}{(1-v^2/c^2)^{1/2}}\,,\quad p_x{}'=\frac{p_x-vE/c^2}{(1-v^2/c^2)^{1/2}}\,,\quad p_y{}'=p_y\,,\quad p_z{}'=p_z\,.$$

For the inverse transformation we set $-v$ in place of $v$.

We notice that the LT mixes energy and momentum, so that the separation of these physical quantities is *frame-dependent* (what appears as energy to one inertial observer may appear as momentum to another, and vice versa). Therefore, as discussed in Sec. 3.3, conservation of energy and conservation of momentum are not separate matters in SR (both energy *and* momentum must be conserved).

**8.** By using the work-energy theorem, justify the expression $E=\gamma(u)mc^2$ for the total relativistic energy of a particle of mass $m$ moving with speed $u$. Does *external* potential energy contribute to the energy $E$ ? How about *internal* potential energy in the case of a composite body?

***Solution:*** As found in Prob. 6, the total force on the particle at time $t$ is given by

$$\vec{F}=\frac{mu}{c^2}\gamma(u)^3\frac{du}{dt}\vec{u}+\gamma(u)m\frac{d\vec{u}}{dt} \tag{1}$$

where $\gamma(u)=(1-u^2/c^2)^{-1/2}$. The work of this force on the particle along the spatial path from $A$ to $B$ is given by the line integral

$$W=\int_A^B \vec{F}\cdot\overrightarrow{dl}=\int_A^B \vec{F}\cdot\vec{u}\,dt \tag{2}$$

where, by (1),

$$\vec{F}\cdot\vec{u}=\frac{mu}{c^2}\gamma(u)^3\frac{du}{dt}\vec{u}\cdot\vec{u}+\gamma(u)m\vec{u}\cdot\frac{d\vec{u}}{dt}\,.$$

But,

$$\vec{u}\cdot\vec{u}=u^2 \quad\text{and}\quad \vec{u}\cdot\frac{d\vec{u}}{dt}=u\frac{du}{dt} \quad \text{(prove!)}$$

so that

$$\vec{F}\cdot\vec{u}=\gamma(u)mu\left[1+\frac{u^2}{c^2}\gamma(u)^2\right]\frac{du}{dt}=\gamma(u)^3mu\frac{du}{dt}\,.$$

Now, as found in Prob. 6,

$$\frac{d\gamma}{dt}=\frac{u}{c^2}\gamma(u)^3\frac{du}{dt} \;\Rightarrow\; \gamma(u)^3 u\frac{du}{dt}=c^2\frac{d\gamma}{dt} \quad .$$

So,

$$\vec{F}\cdot\vec{u}=mc^2\frac{d\gamma}{dt}=\frac{d}{dt}\left(\gamma mc^2\right) \; ,$$

where the mass $m$ is assumed constant. We set

$$E=\gamma(u)mc^2 \tag{3}$$

Then $\vec{F}\cdot\vec{u}=dE/dt$, and therefore (2) yields

$$W=E_B-E_A \tag{4}$$

The quantity $E$ defined in (3) is called the *total relativistic energy* of the particle. For a particle at rest, $u=0\Rightarrow\gamma(u)=1$ and so $E_{\text{rest}}=mc^2$ (*rest energy*). It is thus reasonable to call the difference $E-E_{\text{rest}}$ the *kinetic energy* of the particle, and define

$$T=E-mc^2 \;\Leftrightarrow\; E=mc^2+T \; .$$

Then from (4) we get

$$W=T_B-T_A$$

which formally expresses the *work-energy theorem* in SR. We note that, even if $\vec{F}$ is a conservative force, the corresponding potential energy is not included in $E$, hence does not contribute to the total relativistic energy.

A composite body may be considered as a system of particles (e.g., the molecules of the body) of masses $m_1, m_2, m_3, \ldots$ The relativistic *internal energy* of the body is

$$E_{\text{int}}=\sum m_i c^2+(T+U)_{\text{int}}$$

where $T_{\text{int}}$ and $U_{\text{int}}$ are the *internal kinetic energy* and the *internal potential energy*, respectively, where the latter energy is associated with internal forces. The energy $E_{\text{int}}$ is the total energy of the body in the body's rest frame (or $C$-frame), which is the unique frame in which the center of mass of the system is at rest and the system's total momentum is zero.

The system as a whole may also be viewed as a single "particle" of mass $M$ and rest energy $Mc^2$ equal to $E_{\text{int}}$:

$$Mc^2=\sum m_i c^2+(T+U)_{\text{int}} \;\Rightarrow\; M=\sum m_i+(1/c^2)(T+U)_{\text{int}} \; .$$

To eliminate the usual arbitrariness in the definition of the potential energy $U$ to within an additive constant, we rewrite the last relation in terms of *differences* of physical quantities within a time interval $\Delta t$:

$$\Delta M=(1/c^2)(\Delta T+\Delta U)_{\text{int}} \tag{5}$$

where we have taken into account that the masses $m_i$ are constant ($\Delta m_i=0$).

Now, for a free body, subject to no external forces, the total relativistic energy $E$ must be constant in *any* inertial frame and, in particular, in the body's $C$-frame where $E=E_{int}=Mc^2$. Thus $\Delta E=0 \Rightarrow \Delta M=0$ and, by Eq. (5), $\Delta T_{int}+\Delta U_{int}=0$. This could not be satisfied if we did not include the potential energy $U_{int}$ in the total energy $E$, given that, in general, kinetic energy alone is not conserved. We conclude that internal potential energy *has* to be included in the energy of the body and, therefore, it contributes to the body's mass, which is not the case with potential energy associated with external forces.

## CHAPTER 4

# COVARIANCE IN ELECTRODYNAMICS

### 4.1 The Maxwell Equations

As we know, the *Maxwell equations* describe the behavior (that is, the laws of change in space and time) of the electromagnetic (e/m) field. This field is represented by the pair $(\vec{E},\vec{B})$, where $\vec{E}$ and $\vec{B}$ are the electric and the magnetic field, respectively.

The Maxwell equations are a system of four partial differential equations that is self-consistent, in the sense that these equations are compatible with one another. The self-consistency of the system also implies the satisfaction of two important conditions that are physically meaningful:

- the *equation of continuity*, related to *conservation of charge*; and
- the *e/m wave equation* in its various forms.

We stress that the above conditions are *necessary but not sufficient* for the validity of the Maxwell system. Thus, although every solution $(\vec{E},\vec{B})$ of this system obeys a wave equation separately for the electric and the magnetic field, an arbitrary pair of fields $(\vec{E},\vec{B})$, each field satisfying the corresponding wave equation, does not necessarily satisfy the Maxwell system itself. Also, the principle of conservation of charge *cannot replace* any one of Maxwell's equations. These remarks are justified by the fact that the aforementioned two necessary conditions are derived by *differentiating* the Maxwell system and, in this process, part of the information carried by this system is lost. [Recall, similarly, that cross-differentiation of the Cauchy-Riemann relations of complex analysis yields the Laplace equation, by which, however, we cannot recover the Cauchy-Riemann relations.]

We adopt the following differential form of Maxwell's equations [1,2]:

$$\begin{aligned} &(a)\quad \vec{\nabla}\cdot\vec{E}=\frac{\rho}{\varepsilon_0} &&(c)\quad \vec{\nabla}\times\vec{E}=-\frac{\partial\vec{B}}{\partial t} \\ &(b)\quad \vec{\nabla}\cdot\vec{B}=0 &&(d)\quad \vec{\nabla}\times\vec{B}=\mu_0\vec{J}+\varepsilon_0\mu_0\frac{\partial\vec{E}}{\partial t} \end{aligned} \tag{4.1}$$

where $\rho$, $\vec{J}$ are the charge and current densities, respectively (the "sources" of the e/m field). Both the fields and the sources are functions of the spacetime variables $(x,y,z,t)$. Equations (4.1*a*) and (4.1*b*), which describe the *div* of the e/m field at any moment, constitute *Gauss' law* for the electric and the magnetic field, respectively. In terms of physical content, (4.1*a*) expresses the Coulomb law of electricity, while (4.1*b*) rules out the possibility of existence of magnetic poles analogous to electric charges. Equation (4.1*c*) expresses the *Faraday-Henry law* (law of e/m induction) and Eq. (4.1*d*) expresses the *Ampère-Maxwell law*. Equations (4.1*a*) and (4.1*d*), which contain the sources of the e/m field, constitute the *non-homogeneous* Maxwell equations, while Eqs. (4.1*b*) and (4.1*c*) are the *homogeneous* equations of the system.

By taking the *div* of (4.1*d*) and by using (4.1*a*) we obtain an equation of continuity that physically expresses the principle of conservation of charge:

$$\vec{\nabla}\cdot\vec{J}+\frac{\partial\rho}{\partial t}=0 \tag{4.2}$$

A different kind of differentiation of the Maxwell system (4.1), by taking the *rot* of (*c*) and (*d*), leads to separate wave equations for the electric and the magnetic field:

$$\nabla^2\vec{E}-\varepsilon_0\mu_0\frac{\partial^2\vec{E}}{\partial t^2}=\frac{1}{\varepsilon_0}\vec{\nabla}\rho+\mu_0\frac{\partial\vec{J}}{\partial t} \tag{4.3}$$

$$\nabla^2\vec{B}-\varepsilon_0\mu_0\frac{\partial^2\vec{B}}{\partial t^2}=-\mu_0\vec{\nabla}\times\vec{J} \tag{4.4}$$

The point was made recently [3-5] that the Maxwell equations may be viewed as a *Bäcklund transformation* (BT) relating fields and sources. The conservation of charge and the electromagnetic wave equations then simply express the *integrability* (*consistency*) *conditions* of the BT. The BT property of the Maxwell system further supports the view according to which the four equations (4.1) constitute a set of *independent* equations [6]. This will be analytically discussed in Sec. 5.4.

### 4.2 The Electromagnetic Field Tensor

It can be shown by physical arguments (see, e.g., [2,7]) that, under a Lorentz boost in the *x* direction the fields $\vec{E}$ and $\vec{B}$ transform as follows:

$$\begin{aligned} &E_x{}'=E_x\,,\quad E_y{}'=\gamma(E_y-c\beta B_z),\quad E_z{}'=\gamma(E_z+c\beta B_y)\\ &B_x{}'=B_x\,,\quad B_y{}'=\gamma(B_y+\frac{\beta}{c}E_z),\quad B_z{}'=\gamma(B_z-\frac{\beta}{c}E_y)\end{aligned} \tag{4.5}$$

where $\beta(v)=v/c$, $\gamma(v)=(1-\beta^2)^{-1/2}=(1-v^2/c^2)^{-1/2}$. Moreover, if the densities $\rho$, $\vec{J}$ are related by $\vec{J}=\rho\vec{u}$, where $\vec{u}$ is the local velocity of the moving charge (cf. [1], Chap. 6), and if $\rho_0$ is the charge density in the rest frame of the charge, then

$$\rho=(1-u^2/c^2)^{-1/2}\rho_0\equiv\gamma(u)\rho_0\,,\quad \vec{J}=\rho\vec{u}=\rho_0\,\gamma(u)\vec{u} \tag{4.6}$$

[Careful: $\gamma(v)$ refers to the relative motion of two inertial frames (say, *S* and *S΄*) while $\gamma(u)$ refers to the motion of the charge system in the *S*-frame!] Both the electric charge and the density $\rho_0$ will be considered Lorentz scalars.

We notice that

$$\begin{aligned} &c\rho=\rho_0\,\gamma(u)c=\rho_0U^0\,,\\ &J_k=\rho_0\,\gamma(u)u_k=\rho_0U^k\,,\quad k=1,2,3\end{aligned}$$

where

$$U^\mu\equiv\left(\gamma(u)c\,,\,\gamma(u)\vec{u}\right)\equiv\left(\gamma(u)c\,,\,\gamma(u)u_x\,,\,\gamma(u)u_y\,,\,\gamma(u)u_z\right)$$

is the 4-velocity of the moving charge. Given that $(U^0, U^k)$ ($k$=1,2,3) is a 4-vector while $\rho_0$ is a scalar, it follows that $\rho_0 (U^0, U^k) = (\rho_0 U^0, \rho_0 U^k)$ is a 4-vector. This means that $(c\rho, J_k)$ must be a 4-vector. We thus define a *4-current density* by

$$J^\mu = \rho_0 U^\mu \equiv \left(c\rho, \vec{J}\right) \tag{4.7}$$

Regarding the e/m field $(\vec{E}, \vec{B})$, we notice that it has 3+3=6 independent components, namely, the set of Cartesian components of the electric and the magnetic field. We now ask the question whether these components might be the 6 independent components of some *antisymmetric* tensor $F^{\mu\nu} = -F^{\nu\mu}$. In matrix form,

$$[F^{\mu\nu}] = \begin{bmatrix} 0 & F^{01} & F^{02} & F^{03} \\ -F^{01} & 0 & F^{12} & F^{13} \\ -F^{02} & -F^{12} & 0 & F^{23} \\ -F^{03} & -F^{13} & -F^{23} & 0 \end{bmatrix} \tag{4.8}$$

As mentioned in Sec. 3.6, under an $x$-boost the tensor components transform as follows:

$$\begin{gathered} F^{01}{}' = F^{01}, \quad F^{02}{}' = \gamma (F^{02} - \beta F^{12}), \quad F^{03}{}' = \gamma (F^{03} + \beta F^{31}), \\ F^{23}{}' = F^{23}, \quad F^{31}{}' = \gamma (F^{31} + \beta F^{03}), \quad F^{12}{}' = \gamma (F^{12} - \beta F^{02}) \end{gathered} \tag{4.9}$$

[In case you are worried about an apparent discrepancy with Eq. (3.48) regarding the terms for $F^{03}{}'$ and $F^{31}{}'$, remember that $F^{13} = -F^{31}$ and $F^{13}{}' = -F^{31}{}'$.] On the other hand, the e/m-field transformation relations (4.5) can be rewritten as follows:

$$\left(\frac{E_x}{c}\right)' = \frac{E_x}{c}, \quad \left(\frac{E_y}{c}\right)' = \gamma\left(\frac{E_y}{c} - \beta B_z\right), \quad \left(\frac{E_z}{c}\right)' = \gamma\left(\frac{E_z}{c} + \beta B_y\right)$$

$$B_x' = B_x, \quad B_y' = \gamma\left(B_y + \beta \frac{E_z}{c}\right), \quad B_z' = \gamma\left(B_z - \beta \frac{E_y}{c}\right)$$

Comparison with (4.9) suggests trying the following identification of the $F^{\mu\nu}$:

$$\begin{gathered} F^{01} = -E_x/c, \quad F^{02} = -E_y/c, \quad F^{03} = -E_z/c, \\ F^{12} = -B_z, \quad F^{31} = -F^{13} = -B_y, \quad F^{23} = -B_x \end{gathered} \tag{4.10}$$

Then, $F^{01}{}' = -E_x'/c$, $F^{23}{}' = -B_x'$, etc. The antisymmetric tensor (4.8) then reads

$$[F^{\mu\nu}] = \begin{bmatrix} 0 & -E_x/c & -E_y/c & -E_z/c \\ E_x/c & 0 & -B_z & B_y \\ E_y/c & B_z & 0 & -B_x \\ E_z/c & -B_y & B_x & 0 \end{bmatrix} \tag{4.11}$$

An alternative comparison of (4.5) and (4.9), leading to a different identification of the antisymmetric-tensor components, yields the following result (by renaming $F^{\mu\nu}$ as $G^{\mu\nu}$):

$$G^{01} = -B_x\ , \quad G^{02} = -B_y\ , \quad G^{03} = -B_z\ ,$$
$$G^{12} = E_z/c\ , \quad G^{13} = -E_y/c\ , \quad G^{23} = E_x/c \qquad (4.12)$$

(Check this!) The antisymmetric tensor $G^{\mu\nu}$ then reads

$$[G^{\mu\nu}] = \begin{bmatrix} 0 & -B_x & -B_y & -B_z \\ B_x & 0 & E_z/c & -E_y/c \\ B_y & -E_z/c & 0 & E_x/c \\ B_z & E_y/c & -E_x/c & 0 \end{bmatrix} \qquad (4.13)$$

and is called the *dual tensor* (explicitly, the tensor *dual* to $F^{\mu\nu}$). Note that $G^{\mu\nu}$ can be obtained directly from $F^{\mu\nu}$ by the substitutions $\vec{E}/c \rightarrow \vec{B}$, $\vec{B} \rightarrow -\vec{E}/c$, which operation leaves the transformation relations (4.5) unchanged.

The tensor $G^{\mu\nu}$, also denoted $*F^{\mu\nu}$, is related to $F^{\mu\nu}$ as follows:

$$G^{\mu\nu} \equiv *F^{\mu\nu} = (1/2)\,\varepsilon^{\mu\nu\lambda\rho}\,F_{\lambda\rho} \qquad (4.14)$$

where

$$F_{\lambda\rho} = g_{\lambda\mu}\,g_{\rho\nu}\,F^{\mu\nu} \qquad (4.15)$$

and where $\varepsilon^{\mu\nu\lambda\rho}$ is the *Levi-Civita symbol* in 4 dimensions, equal to 1 or –1 according to whether $(\mu\nu\lambda\rho)$ is an even or odd permutation of (0,1,2,3), respectively. In particular, $\varepsilon^{0123} = 1$ [thus $\varepsilon_{0123} = -1$, since an odd number of spatial indices are lowered by using the metric tensor in a manner similar to (4.15)]. The symbol $\varepsilon^{\mu\nu\lambda\rho}$ is *antisymmetric* in *every* pair of indices (i.e., changes sign when any two indices are interchanged) and vanishes if any two indices are the same. As an example of applying (4.14), let us check the component $G^{01}$:

$$\begin{aligned} G^{01} = *F^{01} &= (1/2)\,\varepsilon^{01\lambda\rho}\,F_{\lambda\rho} = (1/2)\,(\varepsilon^{0123}\,F_{23} + \varepsilon^{0132}\,F_{32}) \\ &= (1/2)\,[F_{23} - (-F_{23})] = F_{23} = g_{2\mu}\,g_{3\nu}\,F^{\mu\nu} \\ &= g_{22}\,g_{33}\,F^{23} = F^{23} = -B_x\ . \end{aligned}$$

## 4.3 Covariant Form of Maxwell's Equations

We now wish to express the Maxwell equations (4.1) in *covariant* form; that is, as relations involving 4-vectors and/or tensors in such a way that these relations be manifestly invariant *in form* under LTs. Examples of such covariant relations were given in Sec. 3.7.

Before we begin, let us recall that the speed $c$ of light in empty space, which is the speed of propagation of *any* form of e/m radiation in general, is a direct prediction of Maxwell's equations and is given by [1]

$$c^2 = (\varepsilon_0 \mu_0)^{-1} \tag{4.16}$$

*Proposition:* (*a*) The inhomogeneous Maxwell equations (4.1*a*) and (4.1*d* ) are represented in covariant form by the set of differential relations

$$\partial_\mu F^{\mu\nu} = \mu_0 J^\nu \quad (\nu = 0,1,2,3) \tag{4.17}$$

where $J^\nu \equiv \left(c\rho,\, \vec{J}\right)$.

(*b*) The homogeneous (source-free) Maxwell equations (4.1*b*) and (4.1*c*) are represented in covariant form by the set of relations

$$\partial_\mu G^{\mu\nu} \equiv \partial_\mu {}^*F^{\mu\nu} = 0 \quad (\nu = 0,1,2,3) \tag{4.18}$$

where $G^{\mu\nu} \equiv {}^*F^{\mu\nu} = (1/2)\, \varepsilon^{\mu\nu\lambda\rho} F_{\lambda\rho}$ and where $F_{\lambda\rho} = g_{\lambda\mu}\, g_{\rho\nu} F^{\mu\nu}$.

(*c*) The homogeneous system (4.18) is equivalent to the set of relations

$$\partial_\lambda F_{\mu\nu} + \partial_\mu F_{\nu\lambda} + \partial_\nu F_{\lambda\mu} = 0 \tag{4.19}$$

*Proof:* (*a*) Note first that, according to Eqs. (3.58) and (3.59), the differential equations (4.17) obey a 4-vector transformation law under LTs. For $\nu$=0 the system (4.17) reads: $\partial_\mu F^{\mu 0} = \mu_0 J^0 \;\Rightarrow$

$$\mu_0 c\rho = \frac{\partial F^{00}}{\partial x^0} + \frac{\partial F^{10}}{\partial x^1} + \frac{\partial F^{20}}{\partial x^2} + \frac{\partial F^{30}}{\partial x^3} = 0 + \frac{1}{c}\left( \frac{\partial E_x}{\partial x} + \frac{\partial E_y}{\partial y} + \frac{\partial E_z}{\partial z} \right) \Rightarrow$$

$$\vec{\nabla} \cdot \vec{E} = \mu_0 c^2 \rho = \mu_0 \frac{1}{\varepsilon_0 \mu_0} \rho = \frac{\rho}{\varepsilon_0}$$

where in the last step we have used (4.16). For $\nu$=1 we have: $\partial_\mu F^{\mu 1} = \mu_0 J^1 \;\Rightarrow$

$$\mu_0 J_x = \frac{\partial F^{01}}{\partial x^0} + \frac{\partial F^{11}}{\partial x^1} + \frac{\partial F^{21}}{\partial x^2} + \frac{\partial F^{31}}{\partial x^3} = -\frac{1}{c^2}\frac{\partial E_x}{\partial t} + 0 + \frac{\partial B_z}{\partial y} - \frac{\partial B_y}{\partial z}$$

$$= \left( -\frac{1}{c^2}\frac{\partial \vec{E}}{\partial t} + \vec{\nabla} \times \vec{B} \right)_x$$

and similarly for $\nu$=2 and $\nu$=3. Therefore, in vector form and by using (4.16),

$$\vec{\nabla} \times \vec{B} = \mu_0 \vec{J} + \frac{1}{c^2}\frac{\partial \vec{E}}{\partial t} = \mu_0 \vec{J} + \varepsilon_0 \mu_0 \frac{\partial \vec{E}}{\partial t} \;.$$

(*b*) We leave the proof of (4.18) to the reader as an exercise.

(*c*) Assume that $\partial_\lambda F_{\mu\nu} + \partial_\mu F_{\nu\lambda} + \partial_\nu F_{\lambda\mu} = 0$ is valid for all choices of $\lambda, \mu, \nu$. We show that this also implies $\partial_\mu G^{\mu\nu} \equiv \partial_\mu {}^*F^{\mu\nu} = 0$. We have:

$$\partial_\lambda G^{\rho\lambda} = (1/2)\,\varepsilon^{\rho\lambda\mu\nu}\,\partial_\lambda F_{\mu\nu} = (1/2)\,\varepsilon^{\rho\mu\nu\lambda}\,\partial_\mu F_{\nu\lambda} = (1/2)\,\varepsilon^{\rho\nu\lambda\mu}\,\partial_\nu F_{\lambda\mu}$$

(all we did was to change the names of repeated "up" and "down" indices). But,

$$\varepsilon^{\rho\lambda\mu\nu} = \varepsilon^{\rho\mu\nu\lambda} = \varepsilon^{\rho\nu\lambda\mu}$$

(even number of permutations leading from one symbol to another). So,

$$\partial_\lambda G^{\rho\lambda} = (1/2)\,\varepsilon^{\rho\lambda\mu\nu}\,\partial_\lambda F_{\mu\nu} = (1/2)\,\varepsilon^{\rho\lambda\mu\nu}\,\partial_\mu F_{\nu\lambda} = (1/2)\,\varepsilon^{\rho\lambda\mu\nu}\,\partial_\nu F_{\lambda\mu}$$
$$= (1/6)\,\varepsilon^{\rho\lambda\mu\nu}\,(\partial_\lambda F_{\mu\nu} + \partial_\mu F_{\nu\lambda} + \partial_\nu F_{\lambda\mu}) = 0 \;\text{ (by assumption) } \Rightarrow$$

$$\partial_\lambda G^{\lambda\rho} = -\,\partial_\lambda G^{\rho\lambda} = 0\ .$$

*Example:* For $\lambda=1, \mu=2, \nu=3$ we have:

$$0 = \partial_1 F_{23} + \partial_2 F_{31} + \partial_3 F_{12} = \partial_1 F^{23} + \partial_2 F^{31} + \partial_3 F^{12} \;\text{ (explain this!)}$$
$$= -\frac{\partial B_x}{\partial x} - \frac{\partial B_y}{\partial y} - \frac{\partial B_z}{\partial z} = -\vec{\nabla}\cdot\vec{B}$$

from which we recover the "no-free-magnetic-poles" equation $\vec{\nabla}\cdot\vec{B} = 0$.

Let us return to the inhomogeneous Maxwell equations (4.17): $\partial_\mu F^{\mu\nu} = \mu_0 J^\nu$. Taking the divergence of both sides, we have:

$$\mu_0\,\partial_\nu J^\nu = \partial_\mu\,\partial_\nu F^{\mu\nu}\ .$$

But, by the antisymmetry of $F^{\mu\nu}$,

$$\partial_\mu\,\partial_\nu F^{\mu\nu} = \partial_\nu\,\partial_\mu F^{\nu\mu} = -\,\partial_\mu\,\partial_\nu F^{\mu\nu} \;\Rightarrow\; \partial_\mu\,\partial_\nu F^{\mu\nu} = 0\ .$$

Hence,

$$\partial_\nu J^\nu \equiv \partial J^\nu / \partial x^\nu = 0 \tag{4.20}$$

Setting

$$J^\nu \equiv \left(c\rho,\, \vec{J}\right) = \left(c\rho,\, J_x\,,\, J_y\,,\, J_z\right)$$

we get:

$$\frac{\partial\rho}{\partial t} + \vec{\nabla}\cdot\vec{J} = 0 \tag{4.21}$$

(Show this.) Equation (4.21) is, of course, the *equation of continuity* for electric charge, expressing *conservation of charge* [1,2].

## 4.4 Relativistic Potentials

As we have seen, the spacetime coordinates $x^\mu \equiv (ct, x, y, z)$ transform as contravariant components of a 4-vector under LTs. We now introduce the *covariant* coordinate vector[1]

$$x_\mu = g_{\mu\nu} x^\nu \equiv (ct, -x, -y, -z) \tag{4.22}$$

To express the $x^\mu$ in terms of the $x_\mu$, we write:

$$x_\lambda = g_{\lambda\rho} x^\rho \;\Rightarrow\; g^{\nu\lambda} x_\lambda = g^{\nu\lambda} g_{\lambda\rho} x^\rho = g^\nu{}_\rho x^\rho$$

where, as mentioned in Sec. 3.7, $g^\nu{}_\rho = \delta^\nu{}_\rho$. Therefore,

$$x^\nu = g^{\nu\lambda} x_\lambda \tag{4.23}$$

We also define the derivatives

$$\partial^\mu = \partial / \partial x_\mu \tag{4.24}$$

By the chain rule of differentiation,

$$\partial / \partial x_\mu = (\partial x^\nu / \partial x_\mu)\, \partial / \partial x^\nu$$

where, by (4.23),

$$\partial x^\nu / \partial x_\mu = g^{\nu\lambda} (\partial x_\lambda / \partial x_\mu) = g^{\nu\lambda}\, \delta^\mu{}_\lambda = g^{\nu\mu} = g^{\mu\nu}\,.$$

Hence,

$$\partial / \partial x_\mu = g^{\mu\nu}\, \partial / \partial x^\nu \quad \text{or} \quad \partial^\mu = g^{\mu\nu}\, \partial_\nu \tag{4.25}$$

in accordance with the definition (3.51) of $\partial^\mu$ given in Sec. 3.7. The latter definition is thus consistent with the definition (4.24) of $\partial^\mu$ given here.

As we know from electrodynamics [1,2], the electric and the magnetic field can be expressed as

$$\vec{E} = -\vec{\nabla} V - \frac{\partial \vec{A}}{\partial t}\;, \quad \vec{B} = \vec{\nabla} \times \vec{A} \tag{4.26}$$

where $V(\vec{r}, t)$ and $\vec{A}(\vec{r}, t)$ are the *e/m potentials* (scalar and vector, respectively). By these expressions the two homogeneous (source-free) Maxwell equations,

$$\vec{\nabla} \cdot \vec{B} = 0\;, \quad \vec{\nabla} \times \vec{E} = -\,\partial \vec{B} / \partial t \tag{4.27}$$

are satisfied automatically.

[1] Note that, then, $ds^2 = g_{\mu\nu}\, dx^\mu dx^\nu = dx^\mu\, dx_\mu$.

We construct the 4-component object (not yet claimed to be a 4-vector!)

$$A^{\mu} \equiv \left( \frac{V}{c} \, , \, \vec{A} \right) \tag{4.28}$$

Then the e/m field tensor $F^{\mu\nu}$ is written in terms of $A^{\mu}$ as follows:

$$F^{\mu\nu} = \partial^{\mu} A^{\nu} - \partial^{\nu} A^{\mu} \equiv \partial A^{\nu} / \partial x_{\mu} - \partial A^{\mu} / \partial x_{\nu} \tag{4.29}$$

(Notice that we differentiate with respect to the *covariant* components of the coordinate vector!) Let us see some examples of using (4.29) to recover the classical relations (4.26):

For $\mu$=0, $\nu$=1, we have:

$$F^{01} = \partial A^{1} / \partial x_{0} - \partial A^{0} / \partial x_{1} = \partial A^{1} / \partial x^{0} + \partial A^{0} / \partial x^{1} \quad \text{(explain!)}$$

from which we find

$$E_{x} = -\frac{\partial V}{\partial x} - \frac{\partial A_{x}}{\partial t} = \left( -\vec{\nabla} V - \frac{\partial \vec{A}}{\partial t} \right)_{x}$$

and similarly for $F^{02}$ and $F^{03}$. In vector form,

$$\vec{E} = -\vec{\nabla} V - \frac{\partial \vec{A}}{\partial t} \; .$$

For $\mu$=1, $\nu$=2, we have:

$$F^{12} = \partial A^{2} / \partial x_{1} - \partial A^{1} / \partial x_{2} = -\partial A^{2} / \partial x^{1} + \partial A^{1} / \partial x^{2} \quad \text{(explain!)}$$

from which we find

$$B_{z} = \frac{\partial A_{y}}{\partial x} - \frac{\partial A_{x}}{\partial y} = \left( \vec{\nabla} \times \vec{A} \right)_{z}$$

and similarly for $F^{13}$ and $F^{23}$. In vector form,

$$\vec{B} = \vec{\nabla} \times \vec{A} \; .$$

Thus $F^{\mu\nu}$ as defined in (4.29) is indeed the e/m field tensor. According to the discussion in Sec. 3.6 [see Eq. (3.49)], by the fact that the $\partial^{\mu}$ transform as (contravariant) components of a 4-vector under LTs, while $F^{\mu\nu}$ is an antisymmetric tensor, it follows that $A^{\mu}$ must be a 4-vector.

We must now show that the expression (4.29) for $F^{\mu\nu}$ satisfies the homogeneous Maxwell equations (4.18) automatically. To this end it is easier to use the equivalent form (4.19) of these equations. Before we begin, however, let us recall that we have used the metric tensor $g_{\mu\nu}$ or $g^{\mu\nu}$ to "lower" or "raise" indices, respectively, of 4-vectors. We can do the same, of course, for indices of tensors of *any* kind. Regarding (4.29) we write:

$$g_{\lambda\mu}\, g_{\rho\nu}\, F^{\mu\nu} = g_{\lambda\mu}\, g_{\rho\nu}\, (\partial^{\mu} A^{\nu} - \partial^{\nu} A^{\mu}) = g_{\lambda\mu}\, \partial^{\mu}\, (g_{\rho\nu} A^{\nu}) - g_{\rho\nu}\, \partial^{\nu}\, (g_{\lambda\mu} A^{\mu}) \;\Rightarrow$$

$$F_{\lambda\rho} = \partial_{\lambda} A_{\rho} - \partial_{\rho} A_{\lambda} = \partial A_{\rho} / \partial x^{\lambda} - \partial A_{\lambda} / \partial x^{\rho} \tag{4.30}$$

which is equivalent to (4.29). By using (4.30) we find, after some straightforward algebra:

$$\partial_{\lambda} F_{\mu\nu} + \partial_{\mu} F_{\nu\lambda} + \partial_{\nu} F_{\lambda\mu} = 0 \ .$$

*Exercise:* Prove this result.

We now come to *gauge transformations*; that is, transformations of the potentials $V$, $\vec{A}$ (or, in 4-vector form, $A^{\mu}$) which do not affect the fields $\vec{E}, \vec{B}$ (or $F^{\mu\nu}$). Classically, a gauge transformation is of the form [1,2]

$$\vec{A}' = \vec{A} + \vec{\nabla}\lambda \ , \quad V' = V - \frac{\partial\lambda}{\partial t} \tag{4.31}$$

where $\lambda(\vec{r}, t)$ is an arbitrary function. Under this transformation,

$$\vec{E} = -\vec{\nabla} V - \frac{\partial \vec{A}}{\partial t} = -\vec{\nabla} V' - \frac{\partial \vec{A}'}{\partial t} \ ,$$

$$\vec{B} = \vec{\nabla} \times \vec{A} = \vec{\nabla} \times \vec{A}'$$

We consider the transformation

$$A^{\mu\,\prime} = A^{\mu} - \partial\lambda / \partial x_{\mu} = A^{\mu} - \partial^{\mu} \lambda \tag{4.32}$$

for arbitrary $\lambda \equiv \lambda(x^{\rho})$. Explicitly,

$$A^{0\,\prime} = A^{0} - \partial\lambda / \partial x_{0} = A^{0} - \partial\lambda / \partial x^{0} \ ,$$

$$A^{k\,\prime} = A^{k} - \partial\lambda / \partial x_{k} = A^{k} + \partial\lambda / \partial x^{k} \quad (k{=}1,2,3) \ .$$

By using the fact that $A^{\mu} \equiv \left( V/c \ , \ \vec{A} \right)$, it can be shown that the transformation (4.32) yields the gauge transformation equations (4.31).

*Exercise:* Show this.

We must now show that the transformation (4.32) leaves the e/m field tensor $F^{\mu\nu}$ invariant. Let $F^{\mu\nu} = \partial^{\mu} A^{\nu} - \partial^{\nu} A^{\mu}$ and $F^{\mu\nu\,\prime} = \partial^{\mu} A^{\nu\,\prime} - \partial^{\nu} A^{\mu\,\prime}$, where $A^{\mu\,\prime} = A^{\mu} - \partial^{\mu} \lambda$. We have:

$$F^{\mu\nu\,\prime} = \partial^{\mu} (A^{\nu} - \partial^{\nu} \lambda) - \partial^{\nu} (A^{\mu} - \partial^{\mu} \lambda) = \partial^{\mu} A^{\nu} - \partial^{\nu} A^{\mu} = F^{\mu\nu} \ .$$

The transformation (4.32) gives us the freedom to choose the $A^{\mu}$ so that the following relation, called the *Lorentz condition*, be satisfied:

$$\partial_{\mu} A^{\mu} \equiv \partial A^{\mu} / \partial x^{\mu} = 0 \tag{4.33}$$

or, explicitly,

$$\vec{\nabla} \cdot \vec{A} + \frac{1}{c^2} \frac{\partial V}{\partial t} = 0 .$$

Let us now consider the inhomogeneous Maxwell equations

$$\partial_\mu F^{\mu\nu} = \mu_0 J^\nu \quad \text{where} \quad J^\nu \equiv \left( c\rho, \vec{J} \right) .$$

For $F^{\mu\nu} = \partial^\mu A^\nu - \partial^\nu A^\mu$, we have:

$$\mu_0 J^\nu = \partial_\mu (\partial^\mu A^\nu - \partial^\nu A^\mu) = \partial_\mu \partial^\mu A^\nu - \partial^\nu (\partial_\mu A^\mu) .$$

By assuming that the Lorentz condition (4.33) is satisfied, the above relation acquires a simpler form:

$$\Box^2 A^\nu \equiv \partial_\mu \partial^\mu A^\nu = \mu_0 J^\nu \tag{4.34}$$

where

$$\Box^2 = \partial_\mu \partial^\mu = \frac{1}{c^2} \frac{\partial^2}{\partial t^2} - \nabla^2$$

is the d'Alembert operator. For $\nu$=0, and by using the fact that $c^2 = (\varepsilon_0 \mu_0)^{-1}$, Eq. (4.34) yields:

$$\Box^2 V \equiv \frac{1}{c^2} \frac{\partial^2 V}{\partial t^2} - \nabla^2 V = \frac{\rho}{\varepsilon_0} \tag{4.35}$$

while for $\nu = 1, 2, 3$ we get, in vector form,

$$\Box^2 \vec{A} \equiv \frac{1}{c^2} \frac{\partial^2 \vec{A}}{\partial t^2} - \nabla^2 \vec{A} = \mu_0 \vec{J} \tag{4.36}$$

*Exercise:* Prove Eqs. (4.35) and (4.36).

## Problems

**1.** Derive the LT ($x$-boost) for the electromagnetic 4-current $J^\mu$.

***Solution:*** The 4-vector for the e/m current is

$$J^\mu = (c\rho, J_x, J_y, J_z) \equiv (J^0, J^1, J^2, J^3)$$

and the LT is

$$J^{0\prime} = \frac{J^0 - (v/c)J^1}{(1-v^2/c^2)^{1/2}}, \quad J^{1\prime} = \frac{J^1 - (v/c)J^0}{(1-v^2/c^2)^{1/2}}, \quad J^{2\prime} = J^2, \quad J^{3\prime} = J^3 .$$

Substituting for $J^\mu$ and $J^{\mu\prime}$ we get

$$\rho' = \frac{\rho - (v/c^2)J_x}{(1-v^2/c^2)^{1/2}}, \quad J_x{}' = \frac{J_x - v\rho}{(1-v^2/c^2)^{1/2}}, \quad J_y{}' = J_y, \quad J_z{}' = J_z .$$

As always, for the inverse transformation we set $-v$ in place of $v$.

**2.** Derive the LT ($x$-boost) for the electromagnetic 4-potential $A^\mu$.

***Solution:*** The 4-vector for the e/m potential is

$$A^\mu = (V/c, A_x, A_y, A_z) \equiv (A^0, A^1, A^2, A^3)$$

and the LT is

$$A^{0\prime} = \frac{A^0 - (v/c)A^1}{(1-v^2/c^2)^{1/2}}, \quad A^{1\prime} = \frac{A^1 - (v/c)A^0}{(1-v^2/c^2)^{1/2}}, \quad A^{2\prime} = A^2, \quad A^{3\prime} = A^3 .$$

This yields

$$V' = \frac{V - vA_x}{(1-v^2/c^2)^{1/2}}, \quad A_x{}' = \frac{A_x - vV/c^2}{(1-v^2/c^2)^{1/2}}, \quad A_y{}' = A_y, \quad A_z{}' = A_z .$$

**3.** Show that, in the non-relativistic limit, the LT of the electric field reduces to

$$\vec{E}' = \vec{E} + (\vec{v} \times \vec{B}) \tag{1}$$

where $\vec{v}$ is the velocity of the inertial observer $O'$ relative to the inertial observer $O$. Justify the above result by physical reasoning.

***Solution:*** With no loss of generality, we assume that the LT relating $O$ with $O'$ is an $x$-boost, so that

$$\vec{v} = (v_x, v_y, v_z) \equiv (v, 0, 0) .$$

The LT for the electric field is given by Eq. (4.5), which we write in the form

$$E_x{}' = E_x , \quad E_y{}' = \gamma(E_y - vB_z), \quad E_z{}' = \gamma(E_z + vB_y) \tag{2}$$

where $\gamma(v)=(1-v^2/c^2)^{-1/2}$. In the non-relativistic limit, $v/c\to 0 \Rightarrow \gamma=1$, and thus relations (2) reduce to

$$E_x{}'=E_x\ ,\quad E_y{}'=E_y-vB_z\ ,\quad E_z{}'=E_z+vB_y \tag{3}$$

On the other hand, by noting that

$$\begin{aligned}\vec{v}\times\vec{B} &\equiv (v_y B_z-v_z B_y\,,\ v_z B_x-v_x B_z\,,\ v_x B_y-v_y B_x)\\ &=(0\,,\,-vB_z\,,\,vB_y)\end{aligned}$$

and by taking the $x$, $y$ and $z$ components of (1), we find again the result (3). We conclude that relation (1) represents the non-relativistic limit of the LT for the electric field.

To see the physics of the situation, we consider an electric charge $q$ that is momentarily at rest relative to $O'$, thus moves with velocity $\vec{v}$ relative to $O$. According to $O$, this charge is subject to a Lorentz force $\vec{F}=q[\vec{E}+(\vec{v}\times\vec{B})]$ by the e/m field. According to $O'$, however, the charge is stationary and thus subject only to an electric force $\vec{F}'=q\vec{E}'$. Now, in the non-relativistic limit the force on a particle is frame-independent; that is, $\vec{F}=\vec{F}'$. Thus, by eliminating $q$ we have that $\vec{E}'=\vec{E}+(\vec{v}\times\vec{B})$.

**4.** It is given that the electric and the magnetic field produced by a charge $q$ moving with velocity $\vec{\upsilon}$ relative to an inertial observer are related by

$$\vec{B}=\frac{1}{c^2}(\vec{\upsilon}\times\vec{E}) \tag{1}$$

By using (1) compare the strengths of the electric and the magnetic interaction between two charges. Comment on the relative strength of the two interactions as we approach the limit of very high speeds of the charges.

***Solution:*** We consider two charges $q$ and $q'$ moving with corresponding velocities $\vec{\upsilon}$ and $\vec{\upsilon}'$ relative to an inertial observer. We regard $q'$ as the source of an e/m field and $q$ as a test charge within this field. We are interested in the force on $q$ due to the e/m field produced by $q'$. Let $(\vec{E}',\vec{B}')$ be the value of this field at the location of $q$. The electric force on $q$ is $\vec{F}_e=q\vec{E}'$ or, in magnitudes, $F_e=qE'$, while the magnetic force on $q$ is $\vec{F}_m=q(\vec{\upsilon}\times\vec{B}')$, where, according to (1),

$$\vec{B}'=\frac{1}{c^2}(\vec{\upsilon}'\times\vec{E}')$$

Therefore,

$$\vec{F}_m=\frac{q}{c^2}[\vec{\upsilon}\times(\vec{\upsilon}'\times\vec{E}')]$$

and, in terms of orders of magnitudes,

$$F_m\approx\frac{q}{c^2}\upsilon\upsilon' E'=\frac{\upsilon\upsilon'}{c^2}F_e\ \Rightarrow\ \frac{F_m}{F_e}\approx\frac{\upsilon\upsilon'}{c^2}\ .$$

We notice that, in the region of low velocities compared to the speed of light (i.e., for $\upsilon \ll c$ and $\upsilon' \ll c$) we have that $F_m \ll F_e$, while $F_m \sim F_e$ when $\upsilon\sim c$ and $\upsilon'\sim c$. This means that, while in the world of low energies (or low speeds) that we experience in our everyday lives the electric interaction between charged particles appears to be much stronger than their magnetic interaction, in the high-energy domain the two interactions become comparable to each other. This is natural in view of the fact that, after all, these interactions are the two "faces" of a single *electromagnetic* interaction. However, the unification of these interactions is truly revealed in high-energy processes and in the framework of a high-energy theory such as SR!

**5.** Consider a charge $q$ in uniform rectilinear motion relative to an inertial observer $O$. By using relativistic arguments, explain why this charge cannot emit e/m radiation.

***Solution:*** Since $q$ moves with constant velocity relative to the inertial observer $O$, it itself defines the origin of an inertial frame of reference. With respect to an observer $O'$ in this frame, $q$ is at rest. Thus the only thing recorded by $O'$ is a *static* electric field, with no presence of any e/m wave (i.e., with no emission of e/m radiation). Let us now assume that, according to $O$, the charge $q$ emits e/m radiation. This means that $O$ records the existence of an e/m wave traveling at speed $c$. But, since the propagation speed of an e/m wave is the same in all inertial frames, it follows that the wave observed by $O$ also propagates at speed $c$ relative to $O'$. This, however, contradicts the fact that $O'$ perceives no e/m wave! We conclude that neither $O$ may perceive an emission of e/m radiation from $q$.

In the case of an *accelerating* charge $q$ the above rationale breaks down since $q$ no longer defines the origin of an inertial frame of reference. Thus an observer $O'$ moving with $q$ is not an inertial observer and her measurements should not be relativistically correlated with those of the inertial observer $O$. The latter observer records emission of e/m radiation and correctly attributes it to the accelerated motion of $q$, in accordance with Maxwell's theory [1]. For the *non-inertial* observer $O'$, however, a seemingly stationary charge appears to emit e/m radiation, contrary to the predictions of electrodynamics. Observer $O'$ thus reaches an erroneous conclusion in an attempt to interpret electromagnetic phenomena in an unsuitable (i.e., non-inertial) frame of reference!

[2] https://www.aemjournal.org/index.php/AEM/article/view/311
[3] https://nausivios.snd.edu.gr/docs/2016C.pdf
[4] https://nausivios.hna.gr/docs/NCH_v8_2022_C1.pdf

## CHAPTER 5

# SPECIAL TOPICS

### 5.1 Lie Groups and Lie Algebras

This section serves as an elementary introduction to Lie groups and Lie algebras. These concepts were introduced in an informal way in Chap. 2, in connection with the Lorentz group. We now present them in more general terms.

A *group* is a set $G=\{a,b,c,...\}$ equipped with an internal "multiplication" operation with the following properties:

1. Closure: $ab\in G,\ \forall\, a,b\in G$ .
2. Associativity: $a(bc)=(ab)c$ .
3. Identity element: $\exists\, e\in G$: $ae=ea,\ \forall\, a\in G$ .
4. Inverse element: $\forall\, a\in G,\ \exists\, a^{-1}\in G$: $aa^{-1}=a^{-1}a=e$ .

A group is *Abelian* (or commutative) if $ab=ba,\ \forall\, a,b\in G$ .

A *subgroup* of $G$ is a subset $H\subseteq G$ that is itself a group under the group operation of $G$. Obviously, $H$ must contain the identity element $e$ of $G$ as well as the inverse of any element of $H$.

A map $\varphi: G\to G'$ from a group $G$ to a group $G'$ is called a *homomorphism* if it preserves group multiplication. That is, for any $a$, $b\in G$, the images $\varphi(a)\in G'$ and $\varphi(b)\in G'$ satisfy the relation

$$\varphi(a)\,\varphi(b)=\varphi(ab)\ .$$

If the homomorphism $\varphi$ is 1-1, it is called an *isomorphism*.

A real *Lie algebra* $\mathcal{L}$ of dimension $n$ is an $n$-dimensional real vector space equipped with an internal *Lie bracket* operation [ , ] that satisfies the following properties:

1. Closure: $[a,b]\in\mathcal{L},\ \forall\, a,b\in\mathcal{L}$ .
2. Linearity: $[\kappa a+\lambda b,\, c]=\kappa[a,c]+\lambda[b,c]\ \ (\kappa,\lambda\in R)$ .
3. Antisymmetry: $[a,b]=-[b,a]$ . Corollary: $[a,a]=0$ .
4. Jacobi identity: $\big[a,[b,c]\big]+\big[b,[c,a]\big]+\big[c,[a,b]\big]=0$ .

A Lie algebra is *Abelian* (or commutative) if $[a,b]=0$ , $\forall\, a,b\in\mathcal{L}$.

A *subalgebra* $S$ of $\mathcal{L}$ is a subspace of $\mathcal{L}$ that itself is a Lie algebra. The algebra $S$ is an *invariant subalgebra* or *ideal* of $\mathcal{L}$ if $[a,b]\in S,\ \forall\, a\in S,\ b\in\mathcal{L}$ . A Lie algebra $\mathcal{L}$ is said to be *simple* if it contains no ideals other than itself; $\mathcal{L}$ is *semisimple* if it contains no *Abelian* ideals.

Examples of Lie algebras:

1. The algebra of $(m\times m)$ matrices, with $[A,B]=AB-BA$ (*commutator*). Diagonal matrices constitute an Abelian subalgebra of this algebra.

2. The algebra of all vectors in 3-dimensional space, with $[\vec{V},\vec{W}]=\vec{V}\times\vec{W}$ (vector product). Vectors parallel to a given axis form an Abelian subalgebra of this algebra.

A map $\psi$: $\mathcal{L}\to\mathcal{L}'$ from a Lie algebra $\mathcal{L}$ to a Lie algebra $\mathcal{L}'$ is a *homomorphism* if it satisfies the following properties:

$$\psi(\kappa a+\lambda b)=\kappa\,\psi(a)+\lambda\,\psi(b)\quad(\kappa,\lambda\in R)\;;$$
$$\psi([a,b])=[\psi(a),\psi(b)]\;.$$

If the map $\psi$ is 1-1, it is called an *isomorphism*. Isomorphic Lie algebras $\mathcal{L}$ and $\mathcal{L}'$ have equal dimensions [1]: $dim\mathcal{L}=dim\mathcal{L}'$.

Let $\{\tau_i\;/\;i=1,2,\dots,n\}$ be a basis of an $n$-dimensional Lie algebra $\mathcal{L}$. Since the Lie bracket of any two basis elements $\tau_i$ and $\tau_j$ is an element of $\mathcal{L}$, it must be a linear combination of the $\{\tau_k\}$. That is,

$$[\tau_i\,,\tau_j]=C_{ij}^k\,\tau_k \tag{5.1}$$

(sum on $k$ from 1 to $n$). By the antisymmetry of the Lie bracket, $C_{ij}^k=-C_{ji}^k$. The real constants $C_{ij}^k$ are called *structure constants* of the Lie algebra $\mathcal{L}$.

*Proposition:* Let $\psi$: $\mathcal{L}\to\mathcal{L}'$ be a Lie algebra isomorphism. If $\{\tau_k\}$ $(k=1,2,\dots,n)$ is a basis of $\mathcal{L}$, then $\{\psi(\tau_k)\}$ is a basis of $\mathcal{L}'$.

*Proof:* Being a basis of $\mathcal{L}$, the $\{\tau_k\}$ are linearly independent; hence no linear combination of them can be zero (unless, of course, all coefficients are trivially zero). Now, by the properties of $\psi$, a linear combination of the $\{\tau_k\}$ is mapped onto a linear combination of the $\{\psi(\tau_k)\}$ with the same coefficients. This means that the latter combination cannot vanish, since it can only be zero if the former one is zero as well; that is, if all coefficients in the combination are zero. We conclude that the $\{\psi(\tau_k)\}$ are linearly independent and may serve as a basis for $\mathcal{L}'$.

*Proposition:* Isomorphic Lie algebras share common structure constants.

*Proof:* Let $\psi$: $\mathcal{L}\to\mathcal{L}'$ be a Lie algebra isomorphism and let $\tau_i$, $\tau_j$ be any two basis elements of $\mathcal{L}$. Then, $\psi(\tau_i)$ and $\psi(\tau_j)$ are basis elements of $\mathcal{L}'$. By the properties of $\psi$,

$$\psi([\tau_i\,,\tau_j])=\psi(C_{ij}^k\,\tau_k)\;\Rightarrow\;[\psi(\tau_i),\psi(\tau_j)]=C_{ij}^k\,\psi(\tau_k)\;;\;\text{ q.e.d.}$$

Roughly speaking, a *Lie group* is a group $G$ whose elements depend on a number of parameters that can be varied in a continuous way. The *dimension n* of $G$ is the number of real parameters parametrizing the elements of $G$. We assume that $dimG=n$ and we let $\{\lambda^1,\lambda^2,\dots,\lambda^n\}$ be the set of $n$ parameters of $G$. We arrange the parameterization of $G$ so that the identity element of $G$ corresponds to $\lambda^k=0$ for all $k=1,2,\dots,n$.

An important class of Lie groups consists of groups of $(m\times m)$ matrices parametrized by $n$ parameters $\lambda^k$ $(k=1,2,...,n)$. Since an $(m\times m)$ matrix produces a linear transformation on an $m$-dimensional Euclidean space, matrix groups are called *linear groups*.

Lie groups are closely related to Lie algebras. Let $G$ be an $n$-dimensional Lie group of $(m\times m)$ matrices $A(\lambda^1, \lambda^2, ..., \lambda^n) \equiv A(\lambda)$ (where by $\lambda$ we collectively denote the set of the $n$ parameters $\lambda^k$). We define the $n$ $(m\times m)$ matrices $\tau_k$ by

$$\tau_k = \frac{\partial A(\lambda)}{\partial \lambda^k}\Big|_{\lambda^1=\lambda^2=\cdots=\lambda^n=0} \tag{5.2}$$

or, in terms of matrix elements,

$$(\tau_k)_{pq} = \frac{\partial A_{pq}}{\partial \lambda^k}\Big|_{\lambda^1=\lambda^2=\cdots=\lambda^n=0}$$

$(k=1,2,...,n\,;\;\; p, q=1,2,...,m)$. The $n$ matrices $\tau_k$ are called *infinitesimal operators* (or generators) of the Lie group $G$ and form the basis of an $n$-dimensional real Lie algebra $\mathcal{L}$ [1]. Thus $[\tau_i, \tau_j] = C_{ij}^k \tau_k$, where the $C_{ij}^k$ are real constants. A general element $a$ of $\mathcal{L}$ is written as a linear combination of the $\tau_k$: $a=\xi^k\tau_k$ (sum on $k$), for real coefficients $\xi^k$. [Note carefully that the matrix elements $(\tau_k)_{pq}$ *themselves* are *not* required to be real numbers!]

Now, let $a=\lambda^k\tau_k$ be the general element of $\mathcal{L}$. The general element $A(\lambda)$ of the Lie group $G$ parametrized by the $\lambda^k$ can then be written as [1]

$$A(\lambda) = e^a = \exp(\lambda^k\tau_k) \tag{5.3}$$

where $e^a$ is the matrix exponential function

$$e^a \equiv \exp a = \sum_{l=0}^{\infty} \frac{a^l}{l!} = 1 + a + \frac{a^2}{2} + \cdots$$

For infinitesimal values of the parameters $\lambda^k$ we may use the approximate expression

$$e^a \simeq 1+a$$

so that

$$A(\lambda) \simeq 1 + \lambda^k\tau_k \tag{5.4}$$

The simplest example of a Lie group is a one-parameter continuous group, such as the group $SO(2)$ of rotations on a plane. A rotation of a vector by an angle $\lambda$ is represented by the $(2\times2)$ orthogonal matrix

$$A(\lambda) = \begin{bmatrix} \cos\lambda & -\sin\lambda \\ \sin\lambda & \cos\lambda \end{bmatrix} \quad (\lambda \in R).$$

(Notice that $A^tA=1$ and $\det A=1$.) Then

$$\frac{dA}{d\lambda} = \begin{bmatrix} -\sin\lambda & -\cos\lambda \\ \cos\lambda & -\sin\lambda \end{bmatrix}$$

and, by Eq. (5.2), the single basis element $\tau$ of the associated Lie algebra is

$$\tau = \frac{dA}{d\lambda}\Big|_{\lambda=0} = \begin{bmatrix} 0 & -1 \\ 1 & 0 \end{bmatrix} .$$

According to (5.3), $A(\lambda)=e^{\lambda\tau}$ and, for infinitesimal $\lambda$, $A(\lambda) \simeq 1+\lambda\tau$. Indeed, by setting $\sin\lambda=\lambda$ and $\cos\lambda=1$, we have:

$$A(\lambda) \simeq \begin{bmatrix} 1 & -\lambda \\ \lambda & 1 \end{bmatrix} = \begin{bmatrix} 1 & 0 \\ 0 & 1 \end{bmatrix} + \lambda \begin{bmatrix} 0 & -1 \\ 1 & 0 \end{bmatrix} = 1 + \lambda\tau .$$

Another single-parameter Lie group is the unitary group $U(1)$ with elements $\{e^{i\lambda}\}$ ($\lambda \in R$), which may be regarded as (1×1) matrices. Consider the map $\varphi$: $U(1) \rightarrow SO(2)$ defined by

$$\varphi\left(e^{i\lambda}\right) = \begin{bmatrix} \cos\lambda & -\sin\lambda \\ \sin\lambda & \cos\lambda \end{bmatrix} .$$

This map is a homomorphism, since

$$\begin{aligned} \varphi\left(e^{i\lambda} \cdot e^{i\lambda'}\right) = \varphi\left(e^{i(\lambda+\lambda')}\right) &= \begin{bmatrix} \cos(\lambda+\lambda') & -\sin(\lambda+\lambda') \\ \sin(\lambda+\lambda') & \cos(\lambda+\lambda') \end{bmatrix} \\ &= \begin{bmatrix} \cos\lambda & -\sin\lambda \\ \sin\lambda & \cos\lambda \end{bmatrix} \begin{bmatrix} \cos\lambda' & -\sin\lambda' \\ \sin\lambda' & \cos\lambda' \end{bmatrix} \\ &= \varphi\left(e^{i\lambda}\right) \cdot \varphi\left(e^{i\lambda'}\right) . \end{aligned}$$

Moreover, it can be shown that the map $\varphi$ is 1-1. Therefore, $\varphi$ is a Lie-group isomorphism.

We finally remark that isomorphic Lie groups have isomorphic Lie algebras [1]. More generally, under certain restrictions, homomorphic Lie groups may have isomorphic Lie algebras. An example of homomorphic Lie groups (with isomorphic Lie algebras) is treated in Sec. 5.2.

## 5.2 Homomorphism of the Lorentz Group with *SL*(2,*C* )

Let $L=SO(3,1)\uparrow$ be the restricted Lorentz group, which is represented by (4×4) real matrices $\Lambda=[\Lambda^{\mu}{}_{\nu}]$ with $\det\Lambda=1$ and $\Lambda^{0}{}_{0} \geq 1$ (this group is also called the proper orthochronous Lorentz group). Also, let $SL(2,C)$ be the group of complex (2×2) matrices with unit determinant. Both $L$ and $SL(2,C)$ are six-parameter Lie groups (that is, the elements of each group depend on 6 *real* parameters). It is thus natural to ask whether a homomorphic relation between these two groups exists.

Both $L$ and $SL(2,C)$ are matrix transformation groups. We seek a correspondence

$$A \in SL(2,C) \rightarrow \Lambda(A) \in L$$

such that $\Lambda(AB)=\Lambda(A)\Lambda(B)$ for any $A, B \in SL(2,C)$. We thus seek a homomorphic mapping associating a transformation produced by an $SL(2,C)$ element $A$ with a corresponding Lorentz transformation (LT) produced by an element $\Lambda(A)$ of $L$. The latter transformation is of the form

$$[x^{\mu}{}'] = \Lambda\,[x^{\mu}] \quad \Leftrightarrow \quad x^{\mu}{}' = \Lambda^{\mu}{}_{\nu}\, x^{\nu} \tag{5.5}$$

($\mu$= 0,1,2,3) and is such that $x^{\mu}{}' x_{\mu}{}' = x^{\mu} x_{\mu}$ , where $x_{\mu}=g_{\mu\nu}\, x^{\nu}$ and $x^{\mu} x_{\mu} = g_{\mu\nu}\, x^{\mu} x^{\nu}$. With the standard metric

$$g = [\,g_{\mu\nu}] = diag\,(1, -1, -1, -1) \quad (\mu, \nu = 0,1,2,3)$$

we have:

$$(x^{0}{}')^2 - (x^{1}{}')^2 - (x^{2}{}')^2 - (x^{3}{}')^2 = (x^{0})^2 - (x^{1})^2 - (x^{2})^2 - (x^{3})^2 \tag{5.6}$$

Here is the plan: We seek a class of complex (2×2) matrices $X$, depending on 4 real parameters – namely, the spacetime coordinates $x^{\mu}$ – and a properly defined action of $A \in SL(2,C)$ on $X$ to produce a new matrix $X\,'$, the parameters $x^{\mu}{}'$ of which are such that the Lorentz invariance condition (5.6) is satisfied. By this process an $SL(2,C)$ transformation will be related to a LT.

A candidate class of 4-parameter complex matrices $X$ is the set of (2×2) *hermitian* (self-adjoint) matrices. These form a linear space with 4-dimensional basis $\{\sigma_{\mu}\}$, where $\sigma_0$=1 (unit matrix) and where $\sigma_i$ ($i$=1,2,3) are the *Pauli matrices*:

$$\sigma_1 = \begin{bmatrix} 0 & 1 \\ 1 & 0 \end{bmatrix} \;,\quad \sigma_2 = \begin{bmatrix} 0 & -i \\ i & 0 \end{bmatrix} \;,\quad \sigma_3 = \begin{bmatrix} 1 & 0 \\ 0 & -1 \end{bmatrix} .$$

A general hermitian matrix $X$ can be written as

$$X = x^{\mu}\sigma_{\mu} = x^0 1 + x^1\sigma_1 + x^2\sigma_2 + x^3\sigma_3 = \begin{bmatrix} x^0 + x^3 & x^1 - i\,x^2 \\ x^1 + i\,x^2 & x^0 - x^3 \end{bmatrix} \tag{5.7}$$

[Notice that $X^{\dagger}=X$ , where $X^{\dagger}= (X^{\,t})^{*}$ .] The determinant of $X$ is

$$\det X = (x^{0})^2 - (x^{1})^2 - (x^{2})^2 - (x^{3})^2 = x^{\mu} x_{\mu} \tag{5.8}$$

For a given $A \in SL(2,C)$, consider now the matrix transformation

$$X' = AXA^{\dagger} \tag{5.9}$$

The matrix $X\,'$ is again hermitian: $(X')^{\dagger} = X'$ . (*Exercise:* Show this by using the general matrix property $(M_1 M_2 M_3)^{\dagger} = M_3{}^{\dagger} M_2{}^{\dagger} M_1{}^{\dagger}$ .) Thus $X\,'$ will be of the form (5.7):

$$X' = x^{\mu\prime}\sigma_\mu \tag{5.10}$$

and, according to (5.8),

$$\det X' = (x^{0\prime})^2 - (x^{1\prime})^2 - (x^{2\prime})^2 - (x^{3\prime})^2 = x^{\mu\prime} x_\mu' .$$

But, by (5.9), $\det X' = \det X$, given that $\det A = 1$. This means that $x^{\mu\prime} x_\mu' = x^\mu x_\mu$ , which implies that the transformation from the $x^\mu$ to the $x^{\mu\prime}$ is a LT.

In conclusion: The matrix transformation (5.9) *induces* a transformation on the coefficients $x^\mu$ of the $\sigma_\mu$ in (5.7), which is equivalent to a LT (5.5): $[x^{\mu\prime}] = \Lambda(A)\,[x^\mu]$, where the transformation matrix $\Lambda$ depends on the particular $SL(2,C)$ matrix $A$.

From (5.9) and by using (5.7), (5.10) and (5.5), we have:

$$x^{\mu\prime}\sigma_\mu = A\,x^\mu\sigma_\mu A^\dagger \;\Rightarrow\; \Lambda^\mu{}_\nu\, x^\nu\sigma_\mu = A\,x^\nu\sigma_\nu A^\dagger .$$

In order for this to be valid independently of the values of the $x^\nu$, the following matrix relation must be satisfied:

$$\Lambda^\mu{}_\nu\,\sigma_\mu = A\sigma_\nu A^\dagger \quad \text{or} \quad [\Lambda(A)]^\mu{}_\nu\,\sigma_\mu = A\sigma_\nu A^\dagger \tag{5.11}$$

where we have emphasized that the (4×4) matrix $\Lambda = [\Lambda^\mu{}_\nu]$ is dependent upon the choice of the (2×2) matrix $A$.

Let us now show that the matrices $\Lambda$ of the Lorentz group $L$ are a representation of the group $SL(2,C)$, i.e., that the correspondence $SL(2,C) \to L$ is a homomorphism. Indeed, let $A_1, A_2 \in SL(2,C) \Rightarrow A_1 A_2 \in SL(2,C)$ . Then, by using (5.11),

$$\begin{aligned}
[\Lambda(A_1A_2)]^\mu{}_\nu\,\sigma_\mu &= (A_1A_2)\,\sigma_\nu\,(A_1A_2)^\dagger = A_1\left(A_2\sigma_\nu A_2^\dagger\right)A_1^\dagger \\
&= A_1\left([\Lambda(A_2)]^\rho{}_\nu\,\sigma_\rho\right)A_1^\dagger = [\Lambda(A_2)]^\rho{}_\nu\,[\Lambda(A_1)]^\mu{}_\rho\,\sigma_\mu \\
&= [\Lambda(A_1)\Lambda(A_2)]^\mu{}_\nu\,\sigma_\mu
\end{aligned}$$

and, given that the matrices $\sigma_\mu$ are linearly independent,

$$[\Lambda(A_1A_2)]^\mu{}_\nu = [\Lambda(A_1)\Lambda(A_2)]^\mu{}_\nu \;\Rightarrow\; \Lambda(A_1A_2) = \Lambda(A_1)\Lambda(A_2) \;, \quad \text{q.e.d.}$$

We now seek an explicit expression for $[\Lambda(A)]^\mu{}_\nu$. We have:

$$\begin{aligned}
&\Lambda^\mu{}_\nu\,\sigma_\mu = A\sigma_\nu A^\dagger \;\Rightarrow\; \Lambda^\mu{}_\nu\,\sigma_\lambda\sigma_\mu = \sigma_\lambda A\sigma_\nu A^\dagger \;\Rightarrow \\
&tr\left(\Lambda^\mu{}_\nu\,\sigma_\lambda\sigma_\mu\right) = \Lambda^\mu{}_\nu\,tr(\sigma_\lambda\sigma_\mu) = tr\left(\sigma_\lambda A\sigma_\nu A^\dagger\right).
\end{aligned}$$

By the properties of the Pauli matrices, $tr(\sigma_\lambda\sigma_\mu) = 2\delta_{\lambda\mu}$. So,

$$\Lambda^\mu{}_\nu\,tr(\sigma_\lambda\sigma_\mu) = 2\,\Lambda^\mu{}_\nu\,\delta_{\lambda\mu} = 2\,\Lambda^\lambda{}_\nu$$

and hence

$$2\,\Lambda^\lambda{}_\nu = tr\left(\sigma_\lambda A\sigma_\nu A^\dagger\right) \quad \text{or} \quad \Lambda^\mu{}_\nu = \frac{1}{2}\,tr\left(\sigma_\mu A\sigma_\nu A^\dagger\right).$$

To make this expression covariant-looking, we introduce the matrices $\tilde{\sigma}^{\mu} \equiv \sigma_{\mu}$ and write, finally,

$$[\Lambda(A)]^{\mu}{}_{\nu} = \frac{1}{2} tr\left(\tilde{\sigma}^{\mu} A \sigma_{\nu} A^{\dagger}\right) \tag{5.12}$$

We notice that $\Lambda(A)=\Lambda(-A)$, which means that the relation $SL(2,C) \to L$ is a *two-to-one homomorphism*. We say that $SL(2,C)$ constitutes a two-valued, 2-dimensional representation of the restricted Lorentz group $L$. Both groups are 6-parameter Lie groups and have isomorphic 6-dimensional real Lie algebras [1].

An analogous homomorphism exists between the subgroups $SU(2)$ and $SO(3)$ of $SL(2,C)$ and $L$, respectively, where $SU(2)$ is the group of (2×2) unitary matrices with unit determinant, while $SO(3)$ is the group of (3×3) real orthogonal matrices with unit determinant [1]. Both $SU(2)$ and $SO(3)$ are 3-parameter Lie groups and their respective Lie algebras are 3-dimensional and isomorphic to each other.

## 5.3 Flat and Curved Spaces

Consider an $n$-dimensional space $S$ with coordinates $(x^1, \ldots, x^n) \equiv (x^k)$. Let $(x^i)$ and $(x^i+dx^i)$ be two neighboring points $P$ and $P'$ of $S$, and let $ds$ denote the distance between these points as measured on $S$. Since the path connecting $P$ and $P'$ is infinitesimal, we may approximately regard this path as an infinitesimal straight-line segment. Moreover, we assume that the value of $ds$ is invariant under any change of coordinates $(x^i) \to (x^{i\,\prime})$ on $S$. Finally, we assume that an $n^2$-component field $g_{ij}(x)$ can be defined on $S$ (where by $x$ we collectively denote the $x^k$) such that

($a$) $\det g \neq 0$ at all points $P \equiv (x^k)$ of $S$, where $g \equiv [g_{ij}(x)]$ is an $(n \times n)$ matrix;

($b$) the squared distance $ds^2$ (*infinitesimal metric form*) can be expressed as

$$ds^2 = g_{ij}(x)\, dx^i dx^j \tag{5.13}$$

The space $S$ is then said to be a *Riemannian space*. If a global coordinate system $(x^k)$ exists on $S$ such that all matrix elements $g_{ij}$ are constants (i.e., independent of the $x^k$), the Riemannian space $S$ is a *flat* space; if no such coordinate system exists, the space $S$ is *curved*.

In tensor analysis [2,3] the field $g_{ij}(x)$ is called the *metric tensor*. Without loss of generality this tensor may be assumed symmetric: $g_{ij}(x)=g_{ji}(x)$. Indeed, assume that

$$ds^2 = h_{ij}(x)\, dx^i dx^j$$

where the $h_{ij}(x)$ have no particular symmetry. Then,

$$ds^2 = (h_{ij}\, dx^i dx^j + h_{ji}\, dx^j dx^i)/2 = [(h_{ij} + h_{ji})/2]\, dx^i dx^j \equiv g_{ij}\, dx^i dx^j$$

where

$$g_{ij} = g_{ji} = (h_{ij} + h_{ji})/2\ .$$

If $g_{ij}(x)=0$ for $i \neq j$, i.e., if the matrix $g$ is diagonal, the coordinate system $(x^k)$ is an *orthogonal* coordinate system. In particular, if $g_{ij}(x)=\delta_{ij}$ so that $g$ is the $(n\times n)$ unit matrix, the coordinate system $(x^k)$ is a *Cartesian* system and the flat space $S$ is a *Euclidean space*. Equation (5.13) then takes the form

$$ds^2 = (dx^1)^2 + (dx^2)^2 + \ldots + (dx^n)^2 \tag{5.14}$$

and expresses the generalized *Pythagorean theorem* in $n$ dimensions.

Given that $ds^2 > 0$, an obvious requirement for the matrix $g$ of the metric of $S$ is that $g_{ij} > 0$ for all $i, j$. This condition is relaxed in Special Relativity, however, where the metric of 4-dimensional flat spacetime (*Minkowski space*) is represented by the $(4\times 4)$ diagonal matrix

$$g = diag(1, -1, -1, -1) \, .$$

Note that $ds^2$ may be positive, negative or zero in this case.

Let us see some examples of metric structures:

1. Consider a 2-dimensional space $S$ with coordinates $(r,\theta)$ and infinitesimal metric form

$$ds^2 = dr^2 + r^2 d\theta^2 \, .$$

Define new coordinates $(x, y)$ by

$$x = r\cos\theta \ , \quad y = r\sin\theta \, .$$

Then, $dx = \cos\theta\, dr - r\sin\theta\, d\theta$, $dy = \sin\theta\, dr + r\cos\theta\, d\theta$, and

$$(dx)^2 + (dy)^2 = dr^2 + r^2 d\theta^2 \quad \text{(show this).}$$

Therefore

$$ds^2 = (dx)^2 + (dy)^2 \, ,$$

which is of the Euclidean form (5.14). A coordinate transformation thus exists that reduces the given metric to that of a plane surface, which surface is a 2-dimensional flat space. In fact, $(r,\theta)$ are polar coordinates on the plane while $(x, y)$ are the usual Cartesian coordinates.

2. Let $S$ be a spherical surface of radius $a$, which is again a 2-dimensional space. In spherical coordinates $(r, \theta, \varphi)$ and for constant $r=a$, the metric form on $S$ is

$$ds^2 = a^2 (d\theta^2 + \sin^2\theta \, d\varphi^2) \, .$$

The matrix $g$ representing the metric on $S$ is

$$g = [g_{ij}(\theta, \varphi)] = \begin{bmatrix} a^2 & 0 \\ 0 & a^2 \sin^2\theta \end{bmatrix} .$$

As can be shown [3] no coordinate transformation $(\theta, \varphi) \to (x^1, x^2)$ on $S$ can reduce $ds^2$ to the form (5.13) with all $g_{ij}$ constant. Thus $S$ is a *curved* space.

3. Let $S$ be a cylindrical surface of radius $a$, which is another example of a 2-dimensional space. The axis of the cylinder coincides with the $z$-axis of a cylindrical system of coordinates $(\rho, \varphi, z)$ and, for constant $\rho = a$, the metric form on $S$ is

$$ds^2 = a^2 d\varphi^2 + dz^2$$

so that

$$g = [g_{ij}(\varphi, z)] = \begin{bmatrix} a^2 & 0 \\ 0 & 1 \end{bmatrix} .$$

Since $g$ is a constant matrix, the surface $S$ is a flat space. Moreover, one may define new coordinates $(x^1, x^2) \equiv (a\varphi, z)$ on $S$, so that

$$ds^2 = (dx^1)^2 + (dx^2)^2$$

which is of the Euclidean form (5.14). We notice that a cylindrical surface *looks locally* like a plane, although *globally* the two surfaces have different topological properties. This local equivalence can be visualized as follows: One may imagine cutting the cylindrical surface $S$ along a line parallel to the $z$-axis and then developing the surface on a plane. This can be done without stretching the surface (if the latter is assumed elastic), so that all lengths on $S$ will be preserved after development on the plane. The coordinates $(x^1, x^2)$ on $S$ will become Cartesian coordinates on the plane.

On the contrary, no such development on a plane is possible for any section of a spherical surface $S$ *without stretching* the surface, i.e., without changing lengths on $S$. Geometrically this reflects the fact that one cannot define Cartesian coordinates on a sphere, which is a *genuinely curved* space (in contrast to a cylindrical surface which is *intrinsically flat*).

## 5.4 On the Independence of Maxwell's Equations

The Maxwell equations for the electromagnetic (e/m) field are written, in differential form,

$$\begin{aligned} &(a) \quad \vec{\nabla} \cdot \vec{E} = \frac{\rho}{\varepsilon_0} &&(c) \quad \vec{\nabla} \times \vec{E} = -\frac{\partial \vec{B}}{\partial t} \\ &(b) \quad \vec{\nabla} \cdot \vec{B} = 0 &&(d) \quad \vec{\nabla} \times \vec{B} = \mu_0 \vec{J} + \varepsilon_0 \mu_0 \frac{\partial \vec{E}}{\partial t} \end{aligned} \tag{5.15}$$

By taking the *div* of (5.15*d*) and by using (5.15*a*) we find the equation of continuity that expresses conservation of charge:

$$\vec{\nabla} \cdot \vec{J} + \frac{\partial \rho}{\partial t} = 0 \tag{5.16}$$

Relation (5.16) places a severe restriction on the charge and current densities that appear on the right-hand sides of (5.15*a*) and (5.15*d*). A different sort of differentiation of the Maxwell system (5.15), by taking the *rot* of (*c*) and (*d*), leads to separate wave equations for the electric and the magnetic field.

In most textbooks on electromagnetism the Maxwell equations (5.15) are treated as a consistent set of four independent partial differential equations (PDEs). A number of authors, however, have doubted the independence of this system. Specifically, they argue that (5.15*a*) and (5.15*b*) – the equations for the *div* of the e/m field, expressing Gauss' law for the corresponding fields – are redundant since they "may be derived" from (5.15*c*) and (5.15*d*) in combination with the equation of continuity (5.16). If this is true, Coulomb's law – the most important experimental law of electricity – loses its status as an independent law and is reduced to a derivative theorem. The same can be said with regard to the non-existence of magnetic poles in Nature. In this section we present some recent ideas in support of the view that the Maxwell equations *do* form a system of independent PDEs [4].

To begin with, let us recall that a part of the "redundant" *div* equations is contained in the covariant equation $\partial_\mu F^{\mu\nu} = \mu_0 J^\nu$, while the other part is contained in the equation $\partial_\mu {}^*F^{\mu\nu} = 0$. Thus, by discarding Eqs. (5.15*a*) and (5.15*b*) we spoil the covariant formulation of Maxwell's equations! But there is more to be said.

As far as we know, the first who doubted the independent status of the two Gauss' laws in electrodynamics was Julius Adams Stratton in his 1941 famous (and, admittedly, very attractive) book [5]. His reasoning may be described as follows:

By taking the *div* of (5.15*c*), the left-hand side vanishes identically while on the right-hand side we may change the order of differentiation with respect to space and time variables. The result is:

$$\frac{\partial}{\partial t}\left(\vec{\nabla}\cdot\vec{B}\right) = 0 \tag{5.17}$$

On the other hand, by taking the *div* of (5.15*d*) and by using the equation of continuity (5.16), we find that

$$\frac{\partial}{\partial t}\left(\vec{\nabla}\cdot\vec{E} - \frac{\rho}{\varepsilon_0}\right) = 0 \tag{5.18}$$

And the line of argument continues as follows: According to (5.17) and (5.18) the quantities $\vec{\nabla}\cdot\vec{B}$ and $(\vec{\nabla}\cdot\vec{E} - \rho/\varepsilon_0)$ are constant in time at every point $(x, y, z)$ of the region $\Omega$ of space that concerns us. *If* we now assume that there has been a period of time during which no e/m field existed in the region $\Omega$, then, in that period,

$$\vec{\nabla}\cdot\vec{B} = 0 \quad \text{and} \quad \vec{\nabla}\cdot\vec{E} - \rho/\varepsilon_0 = 0 \tag{5.19}$$

identically. Later on, although an e/m field did appear in $\Omega$, the left-hand sides in (5.19) continued to vanish everywhere within this region since, as we said above, those quantities are time-constant at every point of $\Omega$. Thus, by the equations for the *rot* of the e/m field and by the principle of conservation of charge – the status of which was elevated from derivative theorem to fundamental law of the theory – we derived Eqs. (5.19) (valid for *all t*), which are precisely the first two Maxwell equations (5.15*a*) and (5.15*b*)!

According to this reasoning, the electromagnetic theory is not based on four independent Maxwell equations but rather on *three* independent equations only; namely, the Faraday-Henry law (5.15*c*), the Ampère–Maxwell law (5.15*d*), and the principle of conservation of charge (5.16).

What makes this view questionable is the assumption that, for *every* region $\Omega$ of space there exists some period of time during which the e/m field in $\Omega$ vanishes. This hypothesis is arbitrary and is not dictated by the theory itself. (It is likely that no such region exists in the Universe!) Therefore, the argument that led from relations (5.17) and (5.18) to relations (5.19) is not convincing since it was based on an arbitrary and, in a sense, artificial initial condition: that the e/m field was zero at some time $t$=0 and before.

Let us assume for the sake of argument, however, that there exists a region $\Omega$ within which the e/m field is zero for $t < t_0$ and nonzero for $t > t_0$. The critical issue is what happens at $t=t_0$; specifically, whether the functions expressing the e/m field are *continuous* at that moment. If they indeed are, the field starts from zero and gradually increases to nonzero values; thus, the line of reasoning that led from (5.17) and (5.18) to (5.19) is acceptable. There are physical situations, however, where the appearance of an e/m field is abrupt. For instance, the moment we connect the ends of a metal wire to a battery, an electric field suddenly appears in the interior of the wire and a magnetic field appears in the exterior. An even more "dramatic" example is the phenomenon of pair production in particle-physics experiments, where a charged particle–antiparticle pair is created and a nonzero e/m field appears at that moment. In such cases the e/m field is *non-continuous* at $t=t_0$ and its time derivative is *not defined* at this instant. Therefore, the line of reasoning that leads from (5.17) and (5.18) to (5.19) again collapses.

Note also a circular reasoning in Stratton's approach. It is assumed that, in a region $\Omega$ where no e/m field exists, the second of relations (5.19) is valid identically. This means that the vanishing of the electric field in $\Omega$ automatically implies the absence of electric charge in that region. This fact, however, follows from Gauss' law (5.15*a*); thus it may not be used *a priori* as a tool for proving the law itself!

In general, conservation laws emerge as *consequences* of the fundamental equations of a theory. In particular, conservation of charge, expressed by the continuity equation (5.16), is derived by *differentiating* the Maxwell system (5.15) and, as is well known, in the process of differentiation of a system of PDEs some part of the information carried by the system is lost. Therefore, the equation of continuity (5.16) cannot be regarded as more fundamental than any equation in the system (5.15) and hence may not replace any equation in this system.

It is thus our view that the Maxwell equations form a system of four independent PDEs that express respective laws of Nature. Moreover, the self-consistency of this system imposes two conditions that physically express the conservation of charge and the wave behavior of the time-dependent e/m field. We now re-examine this issue from the more formal point of view of *Bäcklund transformations* (BTs) [6-8]. To begin with, let us see the simplest, perhaps, example of a BT.

The *Cauchy-Riemann relations* of complex analysis,

$$u_x = v_y \quad (a) \qquad u_y = -v_x \quad (b) \tag{5.20}$$

(where subscripts indicate partial differentiations with respect to the indicated variables) constitute a BT for the *Laplace equation*,

$$w_{xx} + w_{yy} = 0 \tag{5.21}$$

Let us explain this: Suppose we want to solve the system (5.20) for $u$, for a given choice of the function $v(x,y)$. To see if the PDEs (5.20*a*) and (5.20*b*) match for solution for $u$, we must compare them in some way. We thus differentiate (5.20*a*) with respect to $y$ and (5.20*b*) with respect to $x$, and equate the mixed derivatives of $u$. That is, we apply the *integrability condition* (or *consistency condition*) $(u_x)_y = (u_y)_x$ . In this way we eliminate the variable $u$ and we find a condition that must be obeyed by $v(x,y)$:

$$v_{xx} + v_{yy} = 0 .$$

Similarly, by using the integrability condition $(v_x)_y = (v_y)_x$ to eliminate $v$ from the system (5.20), we find the necessary condition in order that this system be integrable for $v$, for a given function $u(x,y)$:

$$u_{xx} + u_{yy} = 0 .$$

We conclude that the integrability of the system (5.20) with respect to either variable requires that the other variable satisfy the Laplace equation (5.21).

Let now $v_0(x,y)$ be a known solution of the Laplace equation (5.21). Substituting $v=v_0$ in the system (5.20) we can integrate this system with respect to $u$. It is not hard to show (by eliminating $v_0$ from the system) that the solution $u$ will also satisfy the Laplace equation. As an example, by choosing the solution $v_0(x,y)=xy$ of (5.21) we find a new solution $u(x,y)= (x^2-y^2)/2 + C$ .

Generally speaking, a Bäcklund transformation is a system of PDEs connecting two functions (say, $u$ and $v$) in such a way that the consistency of the system requires that $u$ and $v$ independently satisfy the respective, higher-order PDEs $F[u]=0$ and $G[v]=0$. Analytically, in order that the system be integrable for $u$, the function $v$ must be a solution of $G[v]=0$; conversely, in order that the system be integrable for $v$, the function $u$ must be a solution of $F[u]=0$. If $F$ and $G$ happen to be functionally identical, as in the example given above, the BT is said to be an *auto-Bäcklund* transformation.

Classically, BTs are useful tools for finding solutions of nonlinear PDEs. In [6-8], however, we suggested that BTs may also be useful for solving *linear systems* of PDEs. The prototype example that we used was the Maxwell equations in empty space:

$$\begin{aligned} &(a)\quad \vec{\nabla}\cdot\vec{E}=0 &&(c)\quad \vec{\nabla}\times\vec{E}=-\frac{\partial\vec{B}}{\partial t}\\ &(b)\quad \vec{\nabla}\cdot\vec{B}=0 &&(d)\quad \vec{\nabla}\times\vec{B}=\varepsilon_0\mu_0\frac{\partial\vec{E}}{\partial t} \end{aligned} \tag{5.22}$$

Here we have a system of four PDEs for two vector fields that are functions of the spacetime coordinates ($x$, $y$, $z$, $t$). We would like to find the integrability conditions necessary for self-consistency of the system (5.22). To this end, we try to uncouple the system to find separate second-order PDEs for $\vec{E}$ and $\vec{B}$, the PDE for each field being a necessary condition in order that the system (5.22) be integrable for the other field. This uncoupling, which eliminates either field (electric or magnetic) in favor of the other, is achieved by properly differentiating the system equations and by using suitable vector identities, in a manner similar in spirit to that which took us from the first-order Cauchy-Riemann system (5.20) to the separate second-order Laplace equa-tions (5.21) for *u* and *v*. As can be shown, the only nontrivial integrability conditions for the system (5.22) are those obtained by using the vector identities

$$\vec{\nabla}\times(\vec{\nabla}\times\vec{E})=\vec{\nabla}(\vec{\nabla}\cdot\vec{E})-\nabla^2\vec{E}$$

and

$$\vec{\nabla}\times(\vec{\nabla}\times\vec{B})=\vec{\nabla}(\vec{\nabla}\cdot\vec{B})-\nabla^2\vec{B} \ .$$

By these we obtain separate wave equations for the electric and the magnetic field:

$$\nabla^2\vec{E} - \varepsilon_0\mu_0\frac{\partial^2\vec{E}}{\partial t^2}=0 \ ,$$

$$\nabla^2\vec{B} - \varepsilon_0\mu_0\frac{\partial^2\vec{B}}{\partial t^2}=0 \ .$$

We conclude that the Maxwell system (5.22) in empty space is a BT relating the e/m wave equations for the electric and the magnetic field, in the sense that the wave equation for each field is an integrability condition for solution of the system (5.22) in terms of the other field.

The case of the full Maxwell equations (5.15) is more complex due to the presence of the source terms $\rho$, $\vec{J}$ in the non-homogeneous equations (5.15*a*) and (5.15*d*). As it turns out, the self-consistency of the BT (5.15) imposes conditions on the terms of non-homogeneity as well as on the fields themselves. The latter conditions are the non-homogeneous wave equations

$$\nabla^2\vec{E} - \varepsilon_0\mu_0\frac{\partial^2\vec{E}}{\partial t^2}=\frac{1}{\varepsilon_0}\vec{\nabla}\rho + \mu_0\frac{\partial\vec{J}}{\partial t} \ ,$$

$$\nabla^2\vec{B} - \varepsilon_0\mu_0\frac{\partial^2\vec{B}}{\partial t^2}=-\mu_0\vec{\nabla}\times\vec{J}$$

while the condition regarding the source terms alone is precisely the continuity equation (5.16) expressing conservation of charge.

In summary: From a mathematical perspective, the Maxwell system (5.15) may be viewed as a Bäcklund transformation (BT) the integrability conditions of which (i.e., the necessary conditions for self-consistency of the system) yield separate (generally non-homogeneous) wave equations for the electric and the magnetic field, as well as the equation of continuity (5.16). These integrability conditions are derived by differentiating the BT (5.15) in different ways, thus they carry less information than the BT itself. Consequently, none of the integrability conditions may replace any equation in the system (5.15). In particular, the continuity equation (5.16) cannot be a partial substitute for Gauss' law (5.15*a*).

[1] https://nausivios.hna.gr/docs/NCH_v8_2022_C1.pdf

[2] http://nausivios.snd.edu.gr/docs/2016C.pdf

[3] http://www.aemjournal.org/index.php/AEM/article/view/311

# SELECTED BIBLIOGRAPHY

## Addendum 1. An analysis of the "twin paradox"

It is shown that, among all possible motions in spacetime, inertial motions are the ones exhibiting maximum proper time. By taking this into account, the so-called twin paradox can be resolved.

### 1. Introduction

The "twin paradox" is a standard paradigm in university-physics textbooks discussing Special Relativity (relativity in flat spacetime); see, e.g., [1-4]. It is certainly the most remarkable example of asymmetry between inertial and non-inertial frames of reference.

The story is well known: The brother stays at home while his twin sister goes on a round trip at very high speed. When she returns home she finds that her brother has grown older than her. Now, the brother can understand why this happened: she has been moving relative to him and hence her clock has been running slower than his. But, from the point of view of his sister, it was *he* that was moving relative to her. Why then isn't she the older one?

The key to resolve the paradox is to realize that the symmetry (reciprocity) of the time dilation effect exists only between *inertial* observers. And, while the brother *is* such an observer, his sister is *not* as she has experienced acceleration in order to perform her round trip. Hence only the brother is in a position to properly analyze the situation.

The resolution of the twin paradox involves the concept of *proper time*, a Lorentz-invariant quantity representing time measured by a clock along its own worldline (spacetime trajectory). As will be shown, among all possible motions between two fixed points (events) in spacetime, inertial motions are the ones exhibiting maximum proper time. It is thus the stationary brother's clock that will record the longest time.

So, let us begin by defining proper time.

### 2. Proper time and time dilation

We consider 4-dimensional flat spacetime. Let $x^\mu \equiv (x^0, x^1, x^2, x^3) \equiv (ct, x, y, z)$ be the spacetime coordinates of an inertial frame of reference $S$ used by an inertial observer (where $c$ is the speed of light in empty space) and let $x^{\mu\,\prime} \equiv (ct', x', y', z')$ be the coordinates of another frame $S'$ used by a different inertial observer (we recall that the speed of light is a frame-independent quantity). According to Special Relativity [5] the two spacetime coordinate systems are related by the linear *Lorentz transformation*

$$x^{\mu\,\prime} = \Lambda^\mu{}_\nu \, x^\nu \qquad (1)$$

(the familiar summation convention for repeated up and down indices is assumed) where the constant (4×4) matrix $\Lambda \equiv [\Lambda^\mu{}_\nu]$ ($\mu, \nu$=0,1,2,3) satisfies the equation

$$\Lambda^t g \Lambda = g \quad \Leftrightarrow \quad \Lambda^\mu{}_\lambda \, g_{\mu\nu} \, \Lambda^\nu{}_\rho = g_{\lambda\rho} \qquad (2)$$

with

$$g \equiv [g_{\mu\nu}] = \begin{bmatrix} 1 & 0 & 0 & 0 \\ 0 & -1 & 0 & 0 \\ 0 & 0 & -1 & 0 \\ 0 & 0 & 0 & -1 \end{bmatrix} = diag\,(1,-1,-1,-1) \qquad (3)$$

Moreover, $\Lambda$ is required to obey the constraints

$$\det\Lambda = +1 \;, \quad \Lambda^0{}_0 \geq 1 \qquad (4)$$

In group-theoretical terms, $\Lambda$ belongs to the *restricted Lorentz group* $SO(3,1)^{\uparrow}$.

By differentiating (1) we get the infinitesimal Lorentz transformation

$$dx^{\mu\,\prime} = \Lambda^{\mu}{}_{\nu}\, dx^{\nu} \qquad (5)$$

(since $\Lambda^{\mu}{}_{\nu}$ is constant). We define the *spacetime interval*

$$\begin{aligned} ds^2 = g_{\mu\nu}\, dx^{\mu} dx^{\nu} &= (dx^0)^2 - (dx^1)^2 - (dx^2)^2 - (dx^3)^2 \\ &= c^2 dt^2 - dx^2 - dy^2 - dz^2 \end{aligned} \qquad (6)$$

As can be shown, $ds^2$ is a *Lorentz scalar*, i.e., a quantity invariant under the transformation (1). This means that

$$ds^2 = g_{\mu\nu}\, dx^{\mu} dx^{\nu} = g_{\mu\nu}\, dx^{\mu\,\prime} dx^{\nu\,\prime} \qquad (7)$$

and, explicitly,

$$c^2 dt^2 - dx^2 - dy^2 - dz^2 = c^2 (dt')^2 - (dx')^2 - (dy')^2 - (dz')^2 \,.$$

Note that $ds^2$ may be positive, negative or zero. In particular, the Lorentz invariance of the relation $ds^2 = 0$, valid along the worldline (spacetime trajectory) of a light ray, is equivalent to the invariance of the speed $c$ of light upon passing from one inertial reference frame to another.

If $ds^2 > 0$ (*timelike interval*) then $ds = (ds^2)^{1/2}$ may be an element of spacetime distance along the worldline of a massive particle, as viewed by an inertial observer using a frame $S$ with coordinates $x^{\mu}$ or $(x, y, z, t)$. The worldline need not be straight, which means that the particle may execute *accelerated* motion relative to the frame $S$. We define the Lorentz scalar $d\tau$ (*proper-time interval*) by

$$d\tau = ds / c \quad \Leftrightarrow \quad ds = c\, d\tau \qquad (8)$$

Proper time represents time as measured by a clock *moving with the particle* (thus moving along the particle's worldline and being at rest relative to the particle). Let us see why.

According to $S$, the speed of the particle and the clock moving with it is

$$u = dl/dt = (dx^2 + dy^2 + dz^2)^{1/2} / dt$$

where $dl$ is the length of an infinitesimal displacement along the spatial trajectory of the particle. From (6) we have that $ds^2 = c^2 dt^2 - dl^2$ and, given that $dl = u\,dt$,

$$ds^2 = (c^2 - u^2)\,dt^2 \;\Rightarrow\; ds = (c^2 - u^2)^{1/2}\,dt\,.$$

By (8) we then have:

$$d\tau = \left(1 - \frac{u^2}{c^2}\right)^{1/2} dt \;\Leftrightarrow\; dt = \left(1 - \frac{u^2}{c^2}\right)^{-1/2} d\tau \equiv \gamma(u)\,d\tau \tag{9}$$

where $\gamma(u) = (1 - u^2/c^2)^{-1/2}$.

Relation (9) is valid for a frame $S$ with respect to which the particle and the associated clock move with speed $u$. Since $d\tau$ is a Lorentz scalar, its value is independent of the particular frame $S$. Consider now a *local* inertial frame $S'$ relative to which the particle and the clock are *momentarily* at rest (obviously, an infinite number of such frames are needed, one for each momentary position of the clock). Then, $u'=0$, $\gamma(u')=1$ and, by (9), $dt' = \gamma(u')\,d\tau = d\tau$. In conclusion:

> *In the rest frame of the particle (and thus of the clock) the proper-time interval $d\tau$ is literally a time interval, as measured by the clock. Proper time is thus actual time measured along the particle's own spacetime trajectory (worldline).*

For any other frame $S$, relative to which $u \neq 0$, we have that $\gamma(u)>1$ and so $dt > d\tau$, hence $dt > dt'$. This expresses the familiar relativistic effect of *time dilation*. Note carefully that, while the *time* interval $dt$ depends on the particular frame $S$, the *proper*-time interval $d\tau$ is, by its very definition (8), a frame-independent quantity! However, $d\tau$ is a genuine time interval in the rest frame of the particle, whereas it is a *spacetime* interval in any other frame relative to which the particle is in motion.

## 3. Time measurements along worldlines; the maximum proper time

To simplify our subsequent analysis, we consider two-dimensional spacetime and an inertial observer using a frame $S$ with coordinates $(x, t)$ (thus we assume that $y=z=0$). The observer observes two clocks (1) and (2) moving along his $x$-axis and draws their worldlines as shown in Fig. 1.

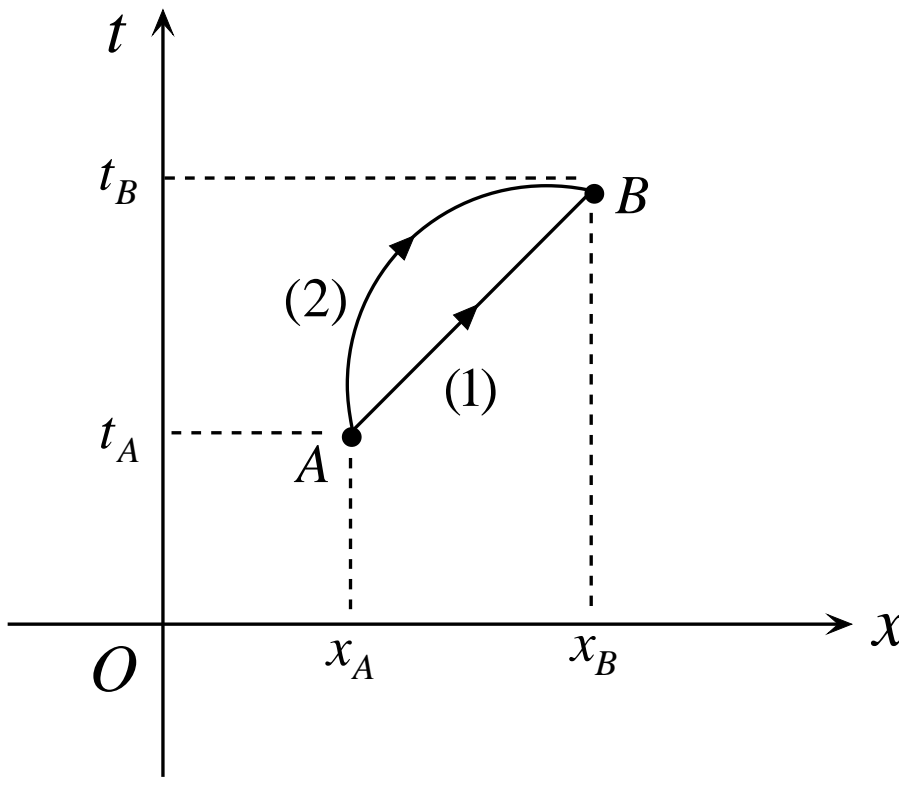

Fig. 1

The clocks start moving together at time $t_A$ , from the same point $x_A$ of the $x$-axis, and meet again at time $t_B$ at point $x_B$ . Thus the worldlines of both clocks connect the spacetime point (*event*) $A$ with the spacetime point $B$. Clock (1) moves *inertially* since its worldline is straight, while the motion of clock (2) is *accelerated* in view of the respective curved worldline. Regarding the curved worldline (2), we can imagine dividing it into an infinite number of infinitesimal linear segments, along each of which the motion of the clock may be considered inertial. Each segment is associated with a local inertial frame relative to which the clock is momentarily at rest.

Since the worldlines (1) and (2) describe real motions at speeds less than $c$, all elementary spacetime intervals along these lines must be *timelike* [5]:

$$ds^2 = c^2 dt^2 - dx^2 > 0 \quad \Rightarrow \quad ds = (ds^2)^{1/2} \in R \ .$$

Moreover, $ds = c\, d\tau \Rightarrow d\tau = ds / c$ , where $d\tau$ is the proper time of the worldline segment, equal to the time measured by a local inertial frame relative to which the corresponding clock is momentarily at rest. So, the *time* interval $d\tau$ measured in the frame of this clock will be equal to the *spacetime* interval $ds / c$ measured in the frame $S$ of the observer relative to whom the two clocks are in motion.

According to either clock the total time between the events $A$ and $B$ is given by the line integral

$$\tau = \int_A^B d\tau = (1/c) \int_A^B ds = (1/c) \int_A^B (c^2 dt^2 - dx^2)^{1/2} \qquad (10)$$

(a Lorentz scalar since $d\tau$ is a frame-independent quantity). Of course, the value of the integral (10) depends on the spacetime path from $A$ to $B$ and will be different for the worldlines (1) and (2). As can be proven (see Appendix), among all possible (timelike) worldlines connecting $A$ and $B$, the unique straight-line path (1) corresponding to *inertial* motion of the associated clock corresponds to the *maximum* proper time $\tau$. The inertial clock (1) will thus measure the *longest* time.

Now, according to the observer using the frame $S$ with coordinates $(x, t)$, the time separation between the events $A$ and $B$ is $\Delta t = t_B - t_A$ . Due to the time dilation effect this time interval appears *longer* than the proper time $\tau$ measured by the inertial clock (1): $\Delta t > \tau$.

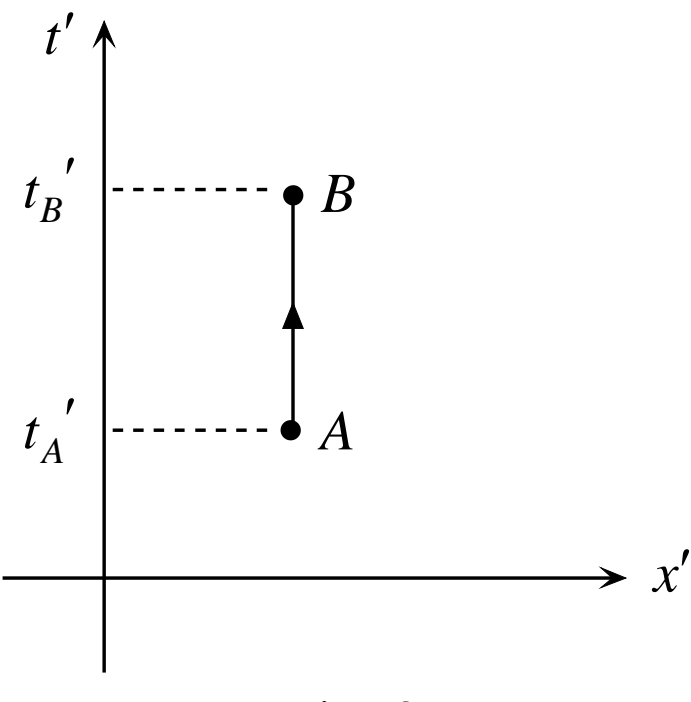

Fig. 2

A particular inertial frame is the frame of the clock (1) itself. Let $S'$ be that frame, with spacetime coordinates $(x', t')$ (see Fig. 2). Since the clock is stationary in $S'$, its worldline in this frame will correspond to $x'$=*const.* (e.g., $x'$=0). In this case the proper time $\tau$ measured by the clock *is* equal to the time difference $\Delta t' = t_B' - t_A'$ . That is, $\Delta t' = \tau$ . Of course, $\Delta t' < \Delta t$ , where $\Delta t$ is the time interval in the frame of Fig. 1.

## 4. The twin paradox

We now come to the "paradox": John and Mary are twins. While John stays at home, his sister goes on a round trip at very high speed (!). When Mary returns home, she finds that John has grown older than her.

Note that John and Mary are *inequivalent* observers, given that John is an *inertial* observer while Mary is *not* (due to the various accelerations she must undergo to perform a round trip). So, John's worldline in an inertial frame will be a straight line, while Mary's worldline will be curved. To simplify matters, we choose this inertial frame to be the rest frame of John himself, with coordinates $(x, y, z, t)$, and we assume that Mary's trip is confined to the $x$-axis. So, in effect we are dealing with a two-dimensional spacetime with coordinates $(x, t)$. We let (1) and (2) be the worldlines of John and Mary, respectively (see Fig. 3).

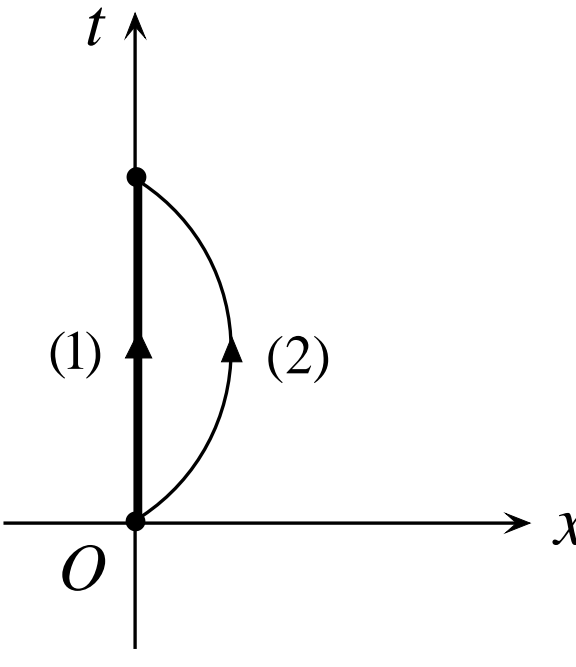

Fig. 3

Let $\tau_1$ and $\tau_2$ be the proper times along the worldlines (1) and (2), equal to the times measured by the clocks of John and Mary, respectively, during Mary's round trip. According to the discussion in Sec. 3, the *longest* proper time corresponds to the clock executing *inertial* motion, which here is John's stationary clock. Therefore $\tau_1 > \tau_2$ , which means that John will record a longer time interval. Hence John will be older than Mary when she returns home.

One might now argue that, from Mary's point of view, it is John who has performed a round trip while she has remained stationary. Why then isn't *she* the oldest one at the moment of John's "return"?

The answer is that the preceding analysis was valid only with respect to an *inertial* frame of reference, like that of John's. A similar analysis of the situation from Mary's rest frame would be improper, given that Mary is not an inertial observer. Thus her "conclusion" that she ought to be the older one would simply be wrong! (See [2] for a more thorough discussion.)

## Appendix

Consider a line integral of the form

$$Q = \int_{t_1}^{t_2} f\,[x(t),\, x'(t),\, t]\, dt \qquad \text{(A.1)}$$

where $f$ is a given function of the indicated variables (with $x'=dx/dt$) and where $x(t)$ is a curve joining two points $(x_1, t_1)$ and $(x_2, t_2)$ on the plane $(x,t)$; that is,

$$x(t_1) = x_1\,, \quad x(t_2) = x_2 \qquad \text{(A.2)}$$

We seek the curve $x(t)$ for which the value of the integral $Q$ is an *extremum* (whether this might be a maximum or a minimum). As shown in the calculus of variations (see, e.g., Chap. 2 of [6]) this particular function $x(t)$ satisfies the *Euler-Lagrange equation*

$$\frac{\partial f}{\partial x} - \frac{d}{dt}\frac{\partial f}{\partial x'} = 0 \qquad \text{(A.3)}$$

with the initial condition (A.2).

Let us now go back to the line integral (10), expressing proper time along the worldline of a particle or a clock in two-dimensional spacetime:

$$\tau = (1/c)\int_A^B (c^2 dt^2 - dx^2)^{1/2}$$

where $c^2 dt^2 - dx^2 > 0$ (timelike interval) and where $A$ and $B$ are fixed spacetime points (events). We have:

$$(c^2 dt^2 - dx^2)^{1/2} = (c^2 - x'^2)^{1/2}\, dt\,.$$

Thus,

$$\tau = (1/c)\int_{t_A}^{t_B} (c^2 - x'^2)^{1/2}\, dt \qquad \text{(A.4)}$$

We seek the spacetime curve $x(t)$ that will make the integral in (A.4) an extremum. To this end, we call

$$(c^2 - x'^2)^{1/2} \equiv f\,(x, x', t)$$

and demand that the differential equation (A.3) be satisfied. Since

$$\frac{\partial f}{\partial x} = 0\,, \qquad \frac{\partial f}{\partial x'} = -\frac{x'}{(c^2 - x'^2)^{1/2}}\,,$$

from (A.3) we have that

$$\frac{x'}{(c^2 - x'^2)^{1/2}} = const. \;\Rightarrow\; x'^2 = const.\,(c^2 - x'^2) \;\Rightarrow\; x'^2 = const.$$

and so $x'(t) = const.$ Thus $x(t)$ is a linear function: $x(t) = \kappa t + \lambda$.

Geometrically, this function describes a *straight* worldline on the plane $(x, t)$ and corresponds to *inertial* motion from $A$ to $B$. This motion thus corresponds to an *extremum* of the integral in (A.4), thus an extremum of the proper time $\tau$ measured by a clock moving from $A$ to $B$. But, is this extremum a maximum or a minimum?

Let us first note that, physically, $x'$ represents the velocity of a (generally accelerated) clock that moves along a (generally curved) worldline from $A$ to $B$, in the inertial frame $S$ with spacetime coordinates $(x,t)$: $x' = dx/dt = u$. We thus rewrite (A.4) as

$$\tau = (1/c)\int_{t_A}^{t_B} (c^2 - u^2)^{1/2}\, dt\,.$$

We have: $(c^2 - u^2)^{1/2} = c\,(1 - u^2/c^2)^{1/2} = c\,\gamma(u)^{-1}$, where $\gamma(u) = (1 - u^2/c^2)^{-1/2}$. Hence,

$$\tau = \int_{t_A}^{t_B} \gamma(u)^{-1}\, dt \qquad \text{(A.5)}$$

Let us assume that the frame $S$ is the frame of the inertial observer whose worldline from $A$ to $B$ is a straight line (this choice of frame will not affect the value of $\tau$ given that this quantity is a Lorentz scalar). In his own frame $S$, the observer (and his clock) is stationary, so that $u$=0, $\gamma(u)$=1 and, by (A.5), $\tau_1 = t_B - t_A$ (see Fig. 4).

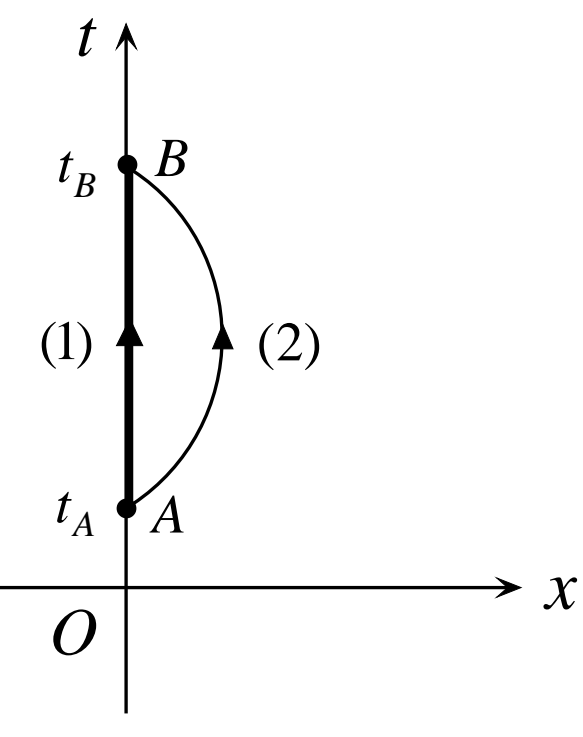

Fig. 4

Any other worldline connecting $A$ with $B$ will necessarily be curved and will represent a clock in accelerated motion. In this case $u \neq 0$ and $\gamma(u)^{-1} < 1$. The integration in (A.5) will thus yield a value $\tau_2 < t_B - t_A$.

*Conclusion:* The inertial observer’s straight worldline corresponds to *maximum* proper time, equal to the actual time measured by the observer’s own clock. This observer, therefore, will record the *longest* time between the events $A$ and $B$.

# Addendum 2. Gravitation and spacetime geometry

Some basic ideas of gravitational theory (both classical and relativistic) are discussed. In particular, the effect of gravity on the geometry of spacetime is explained.

## 1. Introduction

In a recent book on Special Relativity (SR) [1] we have seen that, in the context of this theory, spacetime is regarded as a *flat* pseudo-Euclidean (or *Lorentzian*) space. It was mentioned, however, that, according to General Relativity (GR), spacetime becomes *curved* in the presence of gravity. This means that gravity is intimately related to the intrinsic geometric characteristics of spacetime.

The connection between gravity (a physical concept) and geometry (a mathematical one) is not an easy thing to explain without recourse to some advanced mathematics. In this article, however, we will attempt to describe some fundamental ideas of GR at a mostly conceptual level. For a fuller appreciation of the theory the reader is referred to the many excellent textbooks available in the literature (for a partial list, see, e.g., [2-9], although many other great books exist).

At the heart of GR is the *principle of equivalence* of inertial and gravitational mass, by which principle gravity is seen to be locally equivalent to acceleration. On the other hand, acceleration is intimately related with non-inertial frames of reference, thus with non-Lorentzian systems of spacetime coordinates with non-constant metric, thus with potentially curved spacetime. In rough terms, this line of reasoning explains the connection of gravity with geometry.

Let us begin our discussion with an overview of the basic differences between SR and GR. These ideas will be refined in subsequent sections.

## 2. Special vs. General Relativity

According to both SR and GR it is impossible to define absolute velocity. However, in SR the acceleration continues to have an absolute meaning, in the sense that the physical condition of vanishing (equivalently, non-vanishing) acceleration is Lorentz-invariant. Moreover, in SR we assume the existence of an infinite set of *inertial frames of reference*, all being equivalent to each other with regard to expressing physical laws. As defined, an inertial frame has the property that, relative to it, a *free* particle (a particle subject to no net interaction) moves uniformly (with constant velocity, hence no acceleration). The *Lorentz transformation*, relating inertial frames, guarantees that zero (nonzero) acceleration in one frame means zero (nonzero) acceleration in any other frame [1].

No such preferred class of frames of reference exists in GR, where gravity is taken into account. Indeed, no inertial observer (regarded as a "free body") may exist in the Universe, given that everything possessing mass and/or energy is subject to the universal gravitational interaction.

Geometrically, the spacetime of SR is a *flat* space. It thus allows for a class of Lorentzian coordinate systems which correspond to inertial frames and in which the metric tensor has the standard constant matrix representation $\eta_{\mu\nu}= \pm\, \delta_{\mu\nu}$ or, in explicit diagonal-matrix form, $\eta=diag\,(1,-1,-1,-1)$. These coordinate systems, in turn, admit a special kind of *straight worldlines* representing inertial motions of free particles.

In GR the flat spacetime is replaced by a 4-dimensional *curved* Riemannian space in which there are no preferred coordinate systems. Specifically, there are no *global* Lorentzian frames admitting a globally constant metric. However, the *local flatness* of spacetime allows us to define Lorentzian coordinates in a sufficiently small neighborhood of any spacetime point, as well as a locally constant metric equal to the flat-space metric $\eta_{\mu\nu}$ . To see the physical meaning of these geometric ideas we must first state the *equivalence principle*.

## 3. The principle of equivalence

The principle of equivalence is at the heart of gravitational theory, both classical and relativistic. In this section we discuss the principle within the context of Newtonian gravitational theory. Our conclusions, however, can be readily extended to the relativistic (high-speed) domain.

First, some necessary definitions:

*Gravitational mass* ($m_G$): the "charge" of the gravitational interaction; that is, the physical entity that acts as the source of the interaction while at the same time being subject to it.

*Inertial mass* ($m_I$): the physical quantity entering Newton's law, $F=m_I a$ .

*Freely falling body:* a moving body subject to no forces other than gravity.

*Inertial observer (or frame) inside a gravitational field:* an observer (or frame) subject to non-gravitational forces that exactly counterbalance the gravitational force, thus preventing free fall of the observer (or frame). Example: a stationary observer on the surface of the Earth (assuming the latter to be an "inertial" frame).

Consider now a particle inside a gravitational field $\vec{g}(\vec{r})$, where $\vec{r}$ is the position vector relative to the origin of an inertial frame. The gravitational force on the particle at a given point is $\vec{F}_G(\vec{r}) = m_G \vec{g}(\vec{r})$, where $m_G$ is the gravitational mass of the particle. Assuming that no other forces (of non-gravitational origin) act on the particle, we have that, by Newton's law, $\vec{F}_G(\vec{r}) = m_I \vec{a}(\vec{r})$, where $m_I$ is the inertial mass and $\vec{a}(\vec{r})$ is the local acceleration of the particle. Hence,

$$m_I \vec{a}(\vec{r}) = m_G \vec{g}(\vec{r}) .$$

By postulating that

$$m_I = m_G \tag{1}$$

(an assumption supported by ample experimental evidence; see, e.g., [8]) we have:

$$\vec{a}(\vec{r}) = \vec{g}(\vec{r}) \tag{2}$$

This leads to an important conclusion:

> *All bodies at the same place in a gravitational field experience the same acceleration, equal to the field strength, regardless of their mass or internal composition.*

Relation (1) identifying gravitational with inertial mass expresses the *principle of equivalence* and constitutes one of the most important principles of gravitational theory. Relation (2) then implies an equivalence between *gravity* and *acceleration*. Let us see the physical implications of this equivalence.

An *inertial* observer whose local laboratory is placed in a gravitational field will observe that all bodies that are subject to no other forces except gravity (i.e., all *freely falling* bodies) experience a common acceleration in the laboratory. On the other hand, an *accelerated* observer (relative to an inertial frame) *away from any gravitational field* will notice that, relative to her frame, all *free* particles have a common acceleration. Thus, by assuming the validity of the equivalence principle (1), we come to the conclusion that a local observer cannot tell whether his frame is an inertial one placed inside a gravitational field, or an accelerating one away from any gravitational field. One might say that "gravity is traded for acceleration", as suggested by Eq. (2).

Strictly speaking, the equality (2) concerns a particular point in a gravitational field. If the field varies significantly from one point to another inside the observer's laboratory, then identical test particles will experience different accelerations at different points, which will indicate the presence of an inhomogeneous gravitational field. We will assume that the field varies little (i.e., is almost homogeneous) inside the observer's laboratory. Equivalently, the lab's dimensions are small compared to the distance over which the field varies appreciably so that *tidal effects* [8] become noticeable.

Assume now that the observer's laboratory and all particles inside it are in *free fall* under the sole action of gravity. Then, relative to an inertial frame, the lab and the particles will have a common acceleration equal to the field strength, hence the particles will not accelerate relative to the observer. The latter may thus assume that her lab is an "inertial" frame away from gravity, and all particles in the lab are "free" particles. *In a freely falling frame the effects of gravity are eliminated!* Again, the dimensions of the lab must be sufficiently small compared to the distance of appreciable change of the gravitational field.

We say that *a freely falling laboratory looks locally like an inertial frame of reference*. Of course, no *global* inertial-looking frame can be defined by any freely falling lab in an *inhomogeneous* gravitational field (the case of a globally homogeneous field is purely theoretical).

## 4. Curved spacetime

We now pass to the relativistic point of view. According to SR, an inertial Lorentz frame is equivalent to a system of spacetime coordinates

$$x^\mu \equiv (x^0, x^1, x^2, x^3) \equiv (ct, x, y, z)$$

with metric element (infinitesimal spacetime interval)

$$\begin{aligned} ds^2 = \eta_{\mu\nu}\, dx^\mu dx^\nu &= (dx^0)^2 - (dx^1)^2 - (dx^2)^2 - (dx^3)^2 \\ &= c^2 dt^2 - dx^2 - dy^2 - dz^2 \end{aligned} \tag{3}$$

where the standard summation convention of summing from 0 to 3 over repeated up and down indices has been used, and where $\eta_{\mu\nu}$ is an element of the constant matrix (playing the role of metric tensor in *Minkowski space*)

$$\eta \equiv [\eta_{\mu\nu}] = \begin{bmatrix} 1 & 0 & 0 & 0 \\ 0 & -1 & 0 & 0 \\ 0 & 0 & -1 & 0 \\ 0 & 0 & 0 & -1 \end{bmatrix} = diag\,(1,-1,-1,-1) \qquad (4)$$

The existence of a *global* (pseudo-)Euclidean system of coordinates and a *globally* constant metric tensor $\eta_{\mu\nu}$ is an indication that the spacetime of SR is *flat*.

Coordinate transformations $x^\mu \to x^{\mu\,\prime}$ between Lorentz frames are linear transformations of the form

$$x^{\mu\,\prime} = \Lambda^\mu{}_\nu\, x^\nu \qquad (5)$$

and are such that the metric element $ds^2$ is left invariant:

$$ds^2 = \eta_{\mu\nu}\, dx^\mu dx^\nu = \eta_{\mu\nu}\, dx^{\mu\,\prime} dx^{\nu\,\prime} \qquad (6)$$

It follows that the matrix $\Lambda \equiv [\Lambda^\mu{}_\nu]$ satisfies the relation

$$\Lambda^t \eta\, \Lambda = \eta \quad \Leftrightarrow \quad \Lambda^\mu{}_\lambda\, \eta_{\mu\nu}\, \Lambda^\nu{}_\rho = \eta_{\lambda\rho} \qquad (7)$$

(where $\Lambda^t$ is the transpose of the matrix $\Lambda$). The coordinate transformations defined by Eqs. (5)-(7) constitute the *Lorentz transformations* of SR (see, e.g., [1]).

Transforming from an inertial to an *accelerating* (thus non-inertial) frame is equivalent to transforming to general ("curvilinear") coordinates $\overline{x}^\mu$ that are functions of the $x^\lambda$: $\overline{x}^\mu \equiv \overline{x}^\mu(x^\lambda)$, where by $x^\lambda$ we collectively denote the $(x^0,\ldots,x^3)$. In terms of the new coordinates the spacetime interval is expressed as

$$ds^2 \equiv \eta_{\mu\nu}\, dx^\mu dx^\nu = g_{\mu\nu}(\overline{x}^\lambda)\, d\overline{x}^\mu d\overline{x}^\nu \qquad (8)$$

A transformation of this kind (as, more generally, any transformation relating non-inertial frames) is generally *nonlinear* (see Appendix). In flat spacetime a coordinate transformation may always transform a general metric $g_{\mu\nu}(\overline{x}^\lambda)$ back to the standard constant metric $\eta_{\mu\nu}$ of SR.

In the presence of gravity, only an approximately inertial frame may be *locally* defined in a sufficiently small neighborhood of any spacetime point by considering a freely falling frame, as discussed in Sec. 3. In geometrical terms, the flat spacetime of SR where gravity is absent is analogous to a Euclidean (here, pseudo-Euclidean) space in which a *global* Cartesian (here, Lorentzian) coordinate system and an associated constant metric can be defined. No global inertial frame and Lorentzian coordinate system can be defined, however, in spacetime when gravity is present. Accordingly, no global coordinate transformation from general to Lorentzian coordinates exists that will transform a coordinate-dependent metric $g_{\mu\nu}(\overline{x}^\lambda)$ to the constant metric $\eta_{\mu\nu}$ of SR at *all* spacetime points. This is analogous to the absence of a global Cartesian system of coordinates and a globally constant metric in an intrinsically *curved* space. We thus conclude that, *in the presence of gravity, spacetime must be treated as a curved space* that is only *locally flat*. (Cut a tiny patch off a big spherical surface. It will certainly look almost flat!)

## 5. Geodesics in curved spacetime

As follows from the principle of equivalence, Eq. (1), the trajectory of a freely falling particle (subject only to gravity) is independent of the mass or the internal composition of the particle and is related to the gravitational field itself. This field is determined by the distribution of matter and energy in space and time and, as we have seen, produces an *intrinsic curvature* of spacetime and thus determines the *geometry* of spacetime. It may thus be said that motion under gravity is related to the properties of spacetime itself.

Any freely falling particle follows a *geodesic* [8] of curved spacetime. This may be defined as a *"straightest possible path"* in spacetime, i.e., a curve whose direction is unchanging as one moves along it (in geometrical terms, the tangent vector to the curve is parallel-transported along the curve). Equivalently, given two spacetime points *A* and *B*, a geodesic connecting *A* with *B* is the unique path of extremal length[1] from *A* to *B*.

In particular, even light follows a geodesic of curved spacetime in a gravitational field. Indeed, since light carries energy (consisting of photons) it must also carry inertia, according to the equivalence between these two physical properties. Thus, by the equivalence principle of gravitation, light carries an effective "gravitational mass" and is thus affected by the distribution of matter and energy, hence by the geometry of spacetime.

In flat spacetime (i.e., away from gravity) geodesics are *straight* lines and correspond to inertial motions of free particles and propagation of light rays.

## 6. Local inertial frames

We have seen that, as a consequence of the principle of equivalence, it is possible to *locally* remove the effects of gravity. Moreover, a gravitational field is *locally* equivalent to an accelerated frame of reference. Let us take a closer look at these issues by first treating the problem in the framework of classical Newtonian theory.

Consider a particle of inertial mass $m_I$ and gravitational mass $m_G$ , placed inside a *uniform* gravitational field $\vec{g} = const.$ The gravitational force on the particle is equal to $m_G \vec{g}$ . The particle is also subject to a *non-gravitational* force of the form $\vec{F}'(\vec{R})$, where $\vec{R}$ is the position vector of the particle relative to the physical agent responsible for this force. The motion of the particle relative to an *inertial* frame of reference is determined by Newton's law:

$$m_I \frac{d^2\vec{r}}{dt^2} = m_G \vec{g} + \vec{F}'(\vec{R}) \ ,$$

where $\vec{r}$ is the position vector of the particle relative to the origin of the inertial frame. By postulating that $m_I = m_G \equiv m$ (equivalence principle) we have:

$$m \frac{d^2\vec{r}}{dt^2} = m\vec{g} + \vec{F}'(\vec{R}) \tag{9}$$

[1] Literally: maximum proper time [8].

We now shift to a *freely falling* reference frame inside the uniform gravitational field $\vec{g}$. According to Eq. (2), the frame (as any freely falling object) has acceleration equal to the field strength at any point. Thus the freely falling frame has constant acceleration $\vec{g}$ relative to the inertial frame, executing uniformly accelerated motion.

Let $\vec{r}'$ be the position vector of the particle *m* relative to the origin of the falling frame. The acceleration of *m* relative to this frame is equal to

$$\frac{d^2\vec{r}'}{dt^2} = \frac{d^2\vec{r}}{dt^2} - \vec{g} \ , \quad \text{so that} \quad \frac{d^2\vec{r}}{dt^2} = \frac{d^2\vec{r}'}{dt^2} + \vec{g} \ .$$

Combining this with (9) and eliminating $m\vec{g}$, we find:

$$m\frac{d^2\vec{r}'}{dt^2} = \vec{F}'(\vec{R}) \tag{10}$$

We observe that *the gravitational field has been "washed off" in the freely falling frame*.

Now, a particle *m* that is subject to no forces other than gravity will also be in free fall with acceleration $\vec{g}$ relative to the inertial frame, thus it will have no acceleration relative to the falling frame. Indeed, Eq. (10) with $\vec{F}'(\vec{R}) = 0$ yields

$$\frac{d^2\vec{r}'}{dt^2} = 0 \ ,$$

which means that the particle moves *uniformly* relative to the falling frame, thus describes a straight-line trajectory in that frame. We conclude that, from the point of view of an observer in the falling frame, the particle *m* appears to be a "free" particle moving with constant velocity inside an apparently "inertial" frame *away* from any gravitational field.

Moreover, *in the freely falling frame even light propagates in straight lines*, like any massive freely-falling particle. (As remarked previously, the consequences of the equivalence principle also affect light given that it carries energy, thus inertia, thus an effective gravitational "mass".) However, the path of light will be *curved* relative to an *inertial* frame located somewhere inside the gravitational field. In other words, an inertial observer will notice a *bending of light* due to the gravitational field (see Appendix).

As follows from the above discussion, when performing non-gravitational experiments in a laboratory we cannot tell whether our lab is a truly inertial frame, away from any gravitational field, or a freely falling frame inside a uniform gravitational field.

More generally, we can *locally* remove the effects of a gravitational field even if it is *non-uniform* by choosing a freely falling frame, provided that the size of this frame is small compared to the characteristic length over which the gravitational field varies appreciably.

Generalizing to relativity we may say that, even in the presence of gravity, the curved spacetime is *locally* almost indistinguishable from flat Minkowski space and the laws of Physics are seen to be the usual special relativistic laws in the absence of gravity. Hence, *at each point* of spacetime (or, in a small neighborhood of such a point) we can always define a system of Lorentzian coordinates corresponding to a freely falling frame. This inertial-looking frame is called *local inertial frame*. Of course, a *global* inertial frame and a *global* coordinate system covering the entire spacetime is an impossibility, except in the (unphysical) case of a perfectly uniform gravitational field.

Let us summarize: In the presence of gravity one may define a local inertial frame in a sufficiently small neighborhood of any spacetime point by considering a freely falling frame. In such a frame the non-gravitational laws of Physics are the standard ones of SR (or of classical Physics in the non-relativistic limit). This reflects the fact that spacetime is locally flat, although it is curved on a larger scale due to gravity, the effects of which can only be removed locally (in a small region of spacetime where the gravitational field can be considered almost constant in space and time).

## 7. Concluding remarks

A (generally inhomogeneous) gravitational field adds a geometrical structure to the flat spacetime of SR; namely, it endows spacetime with *curvature*. The spacetime of GR is thus a genuine Riemannian space [8] the specific properties of which are determined by the distribution of matter and energy in space and time.

Since *global* Lorentzian coordinates with a globally constant metric $\eta_{\mu\nu}$ cannot be defined in a curved spacetime, in GR one must allow for general spacetime coordinates $\overline{x}^{\mu}$ as well as a corresponding non-constant metric $g_{\mu\nu}(\overline{x}^{\lambda})$ that cannot be reduced to $\eta_{\mu\nu}$ by any global coordinate transformation, except at any particular spacetime point (or, at best, in a small neighborhood of such a point).

Moreover, the linear Lorentz transformations of SR, relating different Lorentzian systems of coordinates that correspond to genuinely inertial frames, must be replaced by general, *nonlinear* transformations of the $\overline{x}^{\mu}$. Accordingly, the Lorentz invariance of physical laws, assumed in the gravity-free flat space of SR, must be generalized to *general invariance* under the nonlinear coordinate transformations of GR. In order for this to be achieved, the equations expressing physical laws must be re-expressed in forms that comply with the aforementioned requirement of general invariance [8].

The existence of *local inertial frames*, corresponding to local Lorentzian coordinates, is a manifestation of the *local flatness* of a curved Riemannian spacetime. In the special case where a *global* transformation from general spacetime coordinates with non-constant metric, to Lorentzian coordinates with metric $\eta_{\mu\nu}$, exists, the spacetime is flat and gravity is absent. The general coordinates then simply correspond to an *accelerated* frame of reference, relative to the inertial frames allowed by SR.

Finally, the curvature, hence the geometry, of spacetime is determined by the general metric $g_{\mu\nu}(\overline{x}^{\lambda})$ and is dependent on the distribution of matter and energy in spacetime. The relationship between this distribution and the spacetime geometry is mathematically expressed by the *Einstein field equations* (see, e.g., [3,8]), a set of differential equations for the metric components.

# Appendix

## A. Worldline geometry

To simplify our analysis we treat the problem within the framework of classical Newtonian mechanics and Galilean relativity. Thus our "spacetime" will actually be the Newtonian one where time is absolute (frame-independent). Our general conclusions, however, will still be valid in SR (upon necessary modifications regarding the relativity of time).

*Proposition:* The worldline of a *free* particle in a spacetime diagram, as well as the trajectory of this particle in space, are straight lines relative to an inertial frame of reference.

*Proof:* We assume for simplicity that the motion of the free particle is confined to the $xy$-plane ($z$=0). Relative to an inertial frame the particle moves with constant velocity. The equations of motion are thus linear in time, of the form

$$x = \alpha t + \beta \,, \quad y = \gamma t + \delta \tag{A.1}$$

with constant $\alpha$, $\beta$. Relations (A.1) describe a straight worldline in a spacetime diagram. To find the equation of the trajectory we eliminate $t$ between the two equations in (A.1). The result is

$$y = \kappa x + \lambda \quad \text{where} \quad \kappa = \gamma/\alpha \,, \ \lambda = \delta - \beta\gamma/\alpha \,,$$

describing a straight-line path in the $xy$-plane.

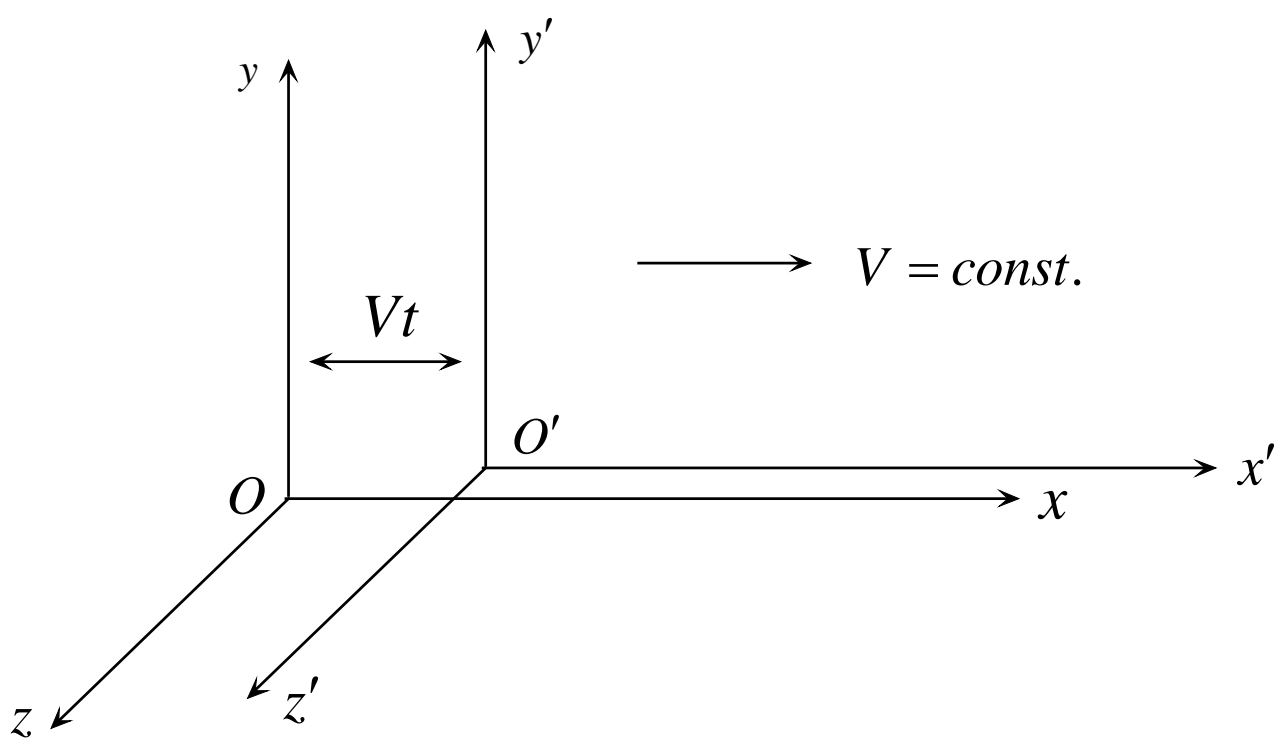

Fig. 1

Let $O$ be the origin of the inertial frame within which the equations of motion (A.1) apply. We consider a second inertial frame with origin $O'$, the axes of which frame are parallel to those of $O$ (see Fig. 1). The two systems of axes coincide at time $t$=0, while for $t$ >0 the frame $O'$ is moving with constant velocity $V$ relative to $O$ along the common $x$-axis of the two frames. The coordinate transformations relating the two frames are

$$x' = x - Vt \,, \quad y' = y \,, \quad t' = t \quad \Leftrightarrow \quad x = x' + Vt' \,, \quad y = y' \,, \quad t = t' \tag{A.2}$$

To find the equations of motion of the free particle relative to the frame $O'$, we substitute $x$, $y$ and $t$ from (A.2) into (A.1):

$$x' = (\alpha - V)t' + \beta\,, \quad y' = \gamma t' + \delta\,,$$

which are linear equations in $t'$. By eliminating $t'$ we can find the equation for the trajectory in the $x'y'$-plane, which is again of the form $y' = \kappa x' + \lambda$. Therefore, a straight-line path in the $O$-frame is transformed into a straight-line path in the $O'$-frame, both frames being inertial.

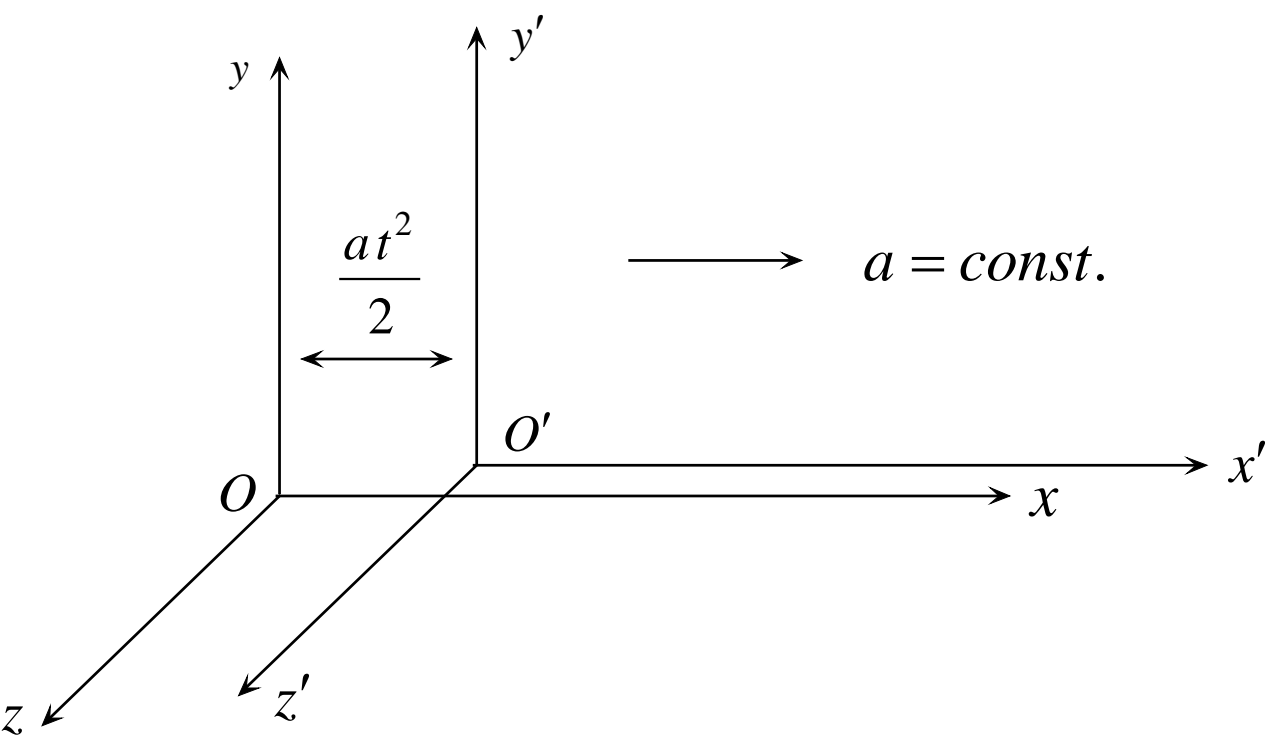

Fig. 2

We next consider the case where the frame $O'$ is accelerating in the $x$-direction relative to the inertial frame $O$, with constant acceleration $a$ (Fig. 2). For simplicity we assume that the initial velocity of $O'$ relative to $O$, at the moment $t$=0 when $O$ and $O'$ coincide, is zero. The coordinate transformations relating the two frames are

$$x' = x - at^2/2\,, \quad y' = y\,, \quad t' = t \quad \Leftrightarrow \quad x = x' + at'^2/2\,, \quad y = y'\,, \quad t = t' \qquad \text{(A.3)}$$

The equations of motion of the free particle relative to the non-inertial frame $O'$ are found by substituting $x$, $y$ and $t$ from (A.3) into (A.1):

$$x' = -\,at'^2/2 + \alpha t' + \beta\,, \quad y' = \gamma t' + \delta\,.$$

By eliminating $t'$ we can find the equation for the trajectory of the free particle in the $x'y'$-plane, which is an equation of a parabola:

$$x' = \kappa y'^2 + \lambda y' + \mu\,.$$

*Conclusion:* A straight path relative to an inertial frame appears *curved* in a non-inertial frame. Equivalently, a straight path in a non-inertial frame appears curved relative to an inertial frame.

The above remarks lead to an explanation of the effect of bending of light in a gravitational field.

## B. Bending of light by gravity

This can be viewed in two ways:

1. In an *inertial* frame *away from gravity*, light travels in straight lines. By the equivalence principle, the same must be true for a *freely falling* frame inside a gravitational field. Indeed, as mentioned previously, since light carries energy it has inertia and therefore possesses an effective "gravitational mass"; it is thus affected by gravity like any material particle. Now, as we saw in Sec. 6, any physical entity (including light) subject to no other interactions except gravity appears to travel in a straight-line path inside a freely falling lab.

On the other hand, relative to an *inertial* observer *inside* a gravitational field (i.e., someone who is *stationary* rather than free-falling), a path that seems straight in a falling lab will appear to be *curved*. Thus the stationary observer (e.g., one on the surface of the Earth) will conclude that the gravitational field has the effect of *bending* the path of light.

2. Again, light travels in straight lines in an inertial frame away from gravity. However, the path of light will appear to be *curved* relative to an *accelerated* frame. On the other hand, the accelerated frame *outside* gravity is locally equivalent to an inertial frame placed *inside* a gravitational field (cf. Sec. 3). Therefore, the observer in the latter frame will conclude that the gravitational field bends the path of light.